\documentclass[12pt,draftclsnofoot,journal,onecolumn]{IEEEtran}
\usepackage{mathtools}
\usepackage{ifpdf}
\usepackage{amsthm}
\usepackage{amsmath}
\usepackage{nicefrac}
\usepackage{cite}
\usepackage{amsfonts}
\usepackage{comment}
 \usepackage{url}
 \usepackage{amssymb}
 \usepackage[paper=letterpaper,margin=0.75in]{geometry}
 \usepackage[ansinew]{inputenc}
\usepackage{lipsum}
\usepackage{hyperref}

\newtheorem{manualtheoreminner}{Theorem}
\newenvironment{mthm}[1]{%
	\renewcommand\themanualtheoreminner{#1}%
	\manualtheoreminner
}{\endmanualtheoreminner}

\theoremstyle{definition}

\makeatletter
\renewenvironment{proof}[1][\proofname] {\par\pushQED{\qed}\normalfont\topsep6\p@\@plus6\p@\relax\trivlist\item[\hskip\labelsep\bfseries#1\@addpunct{.}]\ignorespaces}{\popQED\endtrivlist\@endpefalse}
\makeatother

\begin{document}
\title{Covert Wireless Communication with Artificial Noise Generation}

\author{
           Boulat Bash\IEEEauthorrefmark{2}, Dennis Goeckel\IEEEauthorrefmark{1}, Don Towsley\IEEEauthorrefmark{2}, and Saikat Guha\IEEEauthorrefmark{3}}
           
           \author{Ramin Soltani,~\IEEEmembership{Member,~IEEE,} 
                   Dennis~Goeckel,~\IEEEmembership{Fellow,~IEEE,}
                   Don~Towsley,~\IEEEmembership{Fellow,~IEEE,}
                   Boulat~A.~Bash,~\IEEEmembership{Member,~IEEE,} and~Saikat~Guha,~\IEEEmembership{Senior~Member,~IEEE}
                   %


\thanks{R.~Soltani is with the Electrical and Computer Engineering Department, University of Massachusetts, Amherst, MA (email: soltani@ecs.umass.edu).}%
\thanks{D.~Goeckel is with the Electrical and Computer Engineering Department, University of Massachusetts, Amherst, MA (email: goeckel@ecs.umass.edu).}%
\thanks{D.~Towsley is with the College of Information and Computer Sciences, University of Massachusetts, Amherst, MA (email: towsley@cs.umass.edu).}%
\thanks{B.~Bash is with Raytheon BBN Technologies, Cambridge, MA (email: boulat.bash@raytheon.com).} %
\thanks{S. Guha is with the College of Optical Sciences, University of Arizona, Tucson, AZ (email: saikat@email.arizona.edu).}
\thanks{This work has been supported, in part, by the National Science Foundation under grants CNS-1018464, ECCS-1309573, and CNS-1564067. The preliminary version of this work has been presented at the 52nd Annual Allerton Conference on Communication, Control, and Computing, Allerton, Monticello, IL, October 2014 \cite{soltani2014covert}.}
\thanks{© 2018 IEEE. Personal use of this material is permitted. Permission
	from IEEE must be obtained for all other uses, in any current or future
	media, including reprinting/republishing this material for advertising or
	promotional purposes, creating new collective works, for resale or
	redistribution to servers or lists, or reuse of any copyrighted
	component of this work in other works. DOI: 10.1109/TWC.2018.2865946}
}

\date{}
\maketitle
\thispagestyle{plain}
\pagestyle{plain}
\newtheorem{definition}{Definition}

\begin{abstract}
Covert communication conceals the transmission of the message from an attentive adversary. Recent work on the limits of covert communication in additive white Gaussian noise (AWGN) channels has demonstrated that a covert transmitter (Alice) can reliably transmit a maximum of $\mathcal{O}\left(\sqrt{n}\right)$ bits to a covert receiver (Bob) without being detected by an adversary (Warden Willie) in $n$ channel uses. This paper focuses on the scenario where other ``friendly’’ nodes distributed according to a two-dimensional Poisson point process with density $m$ are present. We propose a strategy where the friendly node closest to the adversary, without close coordination with Alice, produces artificial noise. We show that this method allows Alice to reliably and covertly send $\mathcal{O}(\min\{{n,m^{\gamma/2}\sqrt{n}}\})$ bits to Bob in $n$ channel uses, where $\gamma$ is the path-loss exponent. We also consider a setting where there are $N_{\mathrm{w}}$ collaborating adversaries uniformly and randomly located in the environment and show that in $n$ channel uses, Alice can reliably and covertly send $\mathcal{O}\left(\min\left\{n,\frac{m^{\gamma/2} \sqrt{n}}{N_{\mathrm{w}}^{\gamma}}\right\}\right)$ bits to Bob when $\gamma >2$, and $\mathcal{O}\left(\min\left\{n,\frac{m \sqrt{n}}{N_{\mathrm{w}}^{2}\log^2 {N_{\mathrm{w}}}}\right\}\right)$ when $\gamma = 2$. Conversely, we demonstrate that no higher covert throughput is possible for $\gamma>2$.

\end{abstract}


\section{Introduction}
Covert communication hides the presence of a message from a watchful adversary. This is crucial in scenarios in which the standard method of secrecy, which hides the  message content but not its existence, is not enough; in other words, there are applications where, no matter how strongly the message is protected from being deciphered, the adversary discerning that the communication is taking place results in penalties to the users. Examples of such scenarios include military operations, social unrest, and tracking of people’s daily activities. The Snowden disclosures~\cite{snowden} demonstrate the utility of ``meta-data'' to an observing party and, thus, motivate hiding the presence of the message.

The provisioning of security and privacy has emerged as a critical issue in communication systems~\cite{nichols2001wireless,lopez2008wireless,miller2001facing,arbaugh2003wireless,hadian2016privacy,hadian2018privacy
	,takbiri2017limits
	,takbiri2017fundamental
}. In wireless communications where the signal is not restricted physically to a wire, it is more difficult to hide the existence of the communication. Although spread spectrum approaches have been widely used in the past \cite{simon94ssh}, the fundamental limits of covert communication were only recently established by a subset of the authors \cite{bash_isit2012, bash_jsac2013}, who presented a square root limit on the number of bits that can be transmitted securely from the transmitter (Alice) to the intended receiver (Bob) when there is an additive white Gaussian noise (AWGN) channel between Alice and each of Bob and the adversary (Warden Willie). In particular, by taking advantage of positive noise power at Willie, Alice can reliably transmit $\mathcal{O}(\sqrt{n})$ bits to Bob in $n$ channel uses while lower bounding Willie's error probability $\mathbb{P}_{\mathrm e}^{(\mathrm w)}=\frac{\mathbb{P}_{\mathrm {{FA}}}+\mathbb{P}_{\mathrm {MD}}}{2}\geq \frac{1}{2}-\epsilon $ for any $0<\epsilon<\frac{1}{2}$ where $\mathbb{P}_{\mathrm {{FA}}}$ is the probability of false alarm and $\mathbb{P}_{\mathrm {MD}}$ is the probability of mis-detection. Conversely, if Alice transmits $\omega(\sqrt{n})$ bits in $n$ uses of channel, either Willie detects her or Bob suffers a non-zero probability of decoding error as $n$ goes to infinity. Covert communications recently has been studied in many scenarios such as binary symmetric channels (BSCs)~\cite{jaggi_isit2013}, multi-path noiseless networks~\cite{kadhe2014reliable}, bosonic channels with thermal noise~\cite{ bash_isit2013}, and noisy discrete memoryless channels (DMCs)~\cite{hou2014effective}. Furthermore, higher throughputs are achievable when Alice can leverage Willie's ignorance of her transmission time~\cite{bash_isit2014}, and/or the adversary has uncertainty about channel characteristics~\cite{sobers2017covert}. These works, along with~\cite{bash2015hiding,bloch2016covert}, present a comprehensive characterization of the fundamental limits of covert communications over DMC and AWGN channels and have also motivated studying the fundamental limits of covert techniques for packet channels~\cite{soltani2015covert,soltani2016allerton} and invisible de-anonymization of network flows~\cite{soltani2017towards}.

In this paper, we take necessary steps to answer this question: what is the throughput of covert communication in wireless networks? In particular, we present a single-hop covert communication scheme which can be embedded into a large wireless network to extend the capacity of {\em overt} communication in large wireless networks~\cite{gupta_kumar,francheschetti} to {\em covert} communication. The goal is to establish an analog to the line of work on scalable low probability of intercept communications \cite{jsac_goeckel, vasudevan2010security,capar_infocom,capar_ciss}, which considered the extension of~\cite{gupta_kumar, francheschetti} to the {\em secure} multipair unicast problem in large wireless networks. Here, in analog to~\cite{bash_jsac2013}, we calculate the throughput of single-hop covert communication in the presence of a number of other network nodes: 1) warden Willies which decrease the throughout; 2) friendly nodes which can be employed to increase the throughput. In this paper, we enhance the   throughout of covert communication assuming that Willie knows his channel characteristics, as opposed to~\cite{sobers2017covert} where the throughput of the covert communication is improved by leveraging Willie's ignorance of the channel characteristics in a fading environment or when a jammer with varying power is present.

Assume Alice attempts to communicate covertly with Bob without detection by Willie, but also in the presence of other (friendly) network nodes, which can assist the communication by producing background chatter to inhibit Willie's ability to detect Alice's transmission. We model the locations of the friendly nodes by a two-dimensional Poisson point process of density $m$, and that Alice and Bob share a secret (codebook) unknown to Willie. For this scenario, described in more detail in Section~\ref{prerequisites}, we show in Section~\ref{sec:swillie} that Alice is able to covertly transmit $\mathcal{O}(\min{\{n,m^{\gamma/2} \sqrt{n}\}})$ bits to Bob in $n$ channel uses while keeping Willie's error probability $\mathbb{P}_{\mathrm e}^{(\mathrm w)} \geq \frac{1}{2}-\epsilon$ for any $\epsilon \geq 0$, where $\gamma$ is the path-loss exponent. The construction that enables such a covert throughput is to switch on the closest friendly node to Willie. Conversely, without any restriction on the algorithm for turning on friendly nodes, we show that if Alice attempts to transmit $\omega(m^{\gamma/2} \sqrt{n})$ bits to Bob in $n$ channel uses, there exists a detector that Willie can use to either detect her with arbitrarily low error probability $\mathbb{P}_{\mathrm e}^{(\mathrm w)}$ or prevent Bob from decoding the message with arbitrarily low probability of error. 

Next, we extend the scenario to the case where of multiple Willies, and we show that when $N_{\mathrm{w}}$ collaborating Willies are uniformly and independently distributed in the unit box  (see Fig.~\ref{fig:SysMod}), 
 we can still turn on the closest friendly node to each Willie to improve the covert throughput. However, as $N_{\mathrm{w}} \to \infty$, we observe two effects that reduce the covert throughput: (1) with high probability, there exists a Willie very close to Alice who receives a high signal power from her, thus making Alice employ a lower power to hide the transmission; (2) with high probability, there exists a Willie very close to Bob whose closest friendly node generates additional noise for Bob, hence reducing his ability to decode Alice's message. We explore this scenario in Section~\ref{sec:mwillie} in detail. Finally, we discuss the results in Section~\ref{sec:dis} and present conclusions in Section~\ref{sec:con}.

\section{System Model, Definitions, and Metrics}
\label{prerequisites}

\subsection{System Model}
\begin{figure}
\begin{center}
 \includegraphics[ 
scale=0.6]{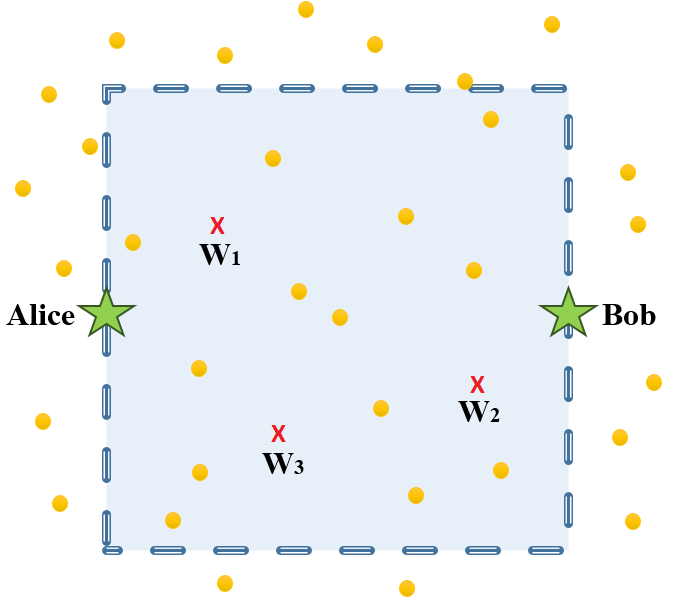}
\end{center}
 \caption{System Configuration: Source node Alice wishes to communicate reliably and without detection to the intended receiver Bob at distance one (normalized) with the assistance of friendly nodes (represented by yellow nodes in the figure) distributed according to a two-dimensional Poisson point process with density $m$  in the presence of adversary nodes  $W_1, W_2, \dots, W_{N_{\mathrm{w}}}$ located in the dashed box ($N_{\mathrm{w}} = 3$ in the figure).}
 \label{fig:SysMod}
 \end{figure}

Consider a source Alice ($A$) wishing to communicate with receiver Bob ($B$) located a unit distance away from her in the presence of adversaries (Warden Willies) $W_1, W_2, \dots, W_{N_{\mathrm{w}}}$, who are distributed independently and uniformly in the unit square (Fig.~\ref{fig:SysMod}) and seek to detect any transmission by Alice. When there is only a single Willie, we omit the subscript and denote it by $\mathrm W$. Also present are friendly nodes $F_1, F_2,\ldots$ allied with Alice and Bob, who help hide Alice's transmission by generating noise. We model the locations of friendly nodes by a two-dimensional Poisson point process with density $m$. The adversaries try to detect whether Alice transmits or not by processing the signals they receive and applying hypothesis testing on them, as discussed in the next subsection. We consider two scenarios: a single Willie ($N_{\mathrm{w}}=1$) and multiple Willies ($N_{\mathrm{w}}>1$). We assume all channels are discrete-time AWGN with real-valued symbols. Alice transmits $n$ real-valued symbols $s_1, s_2, \ldots, s_n$ that are samples of zero-mean Gaussian distribution with variance $P_{\mathrm a}$. Each friendly node is either on or off according to the strategy employed. Let $\theta_j$ denote the state of the $j^{\text{th}}$ friendly node $F_j$; $\theta_j=1$ if $F_j$ is ``on'' (transmits noise) and $\theta_j=0$ (silent) otherwise. If $F_j$ is on, it transmits symbols $\left\{s_i^{(j)} \right\}_{i=1}^{\infty}$, where $\left\{s_i^{(j)} \right\}_{i=1}^{\infty}$ is a collection of independent and identically distributed (i.i.d.) zero-mean Gaussian random variables, each with variance (power) $P_{\mathrm{f}}$. Denote by $\mathcal{J}$ the set of friendly nodes, and by $\mathcal{J}^{\dagger}$ the set of friendly nodes that are on. The locations of all the parties are static and known to everyone. One implication of this assumption is that friendly nodes can determine which friendly node is the closest to each Willie.

Recalling that the distance between Alice and Bob is normalized to unity, 
Bob receives $y_1^{({\mathrm b})}, y_2^{({\mathrm b})}, \ldots, y_n^{(b)}$ where $y_i^{({\mathrm b})} = s_i+z_i^{({\mathrm b})}$ for $1\leq i \leq n$. The noise component is $z_i^{({\mathrm b})}= z_{i,0}^{({\mathrm b})}+\sum_{j=1}^\infty \theta_j   z_{i,j}^{({\mathrm b})}$, where 
$\left\{z_{i,0}^{({\mathrm b})}\right\}_{i=1}^{n}$ is an i.i.d.~sequence representing the background noise of Bob's receiver with $z_{i,0}^{(\mathrm{b})} \sim \mathcal{N}(0,\sigma_{{\mathrm b},0}^2)$ for all $i$, and $\left\{z_{i,j}^{({\mathrm b})}\right\}_{i=1}^{n}$ is an i.i.d.~sequence of zero-mean Gaussian random variables characterizing the chatter from the $j^{\text{th}}$ friendly node when it is ``on'', 
each element of the sequence with variance $\frac{P_{\mathrm{f}}}{d_{{\mathrm b},{\mathrm f}_j}^\gamma}$, where $d_{x,y}$ is the distance between nodes $X$ and $Y$, and $\gamma$ is the path-loss exponent which in most practical cases satisfies $2\leq\gamma\leq4$.

Similarly, the $k^{\text{th}}$ Willie observes  $y_1^{(k)}, y_2^{(k)}, \ldots, y_n^{(k)}$ where  $y_i^{(k)} = \frac{s_i}{d_{{\mathrm a},{\mathrm w}_k}^{\gamma/2}}+z_i^{(k)}$. Here, $z_i^{(k)}= z_{i,0}^{({k})}+\sum_{j=1}^\infty \theta_j z_{i,j}^{({k})}$ where $\left\{z_{i,0}^{(k)}\right\}_{i=1}^{n}$ is an i.i.d.~sequence representing the background noise at Willie's receiver, where $z_{i,0}^{(k)} \sim \mathcal{N}(0,\sigma_{{{\mathrm w}_{k,0}}}^2)$ for all $i$, and $\left\{z_{i,j}^{({k})}\right\}_{i=1}^{n}$ is an i.i.d.~sequence characterizing the chatter from the $j^{\text{th}}$ friendly node when it is ``on''; thus, $\mathcal{N} (0,{P_{\mathrm{f}}}/{d_{{\mathrm w}_k,{\mathrm f}_j}^\gamma})$. For a single Willie scenario, we omit the superscripts on $y_i^{(k)}$, $z_i^{(k)}$, and $z_{i,j}^{(k)}$, and we denote the Willie by $\mathrm W$, and the closest friendly node to Willie by $F$.

We assume Alice and the friendly nodes, while having a common goal, are not able to synchronize their transmissions; that is, the friendly nodes set up a constant power background chatter but are not able to, for example, lower their power at the time Alice transmits. In~\cite{sobers2017covert}, the assumption is that a single jammer with varying power is present or the channel fading leads to uncertainty in Willie's received power when Alice is not transmitting. Such uncertainty is not present here.

In this paper, the density of friendly nodes $m$ and the number of adversaries $N_{\mathrm{W}}$ are functions of the number of channel uses $n$, and $\gamma$ is a constant independent of $n$.
\subsection{Definitions}

Willie's hypotheses are $H_0$ (Alice does not transmit) and $H_1$ (Alice transmits). The parameters that determine Willie's error probabilities (type I and type II errors) are his distance to Alice $d_{\mathrm{a},\mathrm{w}}$ and his noise power $\sigma_{\mathrm{w}}^2$, which are random variables dependent on the locations of the friendly nodes and Willie(s). For given locations of the friendly nodes and Willie, we denote by $\mathbb{P}_{\mathrm {FA}}(\sigma_{\mathrm{w}}^2,d_{\mathrm{a},\mathrm{w}})$ the probability of rejecting $H_0$ when it is true (type I error or false alarm), and $\mathbb{P}_{\mathrm {MD}}(\sigma_{\mathrm{w}}^2,d_{\mathrm{a},\mathrm{w}})$ the probability of rejecting $H_1$ when it is true (type II error or mis-detection). Assuming equal prior probabilities, Willie's error probability given the locations of friendly nodes and Willie(s) is $\mathbb{P}_{\mathrm e}^{(\mathrm w)}(\sigma_{\mathrm{w}}^2,d_{\mathrm{a},\mathrm{w}})=\frac{\mathbb{P}_{\mathrm {FA}}(\sigma_{\mathrm{w}}^2,d_{\mathrm{a},\mathrm{w}}) + \mathbb{P}_{\mathrm {MD}}(\sigma_{\mathrm{w}}^2,d_{\mathrm{a},\mathrm{w}})}{2}$. Willie's type I error, type II error, and probability of error are $\mathbb{P}_{\mathrm {FA}}=\mathbb{E}_{{\mathrm F},{\mathrm W}}\left[\mathbb{P}_{\mathrm {FA}}(\sigma_{\mathrm{w}}^2,d_{\mathrm{a},\mathrm{w}})\right]$, $\mathbb{P}_{\mathrm {MD}}=\mathbb{E}_{{\mathrm F},{\mathrm W}}\left[\mathbb{P}_{\mathrm {MD}}(\sigma_{\mathrm{w}}^2,d_{\mathrm{a},\mathrm{w}})\right]$, and $\mathbb{P}_{\mathrm {e}}^{(\mathrm{w})}=\mathbb{E}_{{\mathrm F},{\mathrm W}}\left[\mathbb{P}_{\mathrm {e}}^{(\mathrm{w})}(\sigma_{\mathrm{w}}^2,d_{\mathrm{a},\mathrm{w}})\right]$, respectively, where $\mathbb{E}_{{\mathrm F},{\mathrm W}}\left[\cdot\right]$ denotes the expectation with respect to the locations of the friendly nodes as well as those of the Willie(s).

We assume that Willie uses classical hypothesis testing and seeks to minimize his probability of error, $\mathbb{P}_{\mathrm e}^{(\mathrm w)}$. The generalization to arbitrarily prior probabilities is available in ~\cite[Section V.B]{bash_jsac2013}.

When there is only a single Willie in the scenario, he applies a hypothesis test to his received signal to determine whether or not Alice is communicating with Bob. For given locations of the friendly nodes and Willie, we denote the probability distribution of Willie's ($W_k$) collection of observations $\left\{y_i^{(k)}\right\}_{i=1}^{n}$ by $\mathbb{P}_1(\sigma_{\mathrm{w}}^2,d_{\mathrm{a},\mathrm{w}})$ when Alice is communicating with Bob, and the distribution of the observations when she does not transmit by $\mathbb{P}_0(\sigma_{\mathrm{w}}^2)$. For a scenario with multiple collaborating Willies (Theorems~\ref{th:mwillie2p} and~\ref{th:mwillie2}), they jointly process the signals they receive to arrive at a single collective decision as to whether Alice transmits or not. In this case, we use $\mathbb{P}_{\mathrm e}^{(\mathrm w)}(\boldsymbol{\sigma_{\mathrm{w}}^2},\boldsymbol{d_{\mathrm{a},\mathrm{w}}}), {\mathbb{P}_{\mathrm {FA}}(\boldsymbol{\sigma_{\mathrm{w}}^2},\boldsymbol{d_{\mathrm{a},\mathrm{w}}}), \mathbb{P}_{\mathrm {MD}}(\boldsymbol{\sigma_{\mathrm{w}}^2},\boldsymbol{d_{\mathrm{a},\mathrm{w}}})},\mathbb{P}_1(\boldsymbol{\sigma_{\mathrm{w}}^2},\boldsymbol{d_{\mathrm{a},\mathrm{w}}})$, and
$\mathbb{P}_0(\boldsymbol{\sigma_{\mathrm{w}}^2})$, where $\boldsymbol{\sigma_{\mathrm{w}}^2}$ and $\boldsymbol{d_{\mathrm{a},\mathrm{w}}}$ are vectors containing  $\sigma_{\mathrm{w}_k}^2$ and $d_{\mathrm{a},\mathrm{w}_k}$, respectively. 

\begin{definition}\label{def:cov} (Covertness) Alice's transmission is covert if and only if she can lower bound Willies' probability  of error ($\mathbb{P}_{\mathrm e}^{(\mathrm w)}=\mathbb{E}_{{\mathrm F},{\mathrm W}}\left[\mathbb{P}_{\mathrm e}^{(\mathrm w)}(\sigma_{\mathrm{w}}^2,d_{\mathrm{a},\mathrm{w}})\right]=\frac{\mathbb{E}_{{\mathrm F},{\mathrm W}}\left[\mathbb{P}_{\mathrm {FA}}(\sigma_{\mathrm{w}}^2,d_{\mathrm{a},\mathrm{w}})+\mathbb{P}_{\mathrm {MD}}(\sigma_{\mathrm{w}}^2,d_{\mathrm{a},\mathrm{w}})\right]}{2}$) by $\frac{1}{2}-\epsilon$ for any $\epsilon>0$, asymptotically~\cite{bash_jsac2013}. The expectation is with respect to the locations of the friendly nodes as well as those of the Willie(s).
\end{definition}

Bob's probability of error depends on his noise power $\sigma_{\mathrm{b}}^2$ which is a random variable dependent on the locations of Willie and friendly nodes. Denote by $\mathbb{P}_{\mathrm e}^{(\mathrm b)}(\sigma_{\mathrm{b}}^2)$ Bob's probability of error for given locations of the friendly nodes and Willie. 
\begin{definition} \label{def:rel} (Reliability) Alice's transmission is reliable if and only if the desired receiver (Bob) can decode her message with arbitrarily low probability of error $\mathbb{P}_{\mathrm e}^{(\mathrm b)}=\mathbb{E}_{{\mathrm F},{\mathrm W}}\left[\mathbb{P}_{\mathrm e}^{(\mathrm b)}(\sigma_{\mathrm{b}}^2)\right]$ at long block lengths. In other words, for any $ \zeta>0$, Bob can achieve $\mathbb{P}_{\mathrm e}^{(\mathrm b)} < \zeta$ as $n \to \infty$.
\end{definition}

In this paper, we use standard Big-O, Little-O, Big-Omega, Little-Omega, and Theta notations~\cite[Ch.~3]{cormen2009introduction}.

\section{Single Warden Scenario}
 
\label{sec:swillie}
In this section, we consider the case where there is only one Willie ($\mathrm W$) located uniformly and randomly on the unit square shown as a dashed box in Fig.~\ref{fig:SysMod}.
We present Theorem~\ref{th:swillie} for $\gamma>2$ in Section~\ref{sec:sgamma2}, and Theorem~\ref{th:swillie2} for $\gamma=2$ in Section~\ref{sec:sgammab2}. We show that Alice is able to covertly transmit $\mathcal{O}(\min\{{n,m^{\gamma/2}\sqrt{n}}\})$ bits to Bob in $n$ channel uses. The construction that enables such a covert throughput is to turn on the closest friendly node to Willie to hide the presence of Alice's transmission. To achieve $\mathbb{P}_{\mathrm e}^{(\mathrm w)} \geq \frac{1}{2}-\epsilon$, Alice transmits codewords with power $P_{\mathrm a}$ which depends on the covertness parameter $\epsilon$. The achievability proof concludes by considering the rate at which reliable decoding is still possible when Alice uses the maximum possible power. In Theorem~\ref{th:swillie}, we present a converse independent of the status of the friendly nodes (being on or off), and in Theorem~\ref{th:swillie2}, we present a converse assuming the closest friendly node to Willie is on.

\subsection{Single Warden Scenario and $\gamma>2$} \label{sec:sgamma2}
\begin{mthm}{1.1} \label{th:swillie}
When there is one warden (Willie) located randomly and uniformly over the unit square
, $m>0$, and $\gamma>0$, Alice can reliably and covertly transmit 
$\mathcal{O}(\min\{{n,m^{\gamma/2}\sqrt{n}}\})$ bits to Bob in $n$ channel uses. Conversely, if Alice attempts to transmit $\omega(m^{\gamma/2} \sqrt{n}) $ bits to Bob in $n$ channel uses, there exists a detector that Willie can use to either detect her with arbitrarily low error probability $\mathbb{P}_{\mathrm e}^{(\mathrm w)}$ or Bob cannot decode the message with arbitrarily low error probability $\mathbb{P}_{\mathrm e}^{(\mathrm b)}$.
\end{mthm}

\begin{proof}

{\it (Achievability)} 

\textbf{Construction}: Alice and Bob share a codebook that is not revealed to Willie. For each message transmission of length $L$ bits, Alice uses a new codebook to encode the message into a codeword of length $n$ at rate $R=\frac{L}{n}$. To build a codebook, we use random coding arguments; that is, codewords $\left\{C(M_l)\right\}_{l=1}^{2^{nR}}$ are associated with messages $\left\{M_l\right\}_{l=1}^{2^{nR}}$, where each codeword $C(M_l)=\{C^{(u)}(M_l)\}_{u=1}^{n}$, for $l=\left\{1,2,\cdots,2^{nR}\right\}$, is an i.i.d. zero-mean Gaussian random sequence; that is, $C^{(u)}(M_l) \sim \mathcal{N}(0,P_{\mathrm{a}})$ where $P_{\mathrm{a}}$ is specified later. Bob employs a maximum-likelihood (ML) decoder to process his observations $\{y_i^{\mathrm (b)}\}_{i=1}^{n}$~\cite{lin1983error}. The decoder picks a codeword $\widehat{C} $ that maximizes $\mathbb{P}(\{y_i^{\mathrm (b)}\}_{i=1}^{n}|\widehat{C})$, i.e., the probability that $\{y_i^{\mathrm (b)}\}_{i=1}^{n}$ was received, given that $\widehat{C}$ was sent.

Alice and Bob turn on the closest friendly node to Willie and keep all other friendly nodes off, whether Alice transmits or not.  Therefore, Willie's observed noise power is given by
\begin{equation}
\nonumber {\sigma_{\mathrm w}^2} = {\sigma_{{\mathrm w},0}^2} + \frac{P_{\mathrm{f}}}{d_{{\mathrm w},{\mathrm f}}^{\gamma}} ,
\end{equation}

\noindent where $\sigma_{{\mathrm w},0}^2$ is Willie's noise power when none of the friendly nodes are transmitting and $d_{{\mathrm w},{\mathrm f}}$ is the (random) distance between Willie and the closest friendly node to him; hence, ${\sigma_{\mathrm w}^2}$ is a random variable that depends on the locations of the friendly nodes.

\textbf{Analysis}: ({\em Covertness}) First, we analyze Willie's error probability conditioned on $\sigma_{\mathrm w}^2$ and $d_{\mathrm{a},\mathrm{w}}$, $\mathbb{P}_{\mathrm e}^{(\mathrm w)}(\sigma_{\mathrm w}^2,d_{\mathrm{a},\mathrm{w}})$, where $d_{{\mathrm a},{\mathrm w}}$ is the distance between Willie to Alice. Then, we lower bound Willie's error probability $ \mathbb{P}_{\mathrm e}^{(\mathrm w)} =  \mathbb{E}_{{\mathrm F},{\mathrm W}}[\mathbb{P}_{\mathrm e}^{(\mathrm b)}(\sigma_{\mathrm w}^2),d_{\mathrm{a},\mathrm{w}}]$. Recall that for given locations of the friendly nodes and Willie, $\mathbb{P}_0(\sigma_{\mathrm{w}}^2)$ is the joint probability density function (pdf) for Willie's observations 
 under the null hypothesis $H_0$ (Alice does not transmit), and $\mathbb{P}_1(\sigma_{\mathrm{w}}^2,d_{\mathrm{a},\mathrm{w}})$ be the joint pdf for corresponding observations under the hypothesis $H_1$ (Alice transmits). Observe
\begin{align}
\nonumber \mathbb{P}_0(\sigma_{\mathrm{w}}^2) &= \mathbb{P}_{\mathrm w}^n(\sigma_{\mathrm{w}}^2),\\
\nonumber \mathbb{P}_1(\sigma_{\mathrm{w}}^2,d_{\mathrm{a},\mathrm{w}}) &= \mathbb{P}_{\mathrm s}^n(\sigma_{\mathrm{w}}^2,d_{\mathrm{a},\mathrm{w}}),
\end{align}
\noindent where $\mathbb{P}_{\mathrm w}(\sigma_{\mathrm{w}}^2) = \mathcal{N}(0,\sigma_{\mathrm w}^2)$ is the pdf for each of Willie's observations when Alice does not transmit, for given locations of friendly nodes and Willie , and $\mathbb{P}_{\mathrm s}(\sigma_{\mathrm{w}}^2,d_{\mathrm{a},\mathrm{w}})=\mathcal{N}\left(0,\sigma_{\mathrm w}^2+\frac{P_{\mathrm{a}}}{d_{{\mathrm a},{\mathrm w}}^{\gamma}}\right)$ is the pdf for each of the corresponding observations when Alice transmits. When Willie applies the optimal hypothesis test to minimize $\mathbb{P}_{\mathrm e}^{(\mathrm w)}(\sigma_{\mathrm{w}}^2,d_{\mathrm{a},\mathrm{w}})$ \cite{bash_jsac2013}:
\begin{align} 
\label{eq:0} \mathbb{P}_{\mathrm e}^{(\mathrm w)}(\sigma_{\mathrm{w}}^2,d_{\mathrm{a},\mathrm{w}}) \geq \frac{1}{2}- \sqrt{\frac{1}{8} \mathcal{D}(\mathbb{P}_1(\sigma_{\mathrm{w}}^2,d_{\mathrm{a},\mathrm{w}}) || \mathbb{P}_0(\sigma_{\mathrm{w}}^2))},
\end{align}
\noindent where $\mathcal{D}(f(x) || g(x))$ is the relative entropy between pdfs $f(x)$ and  $g(x)$. For the given $\mathbb{P}_0$ and $\mathbb{P}_1$ \cite{bash_jsac2013}: 
\begin{align}
\label{eq:thm1d2} \mathcal{D}(\mathbb{P}_1(\sigma_{\mathrm{w}}^2,d_{\mathrm{a},\mathrm{w}}) || \mathbb{P}_0(\sigma_{\mathrm{w}}^2))
  = {n \over 2}\left(   \frac{P_{\mathrm{a}}}{d_{{\mathrm a},{\mathrm w}}^{\gamma} \sigma_{{\mathrm w}}^2} - \ln{\left(1+ \frac{P_{\mathrm{a}}}{d_{{\mathrm a},{\mathrm w}}^{\gamma} \sigma_{{\mathrm w}}^2}\right)} \right)
 \leq  n \left(\frac{P_{\mathrm{a}}}{2 d_{{\mathrm a},{\mathrm w}}^{\gamma}\sigma_{\mathrm w}^2}\right)^2,
\end{align}
\noindent where the last inequality follows from (see the  Appendix~\ref{ap.01})
\begin{align}\label{eq:ineq0}
\ln(1+x) \geq x-\frac{x^2}{2}, \text{ for } x\geq 0.
\end{align}
\noindent By~\eqref{eq:0} and~\eqref{eq:thm1d2}
\begin{align} 
  & \label{eq:expcond1}  \mathbb{P}_{\mathrm e}^{(\mathrm w)}(\sigma_{\mathrm{w}}^2,d_{\mathrm{a},\mathrm{w}})   \geq  \frac{1}{2}- \sqrt{\frac{n}{8}} \frac{P_{\mathrm{a}}}{2 \sigma_{\mathrm w}^2 d_{{\mathrm a},{\mathrm w}}^{\gamma}}.
\end{align}
\noindent If Alice sets her average symbol power 
\begin{align}
\label{eq:power} P_{\mathrm{a}} \leq \frac{c m^{\gamma/2}}{\sqrt{n}},
\end{align}
\noindent where $c = {\epsilon} \left(\frac{\Gamma \left(\gamma/2+1\right)}{4 \sqrt{2}  \psi^{\gamma} P_{\mathrm{f}} \pi^{\gamma/2+1}} \right)^{-1}$ is a constant independent of $n$, $\Gamma(\cdot)$ is the Gamma function, and $\psi =\sqrt{\frac{\epsilon}{2\pi}}$, then~\eqref{eq:expcond1} yields
\begin{align}
& \label{eq:expcond13}  \mathbb{P}_{\mathrm e}^{(\mathrm w)}(\sigma_{\mathrm{w}}^2,d_{\mathrm{a},\mathrm{w}})   \geq  \frac{1}{2}- \sqrt{\frac{1}{8}} \frac{ c m^{\gamma/2}}{2 \sigma_{\mathrm w}^2 d_{{\mathrm a},{\mathrm w}}^{\gamma}}. 
\end{align}
Denote by $\mathbb{E}_{{\mathrm F},{\mathrm W}}\left[\cdot\right]$ the expectation over locations of the friendly nodes ($F_1,F_2,\ldots)$, and the location of Willie ($\mathrm W$). Next, we lower bound $\mathbb{P}_{\mathrm e}^{(\mathrm w)}=\mathbb{E}_{{\mathrm F},{\mathrm W}}\left[\mathbb{P}_{\mathrm e}^{(\mathrm w)}
(\sigma_{\mathrm{w}}^2,d_{\mathrm{a},\mathrm{w}})\right]$. Note that~\eqref{eq:expcond13} contains a singularity at $d_{{\mathrm a},{\mathrm w}}=0$; however, since it occurs with probability measure zero, we can easily show that $\mathbb{E}_{{\mathrm F},{\mathrm W}}\left[\frac{1}{2}- \sqrt{\frac{1}{8}} \frac{ c m^{\gamma/2}}{2 \sigma_{\mathrm w}^2 d_{{\mathrm a},{\mathrm w}}^{\gamma}}\right]$ is bounded. Besides showing that $\mathbb{E}_{{\mathrm F},{\mathrm W}}\left[\frac{1}{2}- \sqrt{\frac{1}{8}} \frac{ c m^{\gamma/2}}{2 \sigma_{\mathrm w}^2 d_{{\mathrm a},{\mathrm w}}^{\gamma}}\right]$ is bounded, we need to show that the bound $\mathbb{E}_{{\mathrm F},{\mathrm W}}\left[\frac{1}{2}- \sqrt{\frac{1}{8}} \frac{ c m^{\gamma/2}}{2 \sigma_{\mathrm w}^2 d_{{\mathrm a},{\mathrm w}}^{\gamma}}\right]>\frac{1}{2}-\epsilon$. To do so, we define the event $d_{{\mathrm a},{\mathrm w}}>\psi$ and we show in Appendix~\ref{ap.015} that
\begin{align} 
\label{eq:th101} \mathbb{E}_{{\mathrm F},{\mathrm W}} \left[\left. \mathbb{P}_{\mathrm e}^{(\mathrm w)}
(\sigma_{\mathrm{w}}^2,d_{\mathrm{a},\mathrm{w}})\right|d_{{\mathrm a},{\mathrm w}}>\psi\right] & \geq \frac{1}{2}-\frac{\epsilon}{2}.
\end{align}
Then, applying the law of total expectation and the fact that $\mathbb{P}(d_{{\mathrm a},{\mathrm w}}>\psi) = 1-\pi {\psi^2}/{2}$, we conclude
\begin{align} 
\nonumber\mathbb{P}_{\mathrm e}^{(\mathrm w)}= \mathbb{E}_{{\mathrm F},{\mathrm W}} [ \mathbb{P}_{\mathrm e}^{(\mathrm w)}
(\sigma_{\mathrm{w}}^2,d_{\mathrm{a},\mathrm{w}})]   &\geq  \mathbb{E}_{{\mathrm F},{\mathrm W}} \left[\left. \mathbb{P}_{\mathrm e}^{(\mathrm w)}
(\sigma_{\mathrm{w}}^2,d_{\mathrm{a},\mathrm{w}})\right|d_{{\mathrm a},{\mathrm w}}>\psi\right] \mathbb{P}(d_{{\mathrm a},{\mathrm w}}>\psi),\\ \label{eq:011} &\geq 
\left( \frac{1}{2}-\frac{\epsilon}{2}\right)\left(1- \frac{\pi \psi^2}{2}\right) 
= \left( \frac{1}{2}-\frac{\epsilon}{2}\right)\left(1- \frac{\epsilon}{4}\right) > \frac{1}{2}-\epsilon.
\end{align}
\noindent Thus, $ \mathbb{P}_{\mathrm e}^{(\mathrm w)}>\frac{1}{2}-\epsilon$ for all $\epsilon>0$, as long as $P_{\mathrm{a}} = \mathcal{O}(\frac{m^{\gamma/2}}{\sqrt{n}})$.

Note that Alice does not use the locations of the friendly nodes nor the location of Willie to select the transmission power (and thus, per below, the corresponding rate). Rather, she selects a power and corresponding rate for a scheme that is covert when averaged over the locations of the friendly nodes.

	({\em Reliability}) First, we analyze Bob's decoding error probability conditioned on $\sigma_{\mathrm b}^2=\sigma_{{\mathrm b},0}^2+\frac{P_{\mathrm{f}}}{d_{{\mathrm b},{\mathrm f}}^{\gamma}}$, which we denote $\mathbb{P}_{\mathrm e}^{(\mathrm b)}(\sigma_{\mathrm b}^2)$, where $d_{{\mathrm b},{\mathrm f}}$ is the distance from Bob to the friendly node closest to Willie. Then, we upper bound Bob's decoding error probability $\nonumber \mathbb{P}_{\mathrm e}^{(\mathrm b)} =  \mathbb{E}_{{\mathrm F},{\mathrm W}}[\mathbb{P}_{\mathrm e}^{(\mathrm b)}(\sigma_{\mathrm b}^2)]$.

Bob's ML decoder results an error when a codeword $\widehat{C}$ other than the transmitted one maximizes  $\mathbb{P}(\{y_i^{\mathrm (b)}\}_{i=1}^{n}|\widehat{C} )$. From an application of~\cite[Eqs. (5)-(9)]{bash_jsac2013}, we can upper bound Bob's decoding error probability averaged over all codebooks for a given $\sigma_{\mathrm b}^2$ by:
\begin{align}
   \label{eq:Pebasic0} \mathbb{P}_{\mathrm e}^{(\mathrm b)} \left(\sigma_{\mathrm b}^2\right)  & \leq 2^{nR- \frac{n}{2} \log_2 \left( 1+ \frac{ P_{\mathrm{a}}}{2 \sigma_{\mathrm b}^2} \right) },\\
\label{eq:Pebasic} &=2^{nR- \frac{n}{2} \log_2 \left( 1+ \frac{ c  m^{\gamma/2} }{2 \sqrt{n} \sigma_{\mathrm b}^2} \right) }.
 \end{align}
\noindent where the last step is obtained by having Alice set $P_{\mathrm{a}} = \frac{c m^{\gamma/2}}{\sqrt{n}}$ to satisfy~\eqref{eq:power}. Let $\phi =\sqrt{\frac{\ln{(2/(2-\zeta))}}{m \pi}}$, where $\zeta>0$ is the reliability parameter (see Definition~\ref{def:rel}). Since the right hand side (RHS) of~\eqref{eq:Pebasic} is a monotonically non-decreasing function of $d_{{\mathrm b},{\mathrm f}}$, when $d_{{\mathrm b},{\mathrm f}}>\phi$
\begin{align}
 \label{eq:th102}\mathbb{P}_{\mathrm e}^{(\mathrm b)} \left(\sigma_{\mathrm b}^2\right) &\leq2^{nR- \frac{n}{2} \log_2 \left( 1+ \frac{ c  m^{\gamma/2} }{2 \sqrt{n} \left(\sigma_{{\mathrm b},0}^2+{P_{\mathrm{f}}}/{\phi^{\gamma}}\right)} \right) }.
 \end{align}
\noindent We set Alice's rate to $R=\min\{1,R_0\}$ where 
\begin{align}
\label{eq:th2rate} R_0= \frac{1}{4} \log_2 \left( 1+ \frac{ c  m^{\gamma/2} }{ 2 \sqrt{n} \left(\sigma_{{\mathrm b},0}^2+{P_{\mathrm{f}}}/{\phi^{\gamma}}\right)} \right).
\end{align}
\noindent By~\eqref{eq:th102},~\eqref{eq:th2rate}, $\mathbb{P}_{\mathrm e}^{(\mathrm b)} \left(\sigma_{\mathrm b}^2\right)\leq 2^{n (R-2R_0)}$. Note that $R\leq R_0$ and thus $R-2R_0\leq -R_0$. Consequently
\begin{align}
  \label{eq:th1bobpe}  \mathbb{P}_{\mathrm e}^{(\mathrm b)} \left(\sigma_{\mathrm b}^2\right) \leq 2^{-n R_0}   =  \left( 1+ \frac{c   m^{\gamma/2} }{2 \sqrt{n} \left(\sigma_{{\mathrm b},0}^2+{ P_{\mathrm{f}}/\phi^\gamma}\right)} \right)^{- \frac{n}{4}} \leq  \left({1+\frac{c  m^{\gamma/2} \sqrt{n} }{8 \left(\sigma_{{\mathrm b},0}^2+{ P_{\mathrm{f}}/\phi^\gamma}\right)}}\right)^{-1},
  \end{align}
\noindent  where~\eqref{eq:th1bobpe} follows from the following inequality provided $n\geq 4$ (proved in the  Appendix~\ref{ap.02})
:
\begin{align} \label{eq:ineq1}
(1+x)^{-r} \leq \left({1+rx}\right)^{-1} \textrm{for any } r\geq 1 \textrm{ and } x>-1.
\end{align}
\noindent Thus,  
\begin{align}
&\mathbb{E}_{{\mathrm F},{\mathrm W}}[\mathbb{P}_{\mathrm e}^{(\mathrm b)}(\sigma_{\mathrm b}^2)|d_{{\mathrm b},{\mathrm f}}>  \phi]     \leq  \left({1+{ \frac{c  m^{\gamma/2} \sqrt{n}}{ 8(\sigma_{{\mathrm b},0}^2+{P_{\mathrm{f}}}/{\phi^{\gamma}})}}}\right)^{-1}.
\label{eq:th2bob1}
\end{align}
\noindent Next, we upper bound Bob's average decoding error probability $\mathbb{P}_{\mathrm e}^{(\mathrm b)}$ using~\eqref{eq:th2bob1}. The law of total expectation yields
\begin{align}
\label{eq:th132} \mathbb{P}_{\mathrm e}^{(\mathrm b)} &=  \mathbb{E}_{{\mathrm F},{\mathrm W}}[\mathbb{P}_{\mathrm e}^{(\mathrm b)}(\sigma_{\mathrm b}^2)]\leq \mathbb{E}_{{\mathrm F},{\mathrm W}}[\mathbb{P}_{\mathrm e}^{(\mathrm b)}(\sigma_{\mathrm b}^2)|d_{{\mathrm b},{\mathrm f}}>  \phi]  +  \mathbb{P}\left(d_{{\mathrm b},{\mathrm f}}\leq   \phi \right).
\end{align}
\noindent Consider the first term on the RHS of~\eqref{eq:th132}. By~\eqref{eq:th2bob1}, $\lim\limits_{n \to \infty} \mathbb{E}_{{\mathrm F},{\mathrm W}}[\mathbb{P}_{\mathrm e}^{(\mathrm b)}(\sigma_{\mathrm b}^2)|d_{{\mathrm b},{\mathrm f}}>  \phi] = 0$. Now, consider the second term on the RHS of~\eqref{eq:th132}. Since the event $\{d_{{\mathrm b},{\mathrm f}}\leq \phi\}$ is a subset of the event that no friendly node is in the circle of radius $\phi$ centered at Bob, $\mathbb{P}\left(d_{{\mathrm b},{\mathrm f}}\leq \phi\right) \leq 1-e^{-m \pi {\phi^2}}  = \zeta/2$, and thus $\lim\limits_{n \to \infty} \mathbb{P}_{\mathrm e}^{(\mathrm b)} \leq \zeta/2 < \zeta$ for any $0< \zeta < 1 $.

({\em Number of Covert Bits}) Now, we calculate $nR$, the number of bits that Bob receives. By~\eqref{eq:th2rate}, if $\frac{c m^{\gamma/2} }{ 2 \sqrt{n} \left(\sigma_{{\mathrm b},0}^2+{ P_{\mathrm{f}}/ \phi^{\gamma}}\right)}\geq 15$ , then $R_0\geq 1$, $R=1$, and thus $n R=n$. Now consider $\frac{c m^{\gamma/2} }{ 2 \sqrt{n} \left(\sigma_{{\mathrm b},0}^2+{ P_{\mathrm{f}}/ \phi^{\gamma}}\right)}< 15$. By~\eqref{eq:th2rate}, $R_0<1$, and thus
\begin{align}
\label{eq:111} nR= \frac{n}{4} \log_2 \left( 1+ \frac{ c  m^{\gamma/2} }{ 2 \sqrt{n} \left(\sigma_{{\mathrm b},0}^2+{ P_{\mathrm{f}}/\phi^\gamma }\right)} \right).
\end{align}
\noindent Consequently, $nR\leq \frac{n}{4} \log_2(1+15)=n$. Now consider $m=o(n^{1/\gamma})$. Note that $\log_2(1+x)\leq x$ with equality when
$x = 0$. Therefore, $nR=\mathcal{O}( m^{\gamma/2}\sqrt{n})$. Thus, Bob receives $\mathcal{O}(\min\{{n,m^{\gamma/2}\sqrt{n}}\})$ bits in $n$ channel uses.

({\em Converse}) We present the converse independent of the status (being on or off) of the friendly nodes. Recall that $ \mathcal{J}^{\dagger}\subset \mathcal{J}$ the set of friendly nodes that are on. Willie uses a power detector on his collection of observations $\left\{y_i\right\}_{i=1}^{n}$ to form $S=\frac{1}{n} \sum_{i=1}^{n}y_i^2$ and performs a hypothesis test based on $S$ and a threshold $t$. If $S<\sigma_{\mathrm w}^2+t$, Willie accepts $H_0$ (Alice does not transmit); otherwise, he accepts $H_1$ (Alice transmits). Recall that when $H_0$ is true, $y_i= z_{i,0}+\sum_{\mathrm{f}_j\in  \mathcal{J}^\dagger}^\infty z_{i,j}$, where $\left\{z_{i,0}\right\}_{i=1}^{n}$ is an i.i.d.~sequence representing the background noise with $z_{i,0} \sim \mathcal{N}(0,\sigma_{{\mathrm w}_{1,0}}^2)$, and $\left\{z_{i,j}\right\}_{i=1}^{n}$ is an i.i.d.~sequence characterizing the chatter from the $j^{\mathrm{th}}$ friendly node with $\mathcal{N} (0,{P_{\mathrm{f}}}/{d_{{\mathrm w},{\mathrm f}_j}^\gamma})$. Since all of the sources of noise are independent, we can model Willie's total noise by a Gaussian noise with $y_i \sim \mathcal{N}(0,\sigma_{\mathrm w}^2)$, where $\sigma_{\mathrm w}^2= \sigma_{{\mathrm w},0}^2+\sum_{\mathrm{f}_j\in  \mathcal{J}^{\dagger}} {P_{\mathrm{f}}}/{d_{{\mathrm w},{\mathrm f}_j}^\gamma}$. Therefore~\cite[Eqs. (12),(13)]{bash_jsac2013}, 
\begin{align}
\nonumber \mathbb{E}_{Y}[S|H_0]&=\sigma_{\mathrm w}^2, \\
\nonumber \mathrm{Var}_{Y}[S|H_0]&=\frac{2\sigma_{\mathrm w}^4}{n},
\end{align}
\noindent where $\mathbb{E}_{Y}[\cdot]$ and $\mathrm{Var}_{Y}[\cdot]$ denote the expectation and variance with respect to Willie's received signal. When $H_1$ is true, Alice transmits a codeword $C(M_l)=\left\{C^{(u)}(M_l)\right\}_{u=1}^{n}$ and Willie observes $\left\{y_i\right\}_{i=1}^{n}$ which contains i.i.d. samples of mean shifted noise $y_i\sim \mathcal{N}\left(\frac{s_i}{d_{{\mathrm a},{\mathrm w}}^{\gamma/2}},\sigma_{{\mathrm w}}^2\right)$, where $s_i$ is the value of Alice's transmitted symbol in the $i^{\text{th}}$ channel use, and each $s_i$ is an instantiation of a Gaussian random variable $\mathcal{N}\left(0, P_{\mathrm{a}}\right)$. Therefore~\cite[Eqs. (14),(15)]{bash_jsac2013},
\begin{align}
\nonumber\mathbb{E}_{Y}[S|H_1]&=\sigma_{\mathrm w}^2+\frac{P_{\mathrm{a}}}{d_{{\mathrm a},{\mathrm w}}^{\gamma}},\\
\nonumber \mathrm{Var}_{Y}[S|H_1]&=\frac{4 \frac{P_{\mathrm{a}}}{d_{{\mathrm a},{\mathrm w}}^{\gamma}} \sigma_{\mathrm w}^2 + 2\sigma_{\mathrm w}^4}{n}.
\end{align}
\noindent We show that Willie can choose the threshold $t$ independent of locations of the friendly nodes, $\sigma_{\mathrm w}^2$, and $ \mathcal{J}^{\dagger}$ such that if Alice transmits $\omega\left(m^{\gamma/2}\sqrt{n}\right)$ bits to Bob, he can achieve arbitrarily small average error probability. Bounding $\mathbb{P}_{\mathrm {FA}}(\sigma_{\mathrm{w}}^2,d_{\mathrm{a},\mathrm{w}})$ by using Chebyshev's inequality yields \cite{bash_jsac2013}: 
\begin{align}
\label{eq:01} \mathbb{P}_{\mathrm {FA}}(\sigma_{\mathrm{w}}^2,d_{\mathrm{a},\mathrm{w}}) \leq \frac{2 \sigma_{\mathrm w}^4}{n t^2}.
\end{align}
\noindent Let  
\begin{align}
\label{eq:eta1}\eta_1 = \sqrt{\frac{\ln {\left(\frac{4}{4-\lambda}\right)}}{m \pi}}.
\end{align}
\noindent Note that $\mathbb{P}_{\mathrm {FA}}=\mathbb{E}_{{\mathrm F},{\mathrm W}} [\mathbb{P}_{\mathrm {FA}}(\sigma_{\mathrm{w}}^2,d_{\mathrm{a},\mathrm{w}})]$. By the law of total expectation: 
\begin{align}
\nonumber \mathbb{P}_{\mathrm {FA}} &= \mathbb{E}_{{\mathrm F},{\mathrm W}}  \left[\left.\mathbb{P}_{\mathrm {FA}}(\sigma_{\mathrm{w}}^2,d_{\mathrm{a},\mathrm{w}})\right|d_{{\mathrm w},{\mathrm f}}\leq \eta_1 \right] \mathbb{P}(d_{{\mathrm w},{\mathrm f}}\leq \eta_1 )+ \mathbb{E}_{{\mathrm F},{\mathrm W}} \left[\left.\mathbb{P}_{\mathrm {FA}}(\sigma_{\mathrm{w}}^2,d_{\mathrm{a},\mathrm{w}})\right|d_{{\mathrm w},{\mathrm f}}>\eta_1 \right] \mathbb{P}(d_{{\mathrm w},{\mathrm f}}>\eta_1 ),\\
\nonumber &\leq \mathbb{P}(d_{{\mathrm w},{\mathrm f}}\leq\eta_1 ) +  \mathbb{E}_{{\mathrm F},{\mathrm W}} \left[\left.\mathbb{P}_{\mathrm {FA}}(\sigma_{\mathrm{w}}^2,d_{\mathrm{a},\mathrm{w}})\right|d_{{\mathrm w},{\mathrm f}}>\eta_1 \right],\\
\nonumber & =   \left(1-e^{- m \pi\eta_1^2}\right) +   \mathbb{E}_{{\mathrm F},{\mathrm W}} \left[\left.\mathbb{P}_{\mathrm {FA}}(\sigma_{\mathrm{w}}^2,d_{\mathrm{a},\mathrm{w}})\right|d_{{\mathrm w},{\mathrm f}}>\eta_1 \right],\\ 
\label{eq:17} &\stackrel{(a)}{=} \frac{\lambda}{4} +    \mathbb{E}_{{\mathrm F},{\mathrm W}} \left[\left.\mathbb{P}_{\mathrm {FA}}(\sigma_{\mathrm{w}}^2,d_{\mathrm{a},\mathrm{w}})\right|d_{{\mathrm w},{\mathrm f}}>\eta_1 \right]= \frac{\lambda}{4} + \frac{2 }{n t^2} \mathbb{E}_{{\mathrm F},{\mathrm W}} \left[\sigma_{\mathrm w}^4|d_{{\mathrm w},{\mathrm f}}>\eta_1 \right], 
\end{align} 
\noindent where $(a)$ follows from \eqref{eq:eta1}, and the last step follows from~\eqref{eq:01}. Let $\sigma_{\mathrm w}^2(r)$ be Willie's noise power considering only the friendly nodes in the circle of radius $r>\eta_1$ centered at Willie, and $N_{\mathrm f}$ be the (random) number of friendly nodes in the area surrounded by the circles of radii $\eta_1$ and $r$ centered at Willie. Then:
\begin{align}
\label{eq:02} \sigma_{\mathrm w}^2(r)&= \sigma_{{\mathrm w},0}^2 + P_{\mathrm{f}} \sum_{ 
	\substack{\eta_1< d_{\mathrm{w},\mathrm{f}_i}\leq r \\ \mathrm{f}_i \in  \mathcal{J}^\dagger}
} \frac{1}{d_{\mathrm{w},\mathrm{f}_i}^{\gamma}} \leq \sigma_{{\mathrm w},0}^2 + P_{\mathrm{f}} \sum_{\eta_1< d_{\mathrm{w},\mathrm{f}_i}\leq r} \frac{1}{d_{\mathrm{w},\mathrm{f}_i}^{\gamma}}, 
\end{align}
\noindent where the inequality in~\eqref{eq:02} becomes equality when all of the friendly nodes in the area surrounded by the circles of radii $\eta_1$ and $r_1$ centered at Willie are on. We show in Appendix~\ref{ap.3.1} that
\begin{align}
\nonumber  \mathbb{E}_{{\mathrm F},{\mathrm W}} \left[\sigma_{\mathrm w}^4(r)|d_{{\mathrm w},{\mathrm f}}>\eta_1 \right]\leq 
\sigma_{{\mathrm w},0}^4 +  2 P_{\mathrm{f}} m \pi r^2 \sigma_{{\mathrm w},0}^2 
\mathbb{E}_{{\mathrm F}} [ {1}/{d_{\mathrm{w},\mathrm{f}_i}^{\gamma}}|d_{{\mathrm w},{\mathrm f}}>\eta_1  ] &+  P_{\mathrm{f}}^2 m \pi r^2 \mathbb{E}_{{\mathrm F}} [{1}/{d_{\mathrm{w},\mathrm{f}_i}^{2 \gamma}}|d_{{\mathrm w},{\mathrm f}}>\eta_1 ],\\
\label{eq:12}&+  P_{\mathrm{f}}^2 m^2 \pi^2 r^4 \mathbb{E}_{{\mathrm F}} [{1}/{d_{\mathrm{w},\mathrm{f}_i}^{\gamma}}|d_{{\mathrm w},{\mathrm f}}>\eta_1 ]^2,
\end{align}
\noindent and in Appendix~\ref{ap.3.2} that for large enough $n$:
\begin{align}
\label{eq:14} \mathbb{E}_{{\mathrm F}} [ {1}/{d_{\mathrm{w},\mathrm{f}_i}^{\gamma}}|d_{{\mathrm w},{\mathrm f}}>\eta_1  ]  &\leq \frac{4 }{\gamma-2} \frac{\eta_1^{2-\gamma}}{r^2 }, \\
\label{eq:15} \mathbb{E}_{{\mathrm F}} [ {1}/{d_{\mathrm{w},\mathrm{f}_i}^{2\gamma}}|d_{{\mathrm w},{\mathrm f}}>\eta_1  ]&\leq \frac{2 }{\gamma-1} \frac{\eta_1^{2-2\gamma}}{r^2 }.
\end{align}
\noindent Since $\eta_1=\Theta(m^{-1/2})$,~\eqref{eq:14},~\eqref{eq:15}, the first four terms on the RHS of~\eqref{eq:12} are $\mathcal{O}(1)$, $\mathcal{O}(m^{\gamma/2})$, $\mathcal{O}(m^{\gamma})$ and $\mathcal{O}(m^{\gamma})$, respectively. Consequently, for large enough $n$:
\begin{align}
\label{eq:04}\mathbb{E}_{{\mathrm F},{\mathrm W}} \left[\sigma_{\mathrm w}^4(r)|d_{{\mathrm w},{\mathrm f}}>\eta_1 \right]\leq \rho^2 m^{\gamma},\end{align} 
\noindent where
\begin{align}
\label{eq:rho} \rho= 2 \pi^{\gamma/2}P_{\mathrm{f}} \sqrt{\frac{\min\{\left({\ln {\left(\frac{4}{4-\lambda}\right)}}\right)^{1-\gamma}, 8 \left({\ln {\left(\frac{4}{4-\lambda}\right)}}\right)^{2-\gamma}\}  }{\gamma-1}} .
\end{align}
\noindent This means that the noise generated by the closest friendly node to Willie dominates the noise generated from other friendly nodes. By~\eqref{eq:04}, $\sigma_{\mathrm w}^4(r_1)\leq \sigma_{\mathrm w}^4(r_2)$ for $\eta_1\leq r_1\leq r_2$. Therefore, the monotone convergence theorem yields: 
\begin{align}
\label{eq:16}\mathbb{E}_{{\mathrm F},{\mathrm W}} \left[\sigma_{\mathrm w}^4|d_{{\mathrm w},{\mathrm f}}>\eta_1 \right]\leq \rho^2 m^{\gamma}.
\end{align} 
Let Willie choose threshold $t = \frac{\sqrt{8 } \rho  m^{\gamma/2}}{\sqrt{n \lambda }} $. By~\eqref{eq:17},
\begin{align} \label{eq:18}
\mathbb{P}_{\mathrm {FA}} \leq  \frac{\lambda}{4}+\frac{\lambda}{4}= \frac{\lambda}{2}.
\end{align}

Next, we upper bound $\mathbb{P}_{\mathrm {MD}}=\mathbb{E}_{{\mathrm F},{\mathrm W}} [\mathbb{P}_{\mathrm {MD}}(\sigma_{\mathrm{w}}^2,d_{\mathrm{a},\mathrm{w}})]$. Since $d_{{\mathrm a},{\mathrm w}}\leq 2$, Willie can achieve~\cite[Eq. (16)]{bash_jsac2013}
\begin{align} \label{eq:th1conv1}
\mathbb{P}_{\mathrm {MD}}(\sigma_{\mathrm{w}}^2,d_{\mathrm{a},\mathrm{w}}) &\leq \frac{4 \frac{P_{\mathrm{a}}}{d_{{\mathrm a},{\mathrm w}}^{\gamma}} \sigma_{\mathrm w}^2 + 2 \sigma_{\mathrm w}^4}{n \left(\frac{P_{\mathrm{a}}}{d_{{\mathrm a},{\mathrm w}}^{\gamma}}-t\right)^2}=\frac{4 \frac{P_{\mathrm{a}}}{d_{{\mathrm a},{\mathrm w}}^\gamma} \sigma_{\mathrm w}^2 + 2 \sigma_{\mathrm w}^4}{n \left(\frac{P_{\mathrm{a}}}{2^{\gamma}}-t\right)^2}.
\end{align}
\noindent Let $\eta_2 = \sqrt{\frac{\ln {\left(\frac{4}{4-\lambda+\lambda^{\prime}}\right)}}{m \pi}}$, where $0<\lambda^{\prime}<\lambda$, and $\eta_3 = \sqrt{\frac{\lambda}{2 \pi}}$. The law of total expectation yields
\begin{align}
\nonumber \mathbb{P}_{\mathrm {MD}}&\leq \mathbb{P}(\{d_{{\mathrm w},{\mathrm f}}\leq\eta_2\} \cup \{d_{{\mathrm a},{\mathrm w}}\leq\eta_3\}) 
\nonumber +  \mathbb{E}_{{\mathrm F},{\mathrm W}} \left[\left.\mathbb{P}_{\mathrm {MD}}(\sigma_{\mathrm{w}}^2,d_{\mathrm{a},\mathrm{w}})\right|\{d_{{\mathrm w},{\mathrm f}}>\eta_2\} \cap \{d_{{\mathrm a},{\mathrm w}}>\eta_3\} \right],\\
&\nonumber \stackrel{(b)}{\leq}  \left(1-e^{- m \pi\eta_2^2}\right) + \frac{\pi}{2} \eta_3^2 +   \mathbb{E}_{{\mathrm F},{\mathrm W}} \left[\left.\mathbb{P}_{\mathrm {MD}}(\sigma_{\mathrm{w}}^2,d_{\mathrm{a},\mathrm{w}})\right|\{d_{{\mathrm w},{\mathrm f}}>\eta_2\} \cap \{d_{{\mathrm a},{\mathrm w}}>\eta_3\} \right],\\
&\nonumber \stackrel{(c)}{=}  \frac{\lambda-\lambda^{\prime}}{4} + \frac{\lambda}{4}  +\mathbb{E}_{{\mathrm F},{\mathrm W}} \left[\left.\mathbb{P}_{\mathrm {MD}}(\sigma_{\mathrm{w}}^2,d_{\mathrm{a},\mathrm{w}})\right|\{d_{{\mathrm w},{\mathrm f}}>\eta_2\} \cap \{d_{{\mathrm a},{\mathrm w}}>\eta_3\} \right],\\
&\label{eq:22} \leq   \frac{\lambda-\lambda^{\prime}}{4} + \frac{\lambda}{4}  +\frac{4 \frac{P_{\mathrm{a}}}{\eta_3^{\gamma}} \mathbb{E}_{{\mathrm F},{\mathrm W}} \left[\sigma_{\mathrm w}^2|d_{{\mathrm w},{\mathrm f}}>\eta_2 \right]}{n \left( \frac{P_{\mathrm{a}}}{2^{\gamma}}-t\right)^2}  +  \frac{2 \mathbb{E}_{{\mathrm F},{\mathrm W}} \left[\sigma_{\mathrm w}^4|d_{{\mathrm w},{\mathrm f}}>\eta_2 \right]}{n \left( \frac{P_{\mathrm{a}}}{2^{\gamma}}-t\right)^2},
\end{align}
\noindent where $(b)$ follows from the union bound, $(c)$ follows from substituting the values of $\eta_2$ and $\eta_3$, and the last step follows from taking the conditional expectation of~\eqref{eq:th1conv1} given $\{d_{{\mathrm w},{\mathrm f}}\leq\eta_2\} \cup \{d_{{\mathrm a},{\mathrm w}}\leq\eta_3\}$ and upper bounding $1/d_{\mathrm{a},\mathrm{w}}^\gamma$ by $1/\eta_3^\gamma$.

Consider $\mathbb{E}_{{\mathrm F},{\mathrm W}} \left[\sigma_{\mathrm w}^2|d_{{\mathrm w},{\mathrm f}}>\eta_2 \right]$ and $\mathbb{E}_{{\mathrm F},{\mathrm W}} \left[\sigma_{\mathrm w}^4|d_{{\mathrm w},{\mathrm f}}>\eta_2 \right]$  in~\eqref{eq:22}. Similar to the arguments leading to~\eqref{eq:16}, we show that $\mathbb{E}_{{\mathrm F},{\mathrm W}} \left[\sigma_{\mathrm w}^4|d_{{\mathrm w},{\mathrm f}}>\eta_2 \right]=\mathcal{O}(m^{\gamma})$. Consequently, Jensen's inequality yields $\mathbb{E}_{{\mathrm F},{\mathrm W}} \left[\sigma_{\mathrm w}^2|d_{{\mathrm w},{\mathrm f}}>\eta_2 \right]=\mathcal{O}(m^{\gamma/2})$. In addition,  $t=\Theta\left(\frac{m^{\gamma/2}}{\sqrt{n}}\right)$. Thus, if Alice sets her average symbol power $P_{\mathrm{a}} = \omega\left(\frac{m^{\gamma/2}}{\sqrt{n}}\right)$, then there exists $n_0>0$ s.t. $\forall n>n_0(\lambda^{\prime})$
\begin{align}\label{eq:20} \mathbb{E}_{{\mathrm F},{\mathrm W}} [\mathbb{P}_{\mathrm {MD}}] \leq \frac{\lambda-\lambda^{\prime}}{4} + \frac{\lambda}{4} + \frac{\lambda^{\prime}}{2} = \frac{\lambda}{2} + \frac{\lambda'}{4}<\lambda.
\end{align}

\noindent By~\eqref{eq:18} and~\eqref{eq:20}, for any $\lambda>0$
\begin{align}
\nonumber \mathbb{P}_{\mathrm e}^{(\mathrm w)} = \frac{\mathbb{P}_{\mathrm {FA}} + \mathbb{P}_{\mathrm {MD}}}{2} \leq \frac{3 \lambda}{4}<\lambda.
\end{align}
\noindent Consequently, Alice cannot send any codeword with average symbol power $\omega\left(\frac{m^{\gamma/2}}{\sqrt{n}}\right)$ covertly. Thus, to avoid detection of a given codeword, she must set the power of that codeword to $P_{\mathcal{U}} = O\left(\frac{m^{\gamma/2}}{\sqrt{n}}\right)$. Suppose that Alice's codebook contains a fraction $\xi>0$ of codewords with power $P_{\mathcal{U}} = O\left(\frac{m^{\gamma/2}}{\sqrt{n}}\right)$. For such low power codewords, we can lower bound Bob's decoding error probability given the locations of the friendly nodes by~\cite[Eq. (20)]{bash_jsac2013}
\begin{align} \label{eq:th1con5}
\mathbb{P}_{\mathrm e}^{\mathcal{U}} (\sigma_{\mathrm b}^2) \geq 1- \frac{\frac{P_{\mathcal{U}}}{2\sigma_{\mathrm b}^2}+\frac{1}{n}}{\frac{\log_2 \xi}{n}+R}\geq  1- \frac{\frac{P_{\mathcal{U}}}{2\sigma_{\mathrm{b},0}^2}+\frac{1}{n}}{\frac{\log_2 \xi}{n}+R}.
\end{align}
\noindent Since Alice's rate is $R=\omega \left(\frac{m^{\gamma/2}}{\sqrt{n}}\right)$ bits/symbol, $\mathbb{P}_{\mathrm e}^{\mathcal{U}}(\sigma_{\mathrm b}^2)$ is bounded away from zero as $n \to \infty$.
\end{proof}

\subsection{Single Warden Scenario and $\gamma=2$} \label{sec:sgammab2}
\begin{mthm}{1.2} \label{th:swillie2}
	When there is one warden (Willie) located randomly and uniformly over the unit square, $m>0$, and $\gamma=2$, Alice can reliably and covertly transmit 
	$\mathcal{O}(\min\{{n,m\sqrt{n}}\})$ bits to Bob in $n$ channel uses. Conversely, if only the closest friendly node to Willie is on and Alice attempts to transmit $\omega(m^{\gamma/2} \sqrt{n})$ bits to Bob in $n$ channel uses, there exists a detector that Willie can use to either detect her with arbitrarily low error probability $\mathbb{P}_{\mathrm e}^{(\mathrm w)}$ or Bob cannot decode the message with arbitrarily low error probability $\mathbb{P}_{\mathrm e}^{(\mathrm b)}$.
\end{mthm}

\begin{proof}

{\it (Achievability)} The achievability (construction and analysis) is the same as that of~\ref{th:swillie}. 

({\em Converse}) For $\gamma>2$, we upper bounded Willie's noise by the received noise power in the worst case scenario where all of the friendly nodes are on, and it was optimal since $\sigma_{\mathrm{w}}^2=\mathcal{O}(m^{\gamma/2})$. However, for $\gamma=2$, noise power for the worst case scenario is $\mathcal{O}(m \log(m))$ which is not optimal. 

We assume only the closest friendly node to Willie is on and Willie knows that. The proof follows from that of $\gamma>2$ with modifications of~\eqref{eq:17} and~\eqref{eq:22}, noting that $\mathbb{E}_{{\mathrm F},{\mathrm W}} \left[\sigma_{\mathrm w}^4|d_{{\mathrm w},{\mathrm f}}>\eta_1 \right]= \mathbb{E}_{{\mathrm F},{\mathrm W}} \left[(\sigma_{\mathrm{w}_0}^2+P_{\mathrm{f}}/d_{{\mathrm w},{\mathrm f}}^{\gamma})^2|d_{{\mathrm w},{\mathrm f}}>\eta_1 \right]\leq (\sigma_{\mathrm{w}_0}^2+P_{\mathrm{f}}/\eta_1^{\gamma})^2$.

\end{proof}

\section{Multiple Collaborating Wardens Scenario}
\label{sec:mwillie}

In this section, we consider the case when there are $N_{\mathrm{w}}$ collaborating Willies located independently and uniformly in the unit square (see Fig.~\ref{fig:SysMod}). We present Theorem~\ref{th:mwillie2p} for $\gamma>2$ in Section~\ref{sec:gamma2p}, and Theorem~\ref{th:mwillie2} for $\gamma=2$ in Section~\ref{sec:gamma2}. Analogous to the single warden scenario, Alice and Bob's strategy is to turn on the closest friendly node to each Willie and keep all other friendly nodes off, whether Alice transmits or not.

\subsection{$\gamma>2$} \label{sec:gamma2p}
\begin{mthm}{2.1} \label{th:mwillie2p}
When friendly nodes are independently distributed according to a two-dimensional Poisson point process with density $m=\omega(1)$, and $N_{\mathrm{w}}={o}\left({m}/{\log{m}}\right)$ collaborating Willies are uniformly and independently distributed over the unit square shown in Fig.~\ref{fig:SysMod}, then Alice can reliably and covertly transmit $\mathcal{O}\left(\min\left\{n,\frac{m^{\gamma/2} \sqrt{n}}{N_{\mathrm{w}}^{\gamma}}\right\}\right)$ bits to Bob in $n$ channel uses. Conversely, if only the closest friendly node to each Willie is on and Alice attempts to transmit $\omega\left(\frac{\sqrt{n} m^{\gamma/2}}{ N_{\mathrm{w}}^\gamma}\right)$ bits to Bob in $n$ channel uses, there exists a detector that Willie can use to either detect her with arbitrarily low error probability $\mathbb{P}_{\mathrm e}^{(\mathrm w)}$ or Bob cannot decode the message with arbitrarily low error probability $\mathbb{P}_{\mathrm e}^{(\mathrm b)}$.
\end{mthm}  
We present the proof assuming $N_{\mathrm{w}}=\omega(1)$, as the proof for a finite $N_{\mathrm{w}}$ follows from it. In addition, according to the statement of Theorem~\ref{th:mwillie2p}, if   $N_{\mathrm{w}}=\Omega\left(n^{\frac{1}{2\gamma}}\sqrt{m}\right)$, then Alice can reliably and covertly transmit $\mathcal{O}\left(1\right)$ bits to Bob in $n$ uses of channel, which is not of interest. Therefore, we present the proof assuming 
$N_{\mathrm{w}}={o}\left(\min\left\{\frac{m}{\log{m}},n^{\frac{1}{2\gamma}}\sqrt{m}\right\}\right)$.
\begin{proof}

{\it (Achievability)} 

\textbf{Construction:} The construction and Bob's decoding are the same as those of Theorems~\ref{th:swillie} and~\ref{th:swillie2}. 

\textbf{Analysis:} ({\em Covertness}) By~\eqref{eq:0}, when Willie applies the optimal hypothesis test to minimize his error probability, 
\begin{align} 
\label{eq:basic2} \mathbb{P}_{\mathrm e}^{(\mathrm w)}(\boldsymbol{\sigma_{\mathrm{w}}^2},\boldsymbol{d_{\mathrm{a},\mathrm{w}}}) \geq \frac{1}{2}- \sqrt{\frac{1}{8} \mathcal{D}(\mathbb{P}_1(\boldsymbol{\sigma_{\mathrm{w}}^2},\boldsymbol{d_{\mathrm{a},\mathrm{w}}}) || \mathbb{P}_0(\boldsymbol{\sigma_{\mathrm{w}}^2}))}.
\end{align}
\noindent Here, $\boldsymbol{\sigma_{\mathrm{w}}^2}$ and $\boldsymbol{d_{\mathrm{a},\mathrm{w}}}$ are vectors containing  $\sigma_{\mathrm{w}_k}^2$ and $d_{\mathrm{a},\mathrm{w}_k}$, respectively, $\mathbb{P}_0(\boldsymbol{\sigma_{\mathrm{w}}^2})=\prod_{i=1}^{n}  \mathbb{P}_{0,i}(\boldsymbol{\sigma_{\mathrm{w}}^2})$ and  $\mathbb{P}_1(\boldsymbol{\sigma_{\mathrm{w}}^2},\boldsymbol{d_{\mathrm{a},\mathrm{w}}})=\prod_{i=1}^{n}  \mathbb{P}_{1,i}(\boldsymbol{\sigma_{\mathrm{w}}^2},\boldsymbol{d_{\mathrm{a},\mathrm{w}}})$ are the joint probability distributions of the Willies' channels observations for the $H_0$ and $H_1$ hypotheses, respectively, where $\mathbb{P}_{0,i}({\boldsymbol{\sigma_{\mathrm{w}}^2}})=\prod_{k=1}^{N_{\mathrm{w}}}\mathbb{P}_{{\mathrm w}_k}^{(k)}({\sigma_{\mathrm{w}_k}^2})$ and $\mathbb{P}_{1,i}(\boldsymbol{\sigma_{\mathrm{w}}^2},\boldsymbol{d_{\mathrm{a},\mathrm{w}}})$ are the joint probability distribution of the $i^{\mathrm{th}}$ channel observation of the Willies for $H_0$ and $H_1$ hypotheses, respectively. The relative entropy between two multivariate normal distributions $\mathbb{P}_1(\boldsymbol{\sigma_{\mathrm{w}}^2},\boldsymbol{d_{\mathrm{a},\mathrm{w}}})$ and  $\mathbb{P}_0(\boldsymbol{\sigma_{\mathrm{w}}^2})$ is~\cite{klref}:
\begin{align} \label{eq:kl}
 \mathcal{D}(\mathbb{P}_1(\boldsymbol{\sigma_{\mathrm{w}}^2},\boldsymbol{d_{\mathrm{a},\mathrm{w}}}) || \mathbb{P}_0(\boldsymbol{\sigma_{\mathrm{w}}^2})) &= { 1 \over 2 } \left(
  \mathrm{tr} \left( \Sigma_0^{-1} \Sigma_1 \right) + \left( \mu_0 - \mu_1\right)^\top \Sigma_0^{-1} ( \mu_0 - \mu_1 ) - \mathrm{dim}\left(\Sigma_0\right) - \ln \left( { | \Sigma_1 | \over | \Sigma_0 | } \right)  \right),
 \end{align}
\noindent where $\mathrm{tr}(\cdot)$, $|\cdot|$, and $\mathrm{dim}(\cdot)$ denote the trace, determinant and dimension of a square matrix respectively, $\mu_0=0$, $\mu_1=0$ are the mean vectors, and $\Sigma_0$, $\Sigma_1$ are nonsingular covariance matrices of $\mathbb{P}_0(\boldsymbol{\sigma_{\mathrm{w}}^2})$ and  $\mathbb{P}_1(\boldsymbol{\sigma_{\mathrm{w}}^2},\boldsymbol{d_{\mathrm{a},\mathrm{w}}})$, respectively, given by
\begin{align}
\nonumber \Sigma_0 & =S\otimes I_{n \times n},\\
\nonumber \Sigma_1 & = \big(S + P_{\mathrm{a}} U U^T\big)\otimes I_{n \times n},
\end{align}
\noindent where $S= \mathrm{diag}(\sigma_{{\mathrm w}_1}^2, \ldots \, ,\sigma_{{\mathrm w}_{N_{\mathrm{w}}}}^2)$, $\otimes$ denotes the Kronecker product between two matrices, $I_{n \times n}$ is the identity matrix of size $n$, and $U$ is a column vector of size $N_{\mathrm{w}}$ given by
\begin{equation}
   \nonumber U= \begin{bmatrix}
    {1}/{d_{{\mathrm a},{\mathrm w}_1}^{\gamma/2}} & {1}/{d_{{\mathrm a},{\mathrm w}_2}^{\gamma/2}} & \dots & {1}/{d_{{\mathrm a},{\mathrm w}_{N_{\mathrm{w}}}}^{\gamma/2}} \
  \end{bmatrix} ^{T}.
\end{equation}
\noindent Next, we calculate the relative entropy in~\eqref{eq:kl}. The first term on the RHS of~\eqref{eq:kl} is:
\begin{align}
\nonumber \mathrm{tr} \left( \Sigma_0^{-1} \Sigma_1 \right)& = n \sum_{k=1}^{N_{\mathrm{w}}} \frac{1}{\sigma_{{\mathrm w}_k}^2}\left( \sigma_{{\mathrm w}_k}^2 + \frac{P_{\mathrm{a}}}{d_{{\mathrm a},{\mathrm w}_k}^{\gamma}}\right) = n N_{\mathrm{w}}+ n \sum_{k=1}^{N_{\mathrm{w}}} \frac{P_{\mathrm{a}}}{d_{{\mathrm a},{\mathrm w}_k}^{\gamma} \sigma_{{\mathrm w}_k}^2}.
\end{align}
\noindent Then,
\begin{align}
\nonumber \left|{\Sigma_0}\right|=  \left|S\otimes I_{n \times n}\right| &\stackrel{(d)}{=}\left|S\right|^n | I_{n \times n}|^{N_{\mathrm{w}}} =\left|S\right|^n = \left(\prod\limits_{k=1}^{N_{\mathrm{w}}} \sigma_{{\mathrm w}_k}^2 \right)^n.
\end{align}
\noindent where $(d)$ is true from the determinant of the Kronecker product property presented in~\cite[p. 279]{abadir2005matrix}. Because $\sigma_{{\mathrm w}_k}^2>0$, $S$ is nonsingular. Therefore,
\begin{align} 
\nonumber \left|{\Sigma_1}\right| =\left|S + P_{\mathrm{a}} U U^T\right|^n | I_{n \times n}|^{N_{\mathrm{w}}} =\left|S + P_{\mathrm{a}} U U^T\right|^n  = \left|S\right|^n \left|I+P_{\mathrm{a}} S^{-1} U U^T \right|^n &\stackrel{(e)}{=} \left|S\right|^n \left(1+P_{\mathrm{a}}  U^T S^{-1} U\right)^n,\\
 \nonumber &= |\Sigma_0| \left(1+\sum_{k=1}^{N_{\mathrm{w}}} \frac{P_{\mathrm{a}}}{d_{{\mathrm a},{\mathrm w}_k}^{\gamma} \sigma_{{\mathrm w}_k }^2} \right)^n,
\end{align}
\noindent where $(e)$ is due to Lemma 1.1 in \cite{matrixDing}. Therefore, 
\begin{align}
\nonumber \ln \left( { | \Sigma_1 | \over | \Sigma_0 | } \right) &= n \ln{\left(1+\sum_{k=1}^{N_{\mathrm{w}}} \frac{P_{\mathrm{a}}}{d_{{\mathrm a},{\mathrm w}_k}^{\gamma} \sigma_{{\mathrm w}_k }^2} \right)}.
\end{align}
\noindent Thus,
\begin{align}\label{eq:th3be}
\mathcal{D}(\mathbb{P}_1(\boldsymbol{\sigma_{\mathrm{w}}^2},\boldsymbol{d_{\mathrm{a},\mathrm{w}}}) || \mathbb{P}_0(\boldsymbol{\sigma_{\mathrm{w}}^2})) = {n \over 2}\left(  \sum_{k=1}^{N_{\mathrm{w}}} \frac{P_{\mathrm{a}}}{d_{{\mathrm a},{\mathrm w}_k}^{\gamma} \sigma_{{\mathrm w}_k}^2} - \ln{\left(1+\sum_{k=1}^{N_{\mathrm{w}}} \frac{P_{\mathrm{a}}}{d_{{\mathrm a},{\mathrm w}_k}^{\gamma} \sigma_{{\mathrm w}_k}^2}\right)} \right).
\end{align}
\noindent Suppose Alice sets her average symbol power so that
\begin{align}\label{eq:th3pf}
P_{\mathrm{a}} \leq \frac{c  m^{\gamma/2}}{\sqrt{n} N_{\mathrm{w}}^{\gamma/2}},
\end{align}
\noindent where 
\begin{align}\label{eq:th3c}
c=  \frac{P_{\mathrm{f}} \epsilon^{\gamma/2} \left(\gamma-2\right)\pi^{\gamma/2}}{2^{\gamma-0.5} \Gamma\left(\gamma/2+1\right)}.
\end{align}
\noindent By~\eqref{eq:ineq0} and~\eqref{eq:th3be},
\begin{align}
\mathcal{D}(\mathbb{P}_1(\boldsymbol{\sigma_{\mathrm{w}}^2},\boldsymbol{d_{\mathrm{a},\mathrm{w}}}) || \mathbb{P}_0(\boldsymbol{\sigma_{\mathrm{w}}^2})) 
\label{eq:dformulw} &  \leq {n \over 4}\left( \sum_{k=1}^{N_{\mathrm{w}}} \frac{P_{\mathrm{a}}}{d_{{\mathrm a},{\mathrm w}_k}^{\gamma} \sigma_{{\mathrm w}_k}^2} \right)^2\leq \frac{c^2  m^{\gamma}}{4 N_{\mathrm{w}}^{\gamma}} \left( \sum_{k=1}^{N_{\mathrm{w}}} \frac{1}{d_{{\mathrm a},{\mathrm w}_k}^{\gamma} \sigma_{{\mathrm w}_k}^2} \right)^2.
\end{align}
\noindent where the last step follows from~\eqref{eq:th3pf}. Similar to the arguments leading to~\eqref{eq:011}, to achieve $\mathbb{P}_{\mathrm e}^{(\mathrm w)}>\frac{1}{2}-\epsilon$, we define the event (see Fig.~\ref{fig:SysMod2})
\begin{align}
\nonumber \mathcal{A} = \bigcap\limits_{i=1}^{N_{\mathrm{w}}} \{d_{{\mathrm a},{\mathrm w}_k}>\kappa \},
\end{align}
\noindent which occurs when all of the Willies are outside of the semicircular region with radius $\kappa=\sqrt{\frac{\epsilon}{4 N_{\mathrm{w}}}}$ around Alice. Then, we show in Appendix~\ref{ap.0} that for any $\epsilon>0$ Alice can achieve:
\begin{figure}
	\begin{center}
		\includegraphics[scale=0.6]{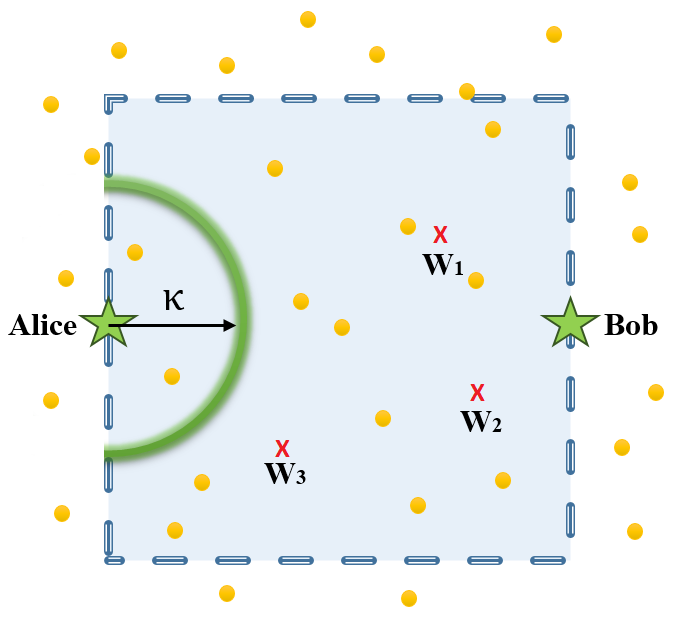}
	\end{center}
	\caption{Event $\mathcal{A}$ is true when there is no Willie in the semicircular region with radius $\kappa$ shown above. Alice is only able to communicate covertly with intended receiver Bob if $\mathcal{A}$ is true.}
	\label{fig:SysMod2}
\end{figure}
\noindent 
\begin{align}\label{eq:basic3}
\mathbb{E}_{{\mathrm F},{\mathrm W}}\left[\mathbb{P}_{\mathrm e}^{(\mathrm w)}(\boldsymbol{\sigma_{\mathrm{w}}^2},\boldsymbol{d_{\mathrm{a},\mathrm{w}}})\right|\mathcal{A}] \geq \frac{1}{2}(1- \epsilon).
\end{align}
\noindent Next, we show that since $\kappa<1/2$,
\begin{align}
\label {eq:th3a5} \mathbb{P}(\mathcal{A}) = \left(1-\frac{\pi \kappa^2}{2}\right)^{N_{\mathrm{w}}}  \stackrel{(f)}{\geq} 1-\frac{\pi N_{\mathrm{w}} \kappa^2}{2} \geq  1-2 N_{\mathrm{w}} \kappa^2= 1-\frac{\epsilon}{2},
\end{align}
\noindent where $(f)$ is true since~\eqref{eq:ineq1} is true. By~\eqref{eq:basic3},~\eqref{eq:th3a5}, and the law of total expectation
\begin{align} 
\nonumber \mathbb{P}_{\mathrm e}^{(\mathrm w)}=\mathbb{E}_{{\mathrm F},{\mathrm W}} [ \mathbb{P}_{\mathrm e}^{(\mathrm w)}(\boldsymbol{\sigma_{\mathrm{w}}^2},\boldsymbol{d_{\mathrm{a},\mathrm{w}}})]   \geq  \mathbb{E}_{{\mathrm F},{\mathrm W}} \left[\left. \mathbb{P}_{\mathrm e}^{(\mathrm w)}(\boldsymbol{\sigma_{\mathrm{w}}^2},\boldsymbol{d_{\mathrm{a},\mathrm{w}}})\right|\mathcal{A}\right]\; \mathbb{P}(\mathcal{A}) = \left(\frac{1}{2}- \frac{ \epsilon}{  2}\right)\left(1- \frac{ \epsilon}{  2}\right)\geq \frac{1}{2}-\epsilon,
\end{align}
\noindent and thus communication is covert as long as $P_{\mathrm{a}} = \mathcal{O}\left(\frac{m^{\gamma/2}}{\sqrt{n} N_{\mathrm{w}}^{\gamma/2}}\right)$.

({\em Reliability}) Next, we calculate the number of bits that Alice can send to Bob covertly and reliably. Consider arbitrarily $\zeta > 0$.  We show that Bob can achieve $\mathbb{P}_{\mathrm e}^{(\mathrm b)} < \zeta$ as $n \to \infty$, where $\mathbb{P}_{\mathrm e}^{(\mathrm b)}$ is Bob's ML decoding error probability averaged over all possible codewords and the locations of friendly nodes and Willies. Bob's noise power is $\sigma_{\mathrm b}^2 \leq \sigma_{{\mathrm b},0}^2 + \sum_{k=1}^{N_{\mathrm{w}}} \frac{P_{\mathrm{f}}}{d_{{\mathrm b},{\mathrm f}_k}^\gamma}$, where $d_{{\mathrm b},{\mathrm f}_k}$ is the distance between Bob and the closest friendly node to the $k^{\mathrm{th}}$ Willie ($W_k$), and the inequality becomes equality when each Willie has a distinct closest friendly node. By~\eqref{eq:Pebasic0} and~\eqref{eq:th3pf},
\begin{align}
   \label{eq:Pebasic1} \mathbb{P}_{\mathrm e}^{(\mathrm b)} \left(\sigma_{\mathrm b}^2\right)  & \leq
2^{nR- \frac{n}{2} \log_2 \left(1+ \frac{ c  m^{\gamma/2} }{2 \sqrt{n} \sigma_{\mathrm b}^2 N_{\mathrm{w}}^{\gamma/2}} \right)}.
 \end{align}
\noindent Suppose Alice sets $R=\min{\{R_0,1\}}$, where
\begin{align}
\label{eq:Alicesrate}R_0&=\frac{1}{4} \log_2\left(1+  \frac{c'  m^{\gamma/2} } {4 N_{\mathrm{w}}^{\gamma} \sqrt{n} } \right),\\
\nonumber c'&=c\frac {  \zeta^{\gamma/2-1} \left(\gamma-2\right)  }{2^{\gamma+3}P_{\mathrm{f}} \pi^{\gamma/2}},
\end{align}
\noindent and $c$ is defined in~\eqref{eq:th3c}. By the law of total expectation,
\begin{align}\label{eq:EPEBobth41}
\mathbb{P}_{\mathrm e}^{(\mathrm b)}= \mathbb{E}_{{\mathrm F},{\mathrm W}}[\mathbb{P}_{\mathrm e}^{(\mathrm b)}\left(\sigma_{\mathrm b}^2\right)] 
\leq \mathbb{E}_{{\mathrm F},{\mathrm W}}\left[\mathbb{P}_{\mathrm e}^{(\mathrm b)}\left( \sigma_{\mathrm b}^2\right)\Big|\frac{c'\sigma_{\mathrm b}^2}{ c N_{\mathrm{w}}^{\gamma/2}} \leq 1 \right] +   \mathbb{P}\left(\frac{c'\sigma_{\mathrm b}^2}{cN_{\mathrm{w}}^{\gamma/2}} > 1\right).
\end{align}
\noindent Consider the first term on the RHS of~\eqref{eq:EPEBobth41}. We show in Appendix~\ref{ap.1} that since $m=\omega(1)$, $N_{\mathrm{w}}=\omega(1)$, and $N_{\mathrm{w}}=o\left(n^{\frac{1}{2\gamma}}\sqrt{m}\right)$,
\begin{align} \label{eq:th3h3}
&\lim\limits_{n \to \infty} \mathbb{E}_{{\mathrm F},{\mathrm W}}\left[\mathbb{P}_{\mathrm e}^{(\mathrm b)}\left(\sigma_{\mathrm b}^2\right)\Big|\frac{c' \sigma_{\mathrm b}^2}{c N_{\mathrm{w}}^{\gamma/2}} \leq 1 \right] = 0.
\end{align}
\noindent Consider the second term on the RHS of~\eqref{eq:EPEBobth41}. To upper bound $\mathbb{P}\left(\frac{c'\sigma_{\mathrm b}^2}{cN_{\mathrm{w}}^{\gamma/2}} > 1\right)$, we define the event
\begin{align}
\nonumber \mathcal{B}= \bigcap_{k=1}^{N_{\mathrm{w}}} \left\{\{ d_{{\mathrm w}_k,{\mathrm f}_k}\leq \delta\} \cap \{ d_{{\mathrm b},{\mathrm w}_k}>2 \delta\}\right\},
\end{align}
\noindent where $\delta = \sqrt{\frac{\zeta}{4\pi N_{\mathrm{w}}}}$. 
\begin{figure}
\begin{center}
 \includegraphics[ scale=0.6]{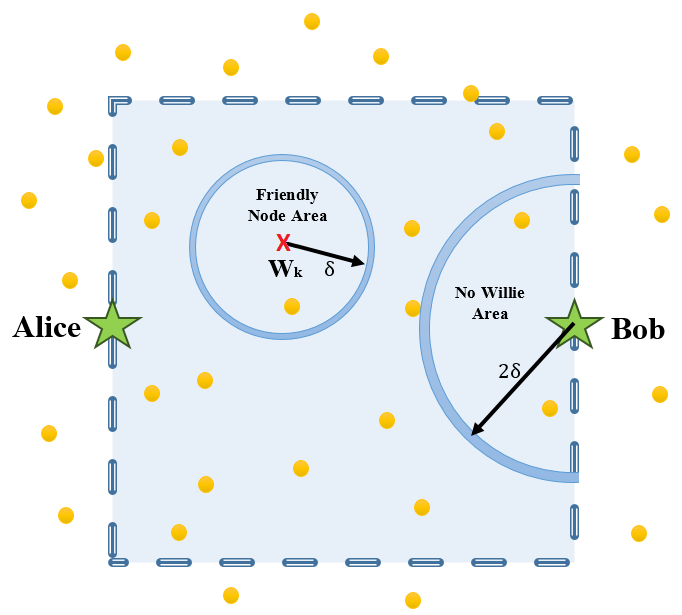}
\end{center}
 \caption{Event $\mathcal{B}$ is true when there is no Willie in the semicircular region with radius $2 \delta$ around Bob, and the distance between each Willie and the closest friendly node to him is smaller than $\delta$, i.e., $\left\{ 2d_{{\mathrm w}_k,{\mathrm f}_k} \leq \delta \}\cap \{d_{{\mathrm b},{\mathrm w}_k}>2 \delta\right\}$ for $1\leq k \leq N_{\mathrm{w}}$.}
 \label{fig:SysMod4}
 \end{figure}
\noindent This event occurs when there is no Willie in the semicircular region with radius $2 \delta$ around Bob, and the distance between each Willie and the closest friendly node to him is smaller than $\delta$ (see Fig.~\ref{fig:SysMod4}).  The law of total probability yields
\begin{align} \nonumber
\mathbb{P}\left(\frac{c' \sigma_{\mathrm b}^2}{c N_{\mathrm{w}}^{\gamma/2}} > 1 \right) &\leq \mathbb{P}\left(\frac{c' \sigma_{\mathrm b}^2}{c  N_{\mathrm{w}}^{\gamma/2}} > 1\bigg|\mathcal{B}\right)
+\mathbb{P}\left(\bar{\mathcal{B}}\right).
\end{align}
\noindent We show in Appendix~\ref{ap.2} that since $N_{\mathrm{w}}=\omega(1)$,
\begin{align}\label{eq:th3h4}
&\lim\limits_{n\to \infty} \mathbb{P}\left( \frac{c'\sigma_{\mathrm b}^2} {c N_{\mathrm{w}}^{\gamma/2}}  >  1\bigg|\mathcal{B}\right)=0,
\end{align}
\noindent and in Appendix~\ref{ap.3} that since $N_{\mathrm{w}}=\omega(1)$ and $N_{\mathrm{w}}=o\left({m}/{\log m}\right)$,
\begin{align}
\label{eq:th3h5}
&\lim\limits_{n\to \infty} \mathbb{P}\left(\bar{\mathcal{B}}\right)  = \zeta/2.
\end{align}
\noindent Thus,~\eqref{eq:EPEBobth41}-\eqref{eq:th3h5} yield $\lim\limits_{n \to \infty} \mathbb{P}_{\mathrm e}^{(\mathrm b)} < \zeta$ for any $0<\zeta<1$.

({\em Number of Covert Bits}) Similar to the analysis of Theorem~\ref{th:swillie}, we can show that when $\gamma>2$, Bob receives $\mathcal{O}\left(\min\left\{n,\frac{m^{\gamma/2} \sqrt{n}}{N_{\mathrm{w}}^{\gamma}}\right\}\right)$ bits in $n$ channel uses.
 
{\it (Converse)} We present the converse assuming that the closest friendly node to each Willie is on and the Willies know this. We show that the signal received by the closest Willie to Alice is sufficient to detect Alice's communication. Intuitively, the Willie closest to Alice has the best signal-to-noise ratio (SNR) and is the best Willie to detect Alice's communication.

Denote Willie with minimum distance to Alice by $W_{1}$. We assume that $W_1$ knows $\sigma_{{\mathrm w}_1}^2$ and the jamming scheme, in particular the distance between the closest friendly node to him and its transmit power. $W_1$ uses a power detector on his collection of observations $\left\{y_i^{(1)}\right\}_{i=1}^{n}$ to form $S=\frac{1}{n} \sum_{i=1}^{n}\left(y_i^{(1)}\right)^2$, picks a threshold $t$, and performs a hypothesis test based on $S$. If $S<\sigma_{{\mathrm w}_1}^2+t$, he chooses $H_0$ (Alice does not transmit), otherwise, $H_1$ (Alice transmits).

Observe
\begin{align}\label{eq:th3williesnoise}
\sigma_{{\mathrm w}_1}^2\leq\sigma_{{\mathrm w}_{1,0}}^2+\sum_{k=1}^{N_{\mathrm{w}}}\frac{P_{\mathrm{f}}}{ d_{{\mathrm w}_1,{\mathrm f}_k}^\gamma},
\end{align}
\noindent where $\sigma_{{\mathrm w}_{1,0}}^2$ is Willie's noise power when all of the friendly nodes are off, i.e., AWGN, and $d_{{\mathrm w}_1,{\mathrm f}_k}$ is the distance between $W_1$ and the closest friendly node to $W_k$. Note that~\eqref{eq:th3williesnoise} becomes equality when all of the Willies have a distinct closest friendly node. 
Similar to the converse in Theorem~\ref{th:swillie}, we show that
\begin{align}
 \label{eq:th374} \mathbb{E}_{Y}\left[S|H_0\right] &=\sigma_{{\mathrm w}_1}^2,\\
\label{eq:th373} \mathrm{Var}_{Y}[S|H_0]&=\frac{2\sigma_{{\mathrm w}_1}^4}{n}, \\
 \label{eq:th375}\mathbb{E}_{Y}[S|H_1]&=\sigma_{{\mathrm w}_1}^2+\frac{P_{\mathrm{a}}}{d_{{\mathrm a},{\mathrm w}_1}^{\gamma}},\\
\label{eq:th201} \mathrm{Var}_{Y}[S|H_1]&= \frac{4 P_{\mathrm{a}} \sigma_{{\mathrm w}_1}^2 }{n d_{{\mathrm a},{\mathrm w}_1}^{\gamma}}  + \frac{2\sigma_{{\mathrm w}_1}^4}{n}.
\end{align}
\noindent If $S<\sigma_{{\mathrm w}_1}^2+t$, $W_1$ accepts $H_0$; otherwise, he accepts $H_1$. In the converse of Theorem~\ref{th:swillie} we upper bounded Willie's noise power by the received noise power when all of the friendly nodes are on. Similar to the arguments leading to~\eqref{eq:18} we show that if we choose $t = \frac{\sqrt{8 } \rho  m^{\gamma/2}}{\sqrt{n \lambda }}$, where $\rho$ is given in~\eqref{eq:rho}, then:
\begin{align} \label{eq:th335}
\mathbb{P}_{\mathrm {FA}} \leq  \frac{\lambda}{2}.
\end{align}

\noindent Now, consider $\mathbb{P}_{\mathrm {MD}}(\boldsymbol{\sigma_{\mathrm{w}}^2},\boldsymbol{d_{\mathrm{a},\mathrm{w}}})$. Similar to the approach leading to~\eqref{eq:th1conv1}, we obtain
\begin{align} \label{eq:th3conv1}
 \mathbb{P}_{\mathrm {MD}}(\boldsymbol{\sigma_{\mathrm{w}}^2},\boldsymbol{d_{\mathrm{a},\mathrm{w}}})\leq \frac{4{ \frac{P_{\mathrm{a}}}{d_{{\mathrm a},{\mathrm w}_1}^\gamma} \sigma_{{\mathrm w}_1}^2 + 2\sigma_{{\mathrm w}_1}^4}
 }{n \left(\frac{P_{\mathrm{a}}}{d_{{\mathrm a},{\mathrm w}_1}^\gamma}-t \right)^2}.
\end{align}
\noindent Define the event  $\mathcal{E} = \{d_{{\mathrm{w}_1},f_1}>\eta_1\} \cap \{\ell \beta' \leq d_{{\mathrm a},{\mathrm w}_1} <  \beta' \}$, where $\eta_1$ is defined in~\eqref{eq:eta1}, and
\begin{align}
\nonumber \beta'&=\sqrt{{2\ln{(8/\lambda)}}/{(\pi N_{\mathrm{w}})}},\\
\label{eq:ell} \ell&=\sqrt{{\ln{(1-\lambda/8)}}/{\ln{(\lambda/8)}}}.
\end{align}
\noindent The law of total expectation yields
\begin{align}
 \label{eq:th3e3}\mathbb{P}_{\mathrm {MD}}=\mathbb{E}_{{\mathrm F},{\mathrm W}} [\mathbb{P}_{\mathrm {MD}}(\boldsymbol{\sigma_{\mathrm{w}}^2},\boldsymbol{d_{\mathrm{a},\mathrm{w}}})] 
&\leq   \mathbb{E}_{{\mathrm F},{\mathrm W}} \left[\left.\mathbb{P}_{\mathrm {MD}}(\boldsymbol{\sigma_{\mathrm{w}}^2},\boldsymbol{d_{\mathrm{a},\mathrm{w}}})\right|\mathcal{E} \right] + \mathbb{P}(\overline{\mathcal{E}}) .
\end{align}
\noindent We show in Appendix~\ref{ap.7.5} since $m=\omega(1)$, and $N_{\mathrm{w}}=\omega(1)$,
\begin{align}
\label{eq:th328} \lim\limits_{n \to \infty} \mathbb{P}(\overline{\mathcal{E}})    &\leq  \lambda/2,
\end{align}
\noindent and, in Appendix~\ref{ap.8} that
\begin{align} 
\label{eq:th333} \mathbb{E}_{{\mathrm F},{\mathrm W}} \left[\left.\mathbb{P}_{\mathrm {MD}}(\boldsymbol{\sigma_{\mathrm{w}}^2},\boldsymbol{d_{\mathrm{a},\mathrm{w}}})\right|\mathcal{E}\right] &\leq \frac{4{ \frac{P_{\mathrm{a}}}{(\ell \beta')^\gamma} \mathbb{E}_{{\mathrm F},{\mathrm W}} [\sigma_{{\mathrm w}_1}^2|d_{{\mathrm{w}_1},f_1}>\eta_1] }
}{n \left(\frac{P_{\mathrm{a}}}{\beta'^\gamma}-t \right)^2}  +  \frac{ 2\mathbb{E}_{{\mathrm F},{\mathrm W}} [\sigma_{{\mathrm w}_1}^4 |d_{{\mathrm{w}_1},f_1}>\eta_1] }{n \left(\frac{P_{\mathrm{a}}}{\beta'^\gamma}-t \right)^2},
\end{align}
\noindent Consider $\mathbb{E}_{{\mathrm F},{\mathrm W}} \left[\sigma_{\mathrm{w}_1}^2|d_{{\mathrm{w}_1},{\mathrm{f}_1}}>\eta_1 \right]$ and $\mathbb{E}_{{\mathrm F},{\mathrm W}} \left[\sigma_{\mathrm{w}_1}^4|d_{{\mathrm{w}_1},{\mathrm{f}_1}}>\eta_1 \right]$  in~\eqref{eq:22}. Similar to the arguments leading to~\eqref{eq:16}, we show that $\mathbb{E}_{{\mathrm F},{\mathrm W}} \left[\sigma_{\mathrm{w}_1}^4|d_{{\mathrm{w}_1},{\mathrm{f}_1}}>\eta_1 \right]=\mathcal{O}(m^{\gamma})$. Consequently, Jensen's inequality yields $\mathbb{E}_{{\mathrm F},{\mathrm W}} \left[\sigma_{\mathrm{w}_1}^2|d_{{\mathrm{w}_1},{\mathrm{f}_1}}>\eta_1 \right]=\mathcal{O}(m^{\gamma/2})$.
Since $t=\Theta({m^{\gamma/2}}/{\sqrt{n}})$, $\beta'=\Theta(1/\sqrt{N_{\mathrm{w}}})$, $m=\omega(1)$, and $N_{\mathrm{w}}=\omega(1)$, if Alice sets her average symbol power $P_{\mathrm{a}}=\omega\left(\frac{m^{\gamma/2}}{\sqrt{n}N_{\mathrm{w}}^{\gamma/2}}\right)$, $\mathbb{E}_{{\mathrm F},{\mathrm W}} [\mathbb{P}_{\mathrm {MD}}(\boldsymbol{\sigma_{\mathrm{w}}^2},\boldsymbol{d_{\mathrm{a},\mathrm{w}}})|\mathcal{E}] = 0$ as $n \to \infty$. By~\eqref{eq:th3e3} and~\eqref{eq:th328}
\begin{align}
\label{eq:th334} \lim\limits_{n \to\infty} \mathbb{P}_{\mathrm {MD}} 
&\leq  \lambda /2.
\end{align}
\noindent Combined with~\eqref{eq:th335},  $\mathbb{P}_{\mathrm {FA}} + \mathbb{P}_{\mathrm {MD}} \leq \lambda$ for any $\lambda>0$.
 
Thus, to avoid detection for a given codeword, Alice must set the power of that codeword to $P_{\mathcal{U}} = \mathcal{O}\left(\frac{m^{\gamma/2}}{\sqrt{n}N_{\mathrm{w}}^{\gamma/2}}\right)$. Suppose that Alice's codebook contains a fraction $\xi>0$ of codewords with power $ P_{\mathcal{U}} = \mathcal{O}\left(\frac{m^{\gamma/2}}{\sqrt{n}N_{\mathrm{w}}^{\gamma/2}}\right)$. Similar to converse of Theorem~\ref{th:swillie}, given the locations of the friendly nodes, Bob's decoding error probability of such low power codewords is lower bounded by (see~\eqref{eq:th1con5})
\begin{align} 
\nonumber \mathbb{P}_{\mathrm e}^{\mathcal{U}}(\sigma_{\mathrm b}^2) &\geq 1- \frac{\frac{P_{\mathcal{U}}}{2\sigma_{\mathrm b}^2}+\frac{1}{n}}{\frac{\log_2  \xi}{n}+R}.
\end{align}
\noindent Denote the closest Willie to Bob by $W_2$. Since Bob's noise is lower bounded by the noise generated from the closest friendly node to $W_2$, $\sigma_{\mathrm b}^2\geq \frac{P_{\mathrm{f}}}{d_{{\mathrm b},{\mathrm f}_2}^\gamma}$,
\begin{align} \nonumber \mathbb{P}_{\mathrm e}^{\mathcal{U}}(\sigma_{\mathrm b}^2)  &\geq 1-\frac{\frac{P_{\mathcal{U}}d_{{\mathrm b},{\mathrm f}_2}^\gamma}{2 P_{\mathrm{f}}}+\frac{1}{n}}{\frac{\log_2 \xi}{n}+R} .
\end{align}
\noindent Define the event $\mathcal{F} = \left\{{ d_{{\mathrm b},{\mathrm f}_2}}<\sqrt{{8 \ln{(1/\tau)}}/{(\pi N_{\mathrm{w}}})} \right\}$, where $0<\tau<1$. The law of total expectation yields
\begin{align}\label{eq:th3c01}
\mathbb{P}_{\mathrm e}^{\mathcal{U}}=\mathbb{E}_{{\mathrm F},{\mathrm W}}\left[\mathbb{P}_{\mathrm e}^{\mathcal{U}}(\sigma_{\mathrm b}^2) \right] \geq \mathbb{E}_{{\mathrm F},{\mathrm W}}\left[\left. \mathbb{P}_{\mathrm e}^{\mathcal{U}}(\sigma_{\mathrm b}^2)\right| \mathcal{F}\right] \mathbb{P}\left(\mathcal{F}\right).
\end{align}
\noindent Consider $\mathbb{P}\left(\mathcal{F}\right)$. We show in Appendix~\ref{ap.9} that since $m=\omega(1)$, $N_{\mathrm{w}}=\omega(1)$, and $N_{\mathrm{w}}=o(m/\log{m})$,
\begin{align}\label{eq:th3c3.6}
 \lim\limits_{n \to \infty} \mathbb{P}\left( \mathcal{F} \right)=1-\tau.
\end{align}
\noindent Now, consider $\mathbb{E}_{{\mathrm F},{\mathrm W}}\left[\left. \mathbb{P}_{\mathrm e}^{\mathcal{U}}\right| \mathcal{F}\right]$ in~\eqref{eq:th3c01}.
\begin{align} \mathbb{E}_{{\mathrm F},{\mathrm W}}\left[\left. \mathbb{P}_{\mathrm e}^{\mathcal{U}}(\sigma_{\mathrm b}^2)\right| \mathcal{F}\right]  
\nonumber \geq 1- \mathbb{E}_{{\mathrm F},{\mathrm W}}\left[\left. \frac{\frac{P_{\mathcal{U}}d_{{\mathrm b},{\mathrm f}_2}^\gamma}{2 P_{\mathrm{f}}}+\frac{1}{n}}{\frac{\log_2 \xi}{n}+R} \right| \mathcal{F}\right] 
 &\stackrel{(g)}{\geq} 1- \mathbb{E}_{{\mathrm F},{\mathrm W}}\left[\left. \frac{\frac{P_{\mathcal{U}}\left(\frac{\frac{2}{\pi} \ln{\frac{1}{\tau}} }{\sqrt{N_{\mathrm{w}}}}\right)^\gamma}{2P_{\mathrm{f}}}+\frac{1}{n}}{\frac{\log_2 \xi}{n}+R} \right| \mathcal{F}\right], \\
\nonumber &= 1- \mathbb{E}_{{\mathrm F},{\mathrm W}}\left[\left. \frac{\frac{P_{\mathcal{U}}}{N_{\mathrm{w}}^{\gamma/2}}\frac{\left({\frac{2}{\pi} \ln{\frac{1}{\tau}} }\right)^\gamma}{2P_{\mathrm{f}}}+\frac{1}{n}}{\frac{\log_2 \xi}{n}+R} \right| \mathcal{F}\right] ,
\end{align}
\noindent where $(g)$ is true since $\mathcal{F}$ occurs. Suppose Alice desires to transmit $\omega \left(\frac{\sqrt{n} m^{\gamma/2}}{N_{\mathrm{w}}^{\gamma}}\right)$ covert bits in $n$ channel uses. Therefore, her rate (bits/symbol) is $R=\omega \left(\frac{ m^{\gamma/2}}{\sqrt{n} N_{\mathrm{w}}^{\gamma}}\right)$. Since $P_{\mathcal{U}} = \mathcal{O}\left(\frac{m^{\gamma/2}}{\sqrt{n}N_{\mathrm{w}}^{\gamma/2}}\right)$, $m=\omega(1)$, and $N_{\mathrm{w}}=\omega(1)$,
\begin{align} \label{eq:th3c1}
\lim\limits_{n \to \infty} \mathbb{E}_{{\mathrm F},{\mathrm W}}\left[\left. \mathbb{P}_{\mathrm e}^{\mathcal{U}}(\sigma_{\mathrm b}^2)\right| \mathcal{F}\right]=1.
\end{align}
\noindent By~\eqref{eq:th3c01},~\eqref{eq:th3c3.6}, and~\eqref{eq:th3c1}, for any $0<\tau<1$, $\lim\limits_{n \to \infty}  \mathbb{P}_{\mathrm e}^{\mathcal{U}}\geq 1-\tau$, and thus $\mathbb{E} \left[ \mathbb{P}_{\mathrm e}^{\mathcal{U}}\right]$ is bounded away from zero.
\end{proof}
\subsection{$\gamma=2$}  \label{sec:gamma2}
\begin{mthm}{2.2} \label{th:mwillie2}
When friendly nodes are independently distributed according to a two-dimensional Poisson point process with density $m=\omega(1)$, and $N_{\mathrm{w}}$ collaborating Willies are uniformly and independently distributed over the unit square shown in Fig.~\ref{fig:SysMod}. If $N_{\mathrm{w}}={o}\left({m}/{\log{m}}\right)$, then Alice can reliably and covertly transmit $\mathcal{O}\left(\min\left\{n,\frac{m \sqrt{n}}{N_{\mathrm{w}}^{2}\log^2 {N_{\mathrm{w}}}}\right\}\right)$ bits to Bob in $n$ channel uses. 
\end{mthm}  

We present the proof assuming $N_{\mathrm{w}}=\omega(1)$, as the proof for a finite $N_{\mathrm{w}}$ follows from it. In addition, according to the statement of Theorem~\ref{th:mwillie2},  $N_{\mathrm{w}}=\Omega\left(\frac{m\sqrt{n}}{\log{(m\sqrt{n})}}\right)$ then Alice can reliably and covertly transmit $\mathcal{O}\left(1\right)$ bits to Bob in $n$ uses of channel, which is not of interest. Therefore, we present the proof assuming $N_{\mathrm{w}}={o}\left(\min\left\{\frac{m}{\log{m}},\frac{m\sqrt{n}}{\log{(m\sqrt{n})}}\right\}\right)$.

\begin{proof} 

{\it (Achievability)} 

\textbf{Construction:} The construction and Bob's decoding are the same as those of Theorem~\ref{th:mwillie2p}.

\textbf{Analysis:} ({\em Covertness}) The difference between the results for $\gamma>2$ and $\gamma=2$ originates from the following integral necessary in the proofs:
 \[\int \frac{dx}{x^{\gamma-1}}=  \begin{cases} 
       {x^{2-\gamma}}/({2-\gamma})+c_0, & \gamma>2 \\
       \ln{x}+c_0', & \gamma=2 
    \end{cases},
 \]
\noindent where $c_0$ and $c_0'$ are constants. Therefore, the analysis for $\gamma=2$ follows similarly with a few minor modifications. Alice sets her average symbol power $P_{\mathrm{a}} \leq \frac{c  m}{\sqrt{n} N_{\mathrm{w}} \ln{N_{\mathrm{w}}}} $ where 
\begin{align}
\label{newc} c=  {4 \sqrt{2}  \epsilon \pi}P_{\mathrm{f}} .
\end{align}
\noindent Next, we modify~\eqref{eq:6} to $\mathbb{E}_{W}\left[\left.  \frac{1}{d_{{\mathrm a},{\mathrm w}_k}^2}\right|d_{{\mathrm a},{\mathrm w}_k} >  \kappa\right]   \leq  \pi \ln{(N_{\mathrm{w}})}$. Then, we show that Alice achieves \eqref{eq:basic3} and thus her communication is covert as long as  $P_{\mathrm{a}}=\mathcal{O}\left(\frac{m}{\sqrt{n} N_{\mathrm{w}} \log{N_{\mathrm{w}}}}\right)$. 

({\em Reliability}) Similar to the approach in the reliability for $\gamma>2$, we can show that if Alice sets $R=\min{\{1,R_0\}}$, where
\begin{align}
\label{eq:Alicesrate2}R_0&=\frac{1}{4} \log_2\left(1+  \frac{c'  m } {4 N_{\mathrm{w}}^{2} (\ln{N_{\mathrm{w}}})^2 \sqrt{n} } \right),\\
\nonumber c'&=\frac {  c }{ 8 \pi P_{\mathrm{f}}},
\end{align}
\noindent and $c$ is defined in~\eqref{newc}, then $m=\omega(1)$, $N_{\mathrm{w}}=\omega(1)$, and $N_{\mathrm{w}}={o}\left(\min\left\{\frac{m}{\log{m}},\frac{m\sqrt{n}}{\log{(m\sqrt{n})}}\right\}\right)$ yield $\lim\limits_{n \to \infty} \mathbb{P}_{\mathrm e}^b < \zeta$ for any $0<\zeta<1$.

({\em Number of Covert Bits}) Similar to the analysis for $\gamma>2$, by~\eqref{eq:Alicesrate2}, Bob receives $\mathcal{O}\left(\min\left\{n,\frac{m^{\gamma/2} \sqrt{n}}{N_{\mathrm{w}}^{2}\log^2 {N_{\mathrm{w}}}}\right\}\right)$ bits in $n$ channel uses. 
\end{proof}

({\em Converse}) The approach used for $\gamma>2$, which involved choosing the closest Willie to Alice to decide whether Alice communicates with Bob or not, does not yield a tight result for $\gamma=2$. Using this approach,  we can show that if Alice sets her average symbol power $P_{\mathrm{a}}=\omega\left(\frac{m}{\sqrt{n}N_{\mathrm{w}}}\right)$, then Willie detects her with arbitrarily small sum of error probabilities. However, from the achievability, we expect that $P_{\mathrm{a}}=\omega\left(\frac{m}{\sqrt{n}N_{\mathrm{w}} \log{N_{\mathrm{w}}}}\right)$ results in detection. This suggests that Willies have to consider their signals received collectively to detect Alice's communication, as we expect for $\gamma=2$ the signal decays slowly with distance.

\section{Discussion}\label{sec:dis}
\subsection{Assumption of $m=\omega(1)$ in Theorems~\ref{th:mwillie2p} and~\ref{th:mwillie2}}

In Theorems~\ref{th:mwillie2p} and~\ref{th:mwillie2}, we assumed $m=\omega(1)$ in order to simplify the proof when $N_w=\omega(1)$, but this condition can be relaxed. When relaxing this assumption, we also have to replace the condition $N_{\mathrm w}=o(m/\log{m})$ with $N_{\mathrm w}\leq \frac{m \zeta}{4 \log{(m \zeta/4)}}$. Furthermore, $m=\omega(1)$ becomes plausible when the single-hop communication scheme presented in this paper is extended to the covert multi-hop communication over large wireless networks~\cite{gupta_kumar,francheschetti} where a collection of nodes work to establish covert communication between a collection of source and destination pairs. In this case, the number of nodes often grows in the region of a single hop of communication~\cite{arkin2015optimal,vasudevan2010security} with the size of the network~\cite{gupta_kumar,francheschetti,wu2012optimum,mukherjee2014principles}. Note that we have allowed a growing number of nodes for both friendly nodes ($m=\omega(1)$) and warden Willies ($N_w=\omega(1))$.

An example of employing artificial noise generation with a growing density of nodes in a large wireless network is presented in~\cite{vasudevan2010security}, where authors analyze the throughput of key-less secure communication in a cell of size $\sqrt{n} \times \sqrt{n}$ and exploit the dynamics of wireless fading channels to achieve secret communication. In particular, transmitter and receiver nodes are distributed according to a Poisson point process with density one in the cell, and each node is allowed to generate artificial noise.

\subsection{Assumption of turning on only the closest friendly node to each Willie}
For the achievability proofs in this paper, our strategy was turning on the closest friendly node to each Willie and keeping other friendly nodes off. For the case of a single Willie and $\gamma>2$, the converse of Theorem~\ref{th:swillie}, which is done over all strategies for turning on the friendly nodes, shows that this was indeed an optimal strategy. However, for the converses of Theorems~\ref{th:swillie2} and~\ref{th:mwillie2p}, we had to restrict ourselves to considering only those strategies that turn on the closest friendly node to each Willie.  Whereas this is a limitation of that converse, it is likely that this strategy is either optimal or close to optimal in practice. In particular, in~\cite{sankararaman2014optimization,Sankararaman:2012:OSP:2248371.2248383}, the authors propose that this strategy is optimal in wireless communication when the jammers (friendly nodes) have the same finite power, and using simulations they show that the noise received from other nodes (second closest node, third closest node, ...) is negligible compared to the noise received from the nearest jammer (friendly node). The optimality of this strategy is also addressed in~\cite{arkin2015optimal}.

Switching on only the closest node to Willie(s) requires knowing the location of Willie(s), collaboration between friendly nodes, and switching off a large number of friendly nodes, which might entail a high cost. However, given the importance of covert communication and the demand for it in specific applications (e.g., military), it is reasonable to pay the cost in these applications to increase the throughput of covert communication to a throughput higher than $\mathcal{O}({\sqrt{n}})$ bits in $n$ channel uses~\cite{bash_jsac2013}. In addition to the this strategy, here we discuss an alternative strategy without these requirements: we only turn off the friendly nodes whose distances to Bob are smaller than $\iota>0$, and we assume that other friendly nodes are on, independently, with probability $p>0$, where $\iota$ and $p$ are independent of $m,n,N_{\mathrm{W}}$. Compared to our previous strategy, $\mathbb{E}_{{\mathrm{F}}}[1/\sigma_{\mathrm w}^2]=\mathcal{O}(m^{-\gamma/2})$ remains the same; however, the conjecture is that $\mathbb{E}_{{\mathrm{F}}}[\sigma_{\mathrm b}^2]$ changes from $\mathcal{O}(1)$ to $\mathcal{O}(m)$, and that Alice can reliably and covertly transmit $\mathcal{O}(\min\{{n,m^{\gamma/2-1}\sqrt{n}}\})$ bits to Bob in $n$ channel uses. Also, it is a conjecture that for a scenario with multiple Willies provided $\gamma>2$, Alice can reliably and covertly transmit $\mathcal{O}\left(\min\left\{n,\frac{m^{\gamma/2-1} \sqrt{n}}{N_{\mathrm{w}}^{\gamma}}\right\}\right)$ bits to Bob in $n$ channel uses. 

\subsection{High probability results}
In this paper, our covertness metric (see Definition~\ref{def:cov}) requires lower bounding the expected value of Willies' probability of error ($\mathbb{P}_e^{(\mathrm{w})}$) over all instantiations of the locations of Willies and friendly nodes, by $\frac{1}{2}-\epsilon$ for all $\epsilon$. In Appendix~\ref{ap.10}, we present an example of the high probability result for the covertness of the single Willie scenario.

\subsection{Assumption of uniform distribution for Willies} 
For spatial modeling of wireless networks, a Poisson point process is the most common choice~\cite{elsawy2013stochastic,andrews2010primer,haenggi2012stochastic}. When a Poisson point process is conditioned on the number of points in an area, the locations of the points in that area become uniformly distributed. In this paper, our goal was to first consider the case of a single Willie and then extend the results to multiple Willies. Therefore, in Theorems~\ref{th:swillie} and~\ref{th:swillie2}, we considered on adversary (Willie) whose location was uniformly distributed on a unit box (see Fig.~\ref{fig:SysMod}). Then, to be consistent with the single Willie scenario, we modeled the locations of the Willies (Theorems~\ref{th:mwillie2p} and~\ref{th:mwillie2}) by a uniform distribution. We do not expect the results to differ if we model the locations of the Willies by a Poisson point process. In Appendix~\ref{ap.11}, we verify this fact by presenting the analysis and the results for the case where the locations of the Willies are modeled by a Poisson process of rate $\lambda_N$ and $\gamma>2$. The results do not differ from that of Theorem~\ref{th:mwillie2p} except for the replacement of $\lambda_N$ with $N_{\mathrm{w}}$.

\section{Conclusion} \label{sec:con}
In this paper, we have considered the first step in establishing covert communications in a network scenario. We establish that Alice can transmit $\mathcal{O}(\min\{n,m^{\gamma/2}\sqrt{n}\})$ bits reliably to the desired recipient, Bob, in $n$ channel uses without detection by an adversary Willie, if randomly distributed system nodes of density $m$ are available to aid in jamming Willie; conversely, no higher covert rate is possible for $\gamma=2$ assuming that the nearest node to Willie is used to jam his receiver, and for $\gamma>2$ without this assumption. The presence of multiple collaborating adversaries inhibits communication in two separate ways: (1) increasing the effective SNR at the adversaries' decision point; and (2) requiring more interference, which inhibits Bob's ability to reliably decode the message. We established that in the presence of $N_{\mathrm{w}}$ Willies, Alice can reliably and covertly send $\mathcal{O}\left(\min\left\{n,\frac{\sqrt{n} m^{\gamma/2}}{ N_{\mathrm{w}}^{\gamma}}\right\}\right)$ bits to Bob when $\gamma >2$, and $\mathcal{O}\left(\min\left\{n,\frac{\sqrt{n} m}{ N_{\mathrm{w}}^2 \log^2 {N_{\mathrm{w}}}}\right\}\right)$ when $\gamma = 2$. Conversely, if the closest friendly node to each adversary transmits noise, no higher covert throughput is possible for $\gamma>2$. Future work consists of proving the converse for $\gamma=2$ and embedding the results of this single-hop formulation into large multi-hop covert networks.
\appendix

\makeatletter
\renewcommand{\theparagraph}{\Alph{paragraph}}
\renewcommand{\@seccntformat}[1]{\csname the#1\endcsname.\quad}
\makeatother

\paragraph{\textbf{Proof of~\eqref{eq:ineq0}}}\label{ap.01}

Consider $x\geq 0$, $f(x)=\ln(1+x)$, and $g(x)=x-\frac{x^2}{2}$. Therefore
\begin{align}
\nonumber f'(x)-g'(x)=  \frac{1}{1+x} -\left(1-x\right) =\frac{x^2}{1+x} \geq 0.
\end{align}
\noindent On the other hand $f(0)=g(0)=0$, therefore
\begin{align}
\nonumber f(x)-g(x)=\int\limits_{0}^{x} \left(f'\left(x\right)-g'\left(x\right)\right) dx  \geq 0.
\end{align}
\noindent Thus,  $\ln(1+x) \geq x-\frac{x^2}{2}$ for $x\geq 0$.

\paragraph{\textbf{Proof of~\eqref{eq:th101}}}\label{ap.015} Taking the conditional expected value of both sides of~\eqref{eq:expcond13} yields:
\begin{align}
\nonumber   \mathbb{E}_{{\mathrm F},{\mathrm W}} [ \mathbb{P}_{\mathrm e}^{(\mathrm w)}
(\sigma_{\mathrm{w}}^2,d_{\mathrm{a},\mathrm{w}})|d_{{\mathrm a},{\mathrm w}}>\psi]  &\geq  \frac{1}{2}- \sqrt{\frac{1}{8}} \mathbb{E}_{{\mathrm F},{\mathrm W}}\left[\left.\frac{c m^{\gamma/2}}{2 \sigma_{\mathrm w}^2 d_{{\mathrm a},{\mathrm w}}^{\gamma}}\right|d_{{\mathrm a},{\mathrm w}}>\psi\right],\\
\label{eq:th2eq1}  &\geq \frac{1}{2}-\frac{c m^{\gamma/2}}{4 \sqrt{2}\psi^{\gamma}}
\mathbb{E}_{{\mathrm F},{\mathrm W}}\left[\left.\frac{1}{\sigma_{\mathrm w}^2}\right|d_{{\mathrm a},{\mathrm w}}>\psi\right] = \frac{1}{2}-\frac{c m^{\gamma/2}}{4 \sqrt{2}\psi^{\gamma}}
\mathbb{E}_{\mathrm F}\left[\frac{1}{\sigma_{\mathrm w}^2}\right],
\end{align}
\noindent where the second inequality is true since when $d_{{\mathrm a},{\mathrm w}}>\psi$, ${1}/{d_{{\mathrm a},{\mathrm w}}^\gamma}\leq {1}/{\psi^\gamma}$, and the equality is true because friendly nodes are distributed according to a Poisson point process over the entire plane, and thus Willie's noise characteristics are independent of his location. The pdf of $d_{{\mathrm w},{\mathrm f}}$ is\cite[p. 10]{moltchanov2012distance}
\begin{equation}
\label{eq:poisson}
{f}_{d_{{\mathrm w},{\mathrm f}}}\left( x \right) = 2 m \pi x \, e^{-m \pi x^2 }.
\end{equation}
\noindent Therefore,
\begin{align}\label{eq:totalex1}
\mathbb{E}_{\mathrm F} \left[\frac{1}{\sigma_{\mathrm w}^2}\right]  =   \mathbb{E}_{\mathrm F}\left[\frac{1}{\sigma_{{\mathrm w},0}^2+{P_{\mathrm{f}}}/{d_{{\mathrm w},{\mathrm f}}^\gamma}}\right]  \leq \frac{\mathbb{E}_{{\mathrm F},{\mathrm W}}\left[d_{{\mathrm w},{\mathrm f}}^\gamma \right]}{P_{\mathrm{f}}} = \frac{2 m \pi}{P_{\mathrm{f}}} \int_{0}^{\infty} x^{\gamma+1} e^{-m \pi x^2 } dx = \frac{\Gamma \left(\gamma/2+1\right)}{2 P_{\mathrm{f}} \pi^{\gamma/2+1} m^{\gamma/2}}.
\end{align}
\noindent By~\eqref{eq:th2eq1},~\eqref{eq:totalex1}, and substituting the value of $c$, we achieve~\eqref{eq:th101}.

\paragraph{\textbf{Proof of~\eqref{eq:ineq1}}}\label{ap.02} Generalized Bernouli's Inequality.
Consider $x>-1$ and $r\geq 1$. If $1+rx\leq 0$, the inequality is trivial. Suppose $1+rx>0$. Since $\log$ function is concave, if $x>-1$ and $r\geq 1$, the Jensen's inequality yields:
\begin{align}
\nonumber \frac{1}{r} \log{\left(1+rx\right)}+\frac{r-1}{r}\log{(1)} &\leq \log{\left(\frac{1}{r}\left(1+rx\right)+\frac{r-1}{r}\right)}=\log{\left(1+rx\right)}.
\end{align}
Therefore, $(1 + x)^{-r} \leq \left(1 + rx\right)^{-1}$ for any $x>-1$ and $r\geq1$.

\paragraph{\textbf{Proof of~\eqref{eq:12}}}\label{ap.3.1}
Let $ \mathcal{H}=\{d_{{\mathrm w},{\mathrm f}}>\eta_1\} \cap \{N_{\mathrm f}\}$ be the event that the distance between Willie and the closest friendly node to him is larger than $\eta_1$ and there are $N_{\mathrm f}$ friendly nodes in the area surrounded by circles of radii $\eta_1$ and $r$ centered at Willie. Squaring both sides of~\eqref{eq:02} and taking the expected value of them, given $\mathcal{H}$ yields:
\begin{align}
\nonumber  \mathbb{E}_{{\mathrm F},{\mathrm W}} \left[\sigma_{\mathrm w}^4(r) |  \mathcal{H} \right] =
\sigma_{{\mathrm w},0}^4 &+  2  P_{\mathrm{f}}\sigma_{{\mathrm w},0}^2 
N_{\mathrm{f}} \mathbb{E}_{{\mathrm F}} [ {1}/{d_{\mathrm{w},\mathrm{f}_i}^{\gamma}}|d_{{\mathrm w},{\mathrm f}}>\eta_1 ] +P_{\mathrm{f}}^2 N_{\mathrm{f}} \mathbb{E}_{{\mathrm F}} [{1}/{d_{\mathrm{w},\mathrm{f}_i}^{2 \gamma}}|d_{{\mathrm w},{\mathrm f}}>\eta_1]\\
\label{eq:05} &+ P_{\mathrm{f}}^2(N^2_{\mathrm{f}}-N_{\mathrm{f}} ) \mathbb{E}_{{\mathrm F}} [{1}/{d_{\mathrm{w},\mathrm{f}_i}^{\gamma}}|d_{{\mathrm w},{\mathrm f}}>\eta_1]^2.
\end{align}
\noindent The expectations on the RHS of~\eqref{eq:05} are only over the locations of the friendly nodes since Willie's noise characteristics are independent of his location. In addition, the conditions on the expectations on the RHS of ~\eqref{eq:05} are reduced from $ \mathcal{H}$ to $d_{{\mathrm w},{\mathrm f}}>\eta_1$. 
Denote by $\mathbb{E}_{N_{\mathrm f}}[\cdot]$ the expectation over values of $N_{\mathrm f}$. By the law of total expectation:
\begin{align}
\label{eq:06}\mathbb{E}_{{\mathrm F},{\mathrm W}} \left[\sigma_{\mathrm w}^4(r)|d_{{\mathrm w},{\mathrm f}}>\eta_1 \right] &= \mathbb{E}_{N_{\mathrm f}}\left[\mathbb{E}_{{\mathrm F},{\mathrm W}} \left[\sigma_{\mathrm w}^4(r) |   \mathcal{H} \right]\right].
\end{align}
\noindent By~\eqref{eq:06},~\eqref{eq:05} becomes:
\begin{align}
\nonumber  \mathbb{E}_{{\mathrm F},{\mathrm W}} \left[\sigma_{\mathrm w}^4(r)|d_{{\mathrm w},{\mathrm f}}>\eta_1 \right]=
\sigma_{{\mathrm w},0}^4 &+  2  P_{\mathrm{f}}\sigma_{{\mathrm w},0}^2 
\mathbb{E}_{N_{\mathrm f}}[N_{\mathrm{f}}] \mathbb{E}_{{\mathrm F}} [ {1}/{d_{\mathrm{w},\mathrm{f}_i}^{\gamma}}|d_{{\mathrm w},{\mathrm f}}>\eta_1 ] + P_{\mathrm{f}}^2 \mathbb{E}_{N_{\mathrm f}}[N_{\mathrm{f}}] \mathbb{E}_{{\mathrm F}} [{1}/{d_{\mathrm{w},\mathrm{f}_i}^{2 \gamma}}|d_{{\mathrm w},{\mathrm f}}>\eta_1]\\
\label{eq:07}&+ P_{\mathrm{f}}^2 \mathbb{E}_{N_{\mathrm f}}[N^2_{\mathrm{f}}-N_{\mathrm{f}}] \mathbb{E}_{{\mathrm F}} [{1}/{d_{\mathrm{w},\mathrm{f}_i}^{\gamma}}|d_{{\mathrm w},{\mathrm f}}>\eta_1]^2.
\end{align}
\noindent Because $N_{\mathrm{f}}$ is a sample of a Poisson distribution with mean $m(\pi r^2 - \pi \eta_1^2)$:
\begin{align} 
&\label{eq:08}\mathbb{E}_{N_{\mathrm f}}[N_{\mathrm{f}}]  = m(\pi r^2 - \pi \eta_1^2)\leq m \pi r^2,\\
&\label{eq:09}\mathbb{E}_{N_{\mathrm f}}[N^2_{\mathrm{f}}-N_{\mathrm{f}}]  = \mathrm{Var}(N_{\mathrm{f}})+ \mathbb{E}_{N_{\mathrm f}}[N_{\mathrm{f}}]^2 - 
\mathbb{E}_{N_{\mathrm f}}[N_{\mathrm{f}}]= m^2(\pi r^2 - \pi \eta_1^2)^2 \leq m^2 \pi^2 r^4.
\end{align}
\noindent Consequently, by~\eqref{eq:07}-\eqref{eq:09},~\eqref{eq:12} is proved.
\paragraph{\textbf{Proofs of~\eqref{eq:14} and~\eqref{eq:15}}}\label{ap.3.2} For $\eta_1\leq x\leq r$, the pdf of $d_{\mathrm{w},\mathrm{f}_i}$ when $\eta_1\leq d_{{\mathrm w},{\mathrm f}}\leq r$ is:
\begin{align}
\frac{d}{dx} \mathbb{P}(\eta_1 \leq d_{\mathrm{w},\mathrm{f}_i} \leq x) = \frac{d}{dx} \left(\frac{\pi x^2 - \pi \eta_1^2}{\pi r^2 - \pi \eta_1^2}\right) = \frac{2  x}{ r^2 -  \eta_1^2}. 
\end{align}
\noindent Since $\gamma>2$, 
\begin{align}
\label{eq:03} \mathbb{E}_{{\mathrm F}} [ {1}/{d_{\mathrm{w},\mathrm{f}_i}^{\gamma}}| d_{{\mathrm w},{\mathrm f}}>\eta_1 ] = \frac{2 }{r^2 -  \eta_1^2} \int_{x=\eta_1}^{r} {x^{1-\gamma}} dx = \frac{2 }{\gamma-2} \frac{\eta_1^{2-\gamma}-r^{2-\gamma}}{r^2 -  \eta_1^2},
\end{align}
\noindent By~\eqref{eq:eta1}, $\eta_1 = \sqrt{\frac{\ln {\left(\frac{4}{4-\lambda}\right)}}{m \pi}}$. For large enough $m$, $r^2 -  \eta_1^2 \geq r^2/2$, and thus~\eqref{eq:03} becomes:
\begin{align}
\nonumber \mathbb{E}_{{\mathrm F}} [ {1}/{d_{\mathrm{w},\mathrm{f}_i}^{\gamma}}|d_{{\mathrm w},{\mathrm f}}>\eta_1 ] \leq  \frac{4 }{\gamma-2} \frac{\eta_1^{2-\gamma}-r^{2-\gamma}}{r^2 } \leq \frac{4 }{\gamma-2} \frac{\eta_1^{2-\gamma}}{r^2 },
\end{align}
\noindent and ~\eqref{eq:14} is proved. Note that the assumption $\gamma>2$ was necessary to obtain~\eqref{eq:14}. Since $2\gamma>2$ when $\gamma>2$, replacing $\gamma$ in~\eqref{eq:14} with $2\gamma$ to yields~\eqref{eq:15}.

\begin{figure}
	\begin{center}
		\includegraphics[scale=0.6]{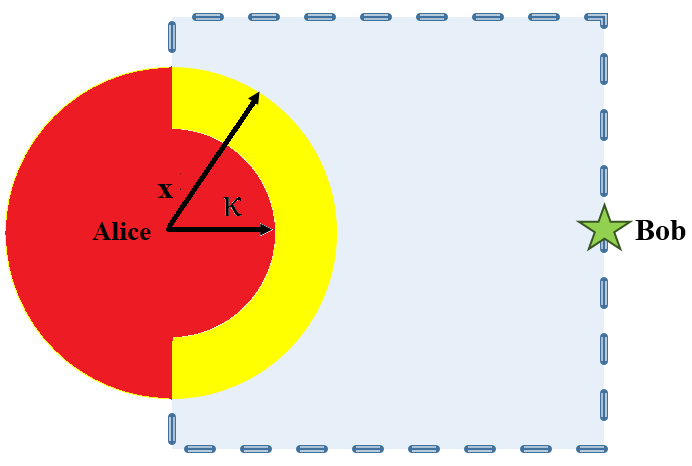}
	\end{center}
	\caption{The circle centered at Alice with radius $x$ is partitioned into the red and the yellow region. The area of the red region is denoted by $h(x)$ and the area of the yellow region is $\mathbb{P}(\kappa \leq d_{{\mathrm a},{\mathrm w}_k}\leq x)$.} 
	\label{fig:pdf1} 
\end{figure}

\paragraph{\textbf{Proof of~\eqref{eq:basic3}}}\label{ap.0} By~\eqref{eq:dformulw},
\begin{align} 
\mathbb{E}_{{\mathrm F},{\mathrm W}}\left[\left.\sqrt{\frac{1}{8} \mathcal{D}\left(\mathbb{P}_1(\boldsymbol{\sigma_{\mathrm{w}}^2},\boldsymbol{d_{\mathrm{a},\mathrm{w}}}) || \mathbb{P}_0(\boldsymbol{\sigma_{\mathrm{w}}^2})\right)}\right|\mathcal{A}\right] &\nonumber  \leq { c \over {4 \sqrt{2} N_{\mathrm{w}}^{\gamma/2} }}  \mathbb{E}_{{\mathrm F},{\mathrm W}}\left[\left.\sum_{k=1}^{N_{\mathrm{w}}} \frac{m^{\gamma/2}}{d_{{\mathrm a},{\mathrm w}_k}^{\gamma} \sigma_{{\mathrm w}_k}^2}\right|\mathcal{A}\right],\\
&\nonumber = { c \over {4 \sqrt{2}  N_{\mathrm{w}}^{\gamma/2} }}  \sum_{k=1}^{N_{\mathrm{w}}} \mathbb{E}_{{\mathrm F},{\mathrm W}}\left[\left. \frac{m^{\gamma/2}}{d_{{\mathrm a},{\mathrm w}_k}^{\gamma} \sigma_{{\mathrm w}_k}^2}\right|\mathcal{A}\right],\\
&\label{eq:th4Er,w0.55} = { c \over {4 \sqrt{2}  N_{\mathrm{w}}^{\gamma/2} }}  \sum_{k=1}^{N_{\mathrm{w}}} \mathbb{E}_{W}\left[\left. \frac{1}{d_{{\mathrm a},{\mathrm w}_k}^{\gamma}}\right|\mathcal{A}\right] \mathbb{E}_{\mathrm F}\left[ \frac{m^{\gamma/2}}{ \sigma_{{\mathrm w}_k}^2}\right],
\end{align}
\noindent where~\eqref{eq:th4Er,w0.55} is true because the locations of friendly nodes are independent of the locations of Willies, and $\mathbb{E}_{W}[\cdot]$ denotes expectation with respect to the locations of Willies. Consider $\mathbb{E}_{\mathrm F}\left[ \frac{m^{\gamma/2}}{ \sigma_{{\mathrm w}_k}^2}\right]$ in~\eqref{eq:th4Er,w0.55}. Similar to the approach leading to~\eqref{eq:totalex1}, we can show that for all $k$, 
\begin{align}
\label{eq:1}\mathbb{E}_{\mathrm F}\left[\frac{m^{\gamma/2}}{\sigma_{\mathrm{w}_k}^2}\right] \leq \frac{\Gamma \left(\gamma/2+1\right)}{2 P_{\mathrm{f}} \pi^{\gamma/2+1}}.
\end{align}
\noindent Now, consider $\mathbb{E}_{W}\left[\left. \frac{1}{d_{{\mathrm a},{\mathrm w}_k}^{\gamma}}\right|\mathcal{A}\right]$ in~\eqref{eq:th4Er,w0.55}. Since Willies are distributed independently,
\begin{align}
\label{eq:3} \sum_{k=1}^{N_{\mathrm{w}}} \mathbb{E}_{W}\left[\left. \frac{1}{d_{{\mathrm a},{\mathrm w}_k}^{\gamma}}\right|\mathcal{A}\right] = \sum_{k=1}^{N_{\mathrm{w}}} \mathbb{E}_{W}\left[\left. \frac{1}{d_{{\mathrm a},{\mathrm w}_k}^{\gamma}}\right|d_{{\mathrm a},{\mathrm w}_k}>\kappa\right]= N_{\mathrm{w}} \mathbb{E}_{W}\left[\left. \frac{1}{d_{{\mathrm a},{\mathrm w}_k}^{\gamma}}\right|d_{{\mathrm a},{\mathrm w}_k}>\kappa\right].
\end{align}
\noindent Next we upper bound the pdf of $d_{{\mathrm a},{\mathrm w}_k}$ given $d_{{\mathrm a},{\mathrm w}_k}>\kappa$, $g(x)$, and then upper bound $\mathbb{E}_{W}\left[\left. \frac{1}{d_{{\mathrm a},{\mathrm w}_k}^{\gamma}}\right|d_{{\mathrm a},{\mathrm w}_k}>\kappa\right]$. Consider a circle of radius $x$ centered at Alice. As shown in Fig.~\ref{fig:pdf1}, we can partition this circle into two regions: the yellow region whose area is $\mathbb{P}(\kappa \leq d_{{\mathrm a},{\mathrm w}_w}\leq x)$ and the red region whose area is denoted by $h(x)$. Note that $h(x)$ is a monotonically increasing function of $x$. Therefore, $\frac{d h(x)}{dx}{}>0$. Consequently, 
\begin{align}
\label{eq:101} g(x)=\frac{d}{dx}\mathbb{P}\left(\kappa \leq d_{{\mathrm a},{\mathrm w}_k}\leq x\right)  = \frac{d}{dx} (\pi x^2 - h(x))  =   2 \pi x - \frac{d h(x)}{dx} \leq  2 \pi x.
\end{align}
\noindent Hence,
\begin{align}
\label{eq:6} \mathbb{E}_{W}\left[\left.  \frac{1}{d_{{\mathrm a},{\mathrm w}_k}^\gamma}\right|d_{{\mathrm a},{\mathrm w}_k} >  \kappa\right]   \leq  \int\limits_{x=\kappa}^{\infty} \frac{2 \pi x}{x^{\gamma}} dx = 2 \pi \frac{\kappa^{2-\gamma}}{\gamma-2}.
\end{align}
\noindent Consequently,~\eqref{eq:3} becomes
\begin{align}
\label{eq:7} \sum_{k=1}^{N_{\mathrm{w}}} \mathbb{E}_{W}\left[\left. \frac{1}{d_{{\mathrm a},{\mathrm w}_k}^{\gamma}}\right|\mathcal{A}\right] &\leq N_{\mathrm{w}} 2\pi \frac{\kappa^{2-\gamma}}{\gamma-2}.
\end{align}
\noindent Thus,~\eqref{eq:th4Er,w0.55},~\eqref{eq:1}, and~\eqref{eq:7} yield
\begin{align} 
\mathbb{E}_{{\mathrm F},{\mathrm W}}\left[\left.\sqrt{\frac{1}{8} \mathcal{D}\left(\mathbb{P}_1(\boldsymbol{\sigma_{\mathrm{w}}^2},\boldsymbol{d_{\mathrm{a},\mathrm{w}}}) || \mathbb{P}_0(\boldsymbol{\sigma_{\mathrm{w}}^2})\right)}\right|\mathcal{A}\right]\label{eq:8} &\leq { c \over {4 \sqrt{2}  N_{\mathrm{w}}^{\gamma/2} }}  \frac{\Gamma \left(\gamma/2+1\right)}{2 P_{\mathrm{f}} \pi^{\gamma/2+1}} N_{\mathrm{w}} \frac{2\pi \kappa^{2-\gamma}}{\gamma-2}  =\frac{\epsilon}{2},
\end{align}
\noindent where the last step is true since $c=  \frac{P_{\mathrm{f}} \epsilon^{\gamma/2} \left(\gamma-2\right)\pi^{\gamma/2}}{2^{\gamma-0.5} \Gamma\left(\gamma/2+1\right)}$ and $\kappa=\sqrt{\frac{\epsilon}{4N_{\mathrm{w}}}}$. By~\eqref{eq:basic2} and~\eqref{eq:8},~\eqref{eq:basic3} is proved.
\paragraph{\textbf{Proof of~\eqref{eq:th3h3}}}\label{ap.1} Assume $\frac{c' \sigma_{\mathrm b}^2}{c N_{\mathrm{w}}^{\gamma/2}}\leq 1$. Since the RHS of~\eqref{eq:Pebasic1} is a monotonically increasing function of $\sigma_{\mathrm b}^2$,~\eqref{eq:Pebasic1} yields 
\begin{align}
\label{eq:th3411}\mathbb{P}_{\mathrm e}^{(\mathrm b)} \left(\sigma_{\mathrm b}^2\right)  \leq
2^{nR- \frac{n}{2} \log_2 \left(1+ \frac{ c'  m^{\gamma/2} }{2 \sqrt{n}  N_{\mathrm{w}}^{\gamma}} \right)}.
\end{align}
\noindent By~\eqref{eq:Alicesrate} and~\eqref{eq:th3411}, $\mathbb{P}_{\mathrm e}^{(\mathrm b)} \left(\sigma_{\mathrm b}^2\right)  \leq
2^{nR- 2n R_0 }$. Since $R=\min\{1,R_0\}\leq R_0$,~\eqref{eq:th3411} becomes: 
\begin{align} 
\label{eq:th3h1}  \mathbb{P}_{\mathrm e}^{(\mathrm b)} \left(\sigma_{\mathrm b}^2\right)  \leq
2^{ -n R_0 }  &\leq  2^{- \frac{n}{4} \log_2\left(1+  \frac{c'  m^{\gamma/2} } {4 N_{\mathrm{w}}^{\gamma} \sqrt{n} } \right)}
=   \left( {1+ \frac{ c'  m^{\gamma/2} }{2 N_{\mathrm{w}}^{\gamma} \sqrt{n} } }\right)^{ -\frac{n}{4}}.
\end{align}
\noindent By~\eqref{eq:th3h1}, $N_{\mathrm{w}}=o\left(n^{\frac{1}{2\gamma}}\sqrt{m}\right)$, $m=\omega(1)$, and $N_{\mathrm{w}}=\omega(1)$, 
\begin{align}
\label{eq:th3100} \mathbb{E}_{{\mathrm F},{\mathrm W}}\left[\mathbb{P}_{\mathrm e}^{(\mathrm b)}(\sigma_{\mathrm b}^2)\Big|\frac{c' \sigma_{\mathrm b}^2}{c N_{\mathrm{w}}^{\gamma/2}} \leq 1\right]   &\leq \left( {1+ \frac{ c'  m^{\gamma/2} }{2 N_{\mathrm{w}}^{\gamma} \sqrt{n} } }\right)^{ -\frac{n}{4}} \stackrel{(h)}{\leq} \left( {1+ \frac{ c' \sqrt{n} m^{\gamma/2} }{8 N_{\mathrm{w}}^{\gamma}  } }\right)^{ -1} \to 0 \text{ as } n \to \infty ,
\end{align}
\noindent where $(h)$ is true since~\eqref{eq:ineq1} is true.
\paragraph{\textbf{Proof of~\eqref{eq:th3h4}}}\label{ap.2} When $\mathcal{B}$ is true, $d_{{\mathrm b},{\mathrm w}_k}>2 \delta$ and $2 \delta >2 d_{{\mathrm w}_k,{\mathrm f}_k}$. Thus, $ - d_{{\mathrm w}_k,{\mathrm f}_k} >  -\frac{d_{{\mathrm b},{\mathrm w}_k}}{2} $. On the other hand, the triangle inequality yields  $d_{{\mathrm b},{\mathrm f}_k} \geq d_{{\mathrm b},{\mathrm w}_k}-d_{{\mathrm w}_k,{\mathrm f}_k}$. Thus, 
\begin{align}
\label{ineq:03}  d_{{\mathrm b},{\mathrm f}_k} > \frac{ d_{{\mathrm b},{\mathrm w}_k}}{2}.
\end{align}
\noindent Now, consider $ \frac{c'\sigma_{\mathrm b}^2} {c N_{\mathrm{w}}^{\gamma/2}}$. Recall that $\sigma_{\mathrm b}^2  \leq \sigma_{{\mathrm b},0}^2 + \sum_{k=1}^{N_{\mathrm{w}}}\frac{P_{\mathrm{f}}}{d_{{\mathrm b},{\mathrm f}_k}^\gamma}$. When $\mathcal{B}$ is true, 
\begin{align}
\label{eq:1.3} \frac{c'} {c N_{\mathrm{w}}^{\gamma/2}}  \sigma_{\mathrm b}^2 \leq  \frac{c' \sigma_{{\mathrm b},0}^2} {c N_{\mathrm{w}}^{\gamma/2}} + \frac{c'} {c N_{\mathrm{w}}^{\gamma/2}} \sum_{k=1}^{N_{\mathrm{w}}}\frac{P_{\mathrm{f}}}{d_{{\mathrm b},{\mathrm f}_k}^\gamma}
 &< \frac{c' \sigma_{{\mathrm b},0}^2} {c N_{\mathrm{w}}^{\gamma/2}} + \frac{c'} {c N_{\mathrm{w}}^{\gamma/2}} \sum_{k=1}^{N_{\mathrm{w}}}\frac{P_{\mathrm{f}} 2^{\gamma}}{d_{{\mathrm b},{\mathrm w}_k}^\gamma},\\
\label{eq:1.5}&= \frac{c' \sigma_{{\mathrm b},0}^2} {c N_{\mathrm{w}}^{\gamma/2}} + \frac{\gamma-2} {2^{5-\gamma} \pi }\frac{1}{N_{\mathrm{w}}} \sum_{k=1}^{N_{\mathrm{w}}}\frac{\delta^{\gamma-2} }{d_{{\mathrm b},{\mathrm w}_k}^\gamma},
\end{align}
\noindent where~\eqref{eq:1.3} is true since $\mathcal{B}$ implies~\eqref{ineq:03}, and~\eqref{eq:1.5} is true since $c'=c\frac {  \zeta^{\gamma/2-1} \left(\gamma-2\right)  }{2^{\gamma+3}P_{\mathrm{f}} \pi^{\gamma/2}}$ and $\delta=\sqrt{\frac{ \zeta}{4\pi N_{\mathrm{w}}}}$. By~\eqref{eq:1.5},
\begin{align}
\mathbb{P}\left( \frac{c'\sigma_{\mathrm b}^2} {c N_{\mathrm{w}}^{\gamma/2}}  >  1\bigg|\mathcal{B}\right) &\leq \mathbb{P}\left( \frac{c' \sigma_{{\mathrm b},0}^2} {c N_{\mathrm{w}}^{\gamma/2}} +  \frac{\gamma-2} {2^{5-\gamma} \pi }\frac{1}{N_{\mathrm{w}}} \sum_{k=1}^{N_{\mathrm{w}}}\frac{\delta^{\gamma-2} }{d_{{\mathrm b},{\mathrm w}_k}^\gamma} >  1\bigg|\mathcal{B}\right).
\end{align}
\noindent Consider $\frac{c' \sigma_{{\mathrm b},0}^2} {c N_{\mathrm{w}}^{\gamma/2}}$ in the above equation. Since $N_{\mathrm{w}}=\omega(1)$, for large enough $n$, $\frac{c' \sigma_{{\mathrm b},0}^2} {c N_{\mathrm{w}}^{\gamma/2}}\leq \frac{1}{2}$. Thus, 
\begin{align}
\nonumber \lim\limits_{n\to \infty}\mathbb{P}\left( \frac{c'\sigma_{\mathrm b}^2} {c N_{\mathrm{w}}^{\gamma/2}}  >  1\bigg|\mathcal{B}\right) &\leq \lim\limits_{n\to \infty} \mathbb{P}\left( \frac{1}{2} + \frac{\gamma-2} {2^{5-\gamma} \pi }\frac{1}{N_{\mathrm{w}}} \sum_{k=1}^{N_{\mathrm{w}}}\frac{\delta^{\gamma-2} }{d_{{\mathrm b},{\mathrm w}_k}^\gamma} >  1\bigg|\mathcal{B}\right),\\
\label{eq:11} &= \lim\limits_{n\to \infty} \mathbb{P}\left( \frac{\gamma-2} {2^{5-\gamma} \pi }\frac{1}{N_{\mathrm{w}}} \sum_{k=1}^{N_{\mathrm{w}}}\frac{\delta^{\gamma-2} }{d_{{\mathrm b},{\mathrm w}_k}^\gamma} > \frac{1}{2}\bigg|\mathcal{B}\right)= \lim\limits_{n\to \infty} \mathbb{P}\left(\frac{1}{N_{\mathrm{w}}} \sum_{k=1}^{N_{\mathrm{w}}}\frac{\delta^{\gamma-2} }{d_{{\mathrm b},{\mathrm w}_k}^\gamma} >  \frac{\pi 2^{4-\gamma}  }{\gamma-2} \bigg|\mathcal{B}\right).
\end{align}
\indent Next, we upper bound $\alpha=\mathbb{E}_{{\mathrm F},{\mathrm W}}\left[ \frac{\delta^{\gamma-2}}{d_{{\mathrm b},{\mathrm w}_k}^\gamma} \bigg| \mathcal{B}\right]$ and then apply the weak law of large numbers (WLLN) to show that~\eqref{eq:11} is equal to zero. Since the locations of Willies are independent of the locations of friendly nodes,
\begin{align}
\label{eq:th3p2014} \alpha= \mathbb{E}_{{\mathrm F},{\mathrm W}}\left[ \frac{\delta^{\gamma-2}}{d_{{\mathrm b},{\mathrm w}_k}^\gamma} \bigg| d_{{\mathrm w}_k,{\mathrm f}_k}\leq \delta \cap  d_{{\mathrm b},{\mathrm w}_k}>2 \delta\right] &= \mathbb{E}_{{\mathrm F},{\mathrm W}}\left[ \frac{\delta^{\gamma-2}}{d_{{\mathrm b},{\mathrm w}_k}^\gamma} \bigg|   d_{{\mathrm b},{\mathrm w}_k}>2 \delta\right]\leq   \frac{\pi 2^{3-\gamma}}{\gamma-2} 
\end{align}
\noindent where the last step follows from the arguments leading to~\eqref{eq:6}. Thus, $\alpha$ is finite. By the WLLN and $N_{\mathrm{w}}=\omega(1)$, for all $\epsilon'>0$, $\mathbb{P}\left(\frac{1}{N_{\mathrm{w}}} \sum_{k=1}^{N_{\mathrm{w}}}  \frac{\delta^{\gamma-2} }{d_{{\mathrm b},{\mathrm w}_k}^\gamma}-\alpha\geq \epsilon' \Bigg| \mathcal{B}\right)=0$, as ${n \to \infty}$. Let $\epsilon'=\alpha$,
\begin{align}
\label{eq:th450} \lim\limits_{n \to \infty}\mathbb{P}\left(\frac{1}{N_{\mathrm{w}}} \sum_{k=1}^{N_{\mathrm{w}}}  \frac{\delta^{\gamma-2} }{d_{{\mathrm b},{\mathrm w}_k}^\gamma}\geq 2 \alpha \Bigg| \mathcal{B}\right)=0.
\end{align}
\noindent Using the upper bound on $\alpha$ presented in~\eqref{eq:th3p2014},~\eqref{eq:th450} yields
\begin{align}
\label{eq:10}\lim\limits_{n \to \infty}\mathbb{P}\left(\frac{1}{N_{\mathrm{w}}} \sum_{k=1}^{N_{\mathrm{w}}}  \frac{\delta^{\gamma-2} }{d_{{\mathrm b},{\mathrm w}_k}^\gamma}\geq  \frac{\pi 2^{4-\gamma}}{\gamma-2} \Bigg| \mathcal{B}\right)=0.
\end{align}
\noindent By~\eqref{eq:11} and~\eqref{eq:10},~\eqref{eq:th3h4} is proved.

\paragraph{\textbf{Proof of~\eqref{eq:th3h5}}}\label{ap.3} Since  $\overline{\mathcal{B}}$ is the union of $\bigcup_{k=1}^{k=N_{\mathrm{w}}}  \{d_{{\mathrm b},{\mathrm w}_k}\leq 2 \delta\}$ and $\bigcup_{k=1}^{k=N_{\mathrm{w}}}  \{d_{{\mathrm w}_k,{\mathrm f}_k}> \delta\}$, 
\begin{align}
\mathbb{P}\left(\overline{\mathcal{B}}\right) 
\label{eq:19} \leq \sum\limits_{k=1}^{N_{\mathrm{w}}} \mathbb{P}\left(d_{{\mathrm b},{\mathrm w}_k}\leq 2 \delta\right) + \sum\limits_{k=1}^{N_{\mathrm{w}}} \mathbb{P}\left(d_{{\mathrm w}_k,{\mathrm f}_k}> \delta\right) 
&=  {N_{\mathrm{w}}} \mathbb{P}\left(d_{{\mathrm b},{\mathrm w}_k}\leq 2 \delta\right) + {N_{\mathrm{w}}} \mathbb{P}\left(d_{{\mathrm w}_k,{\mathrm f}_k}> \delta\right).
\end{align}
\noindent Because Willies are distributed uniformly, $\mathbb{P}\left(d_{{\mathrm b},{\mathrm w}_k}\leq 2 \delta\right) \leq 2\pi  \delta^2$, and by~\eqref{eq:poisson}, $\mathbb{P}\left(d_{{\mathrm w}_k,{\mathrm f}_k}> \delta\right) = e^{-m \pi \delta^2}$. Therefore,~\eqref{eq:19} becomes $\mathbb{P}\left(\overline{\mathcal{B}}\right) 
\leq  2\pi {N_{\mathrm{w}}}   \delta^2+ {N_{\mathrm{w}}} e^{-m \pi \delta^2}$. Since  $\delta=\sqrt{\frac{\zeta}{4 \pi  N_{\mathrm{w}}}}$,
\begin{align}
\mathbb{P}\left(\overline{\mathcal{B}}\right) 
&\leq \zeta/2+ {N_{\mathrm{w}}} e^{- {\frac{m \zeta}{4   N_{\mathrm{w}}}}} = \zeta/2+  e^{\ln{N_{\mathrm{w}}}- {\frac{m \zeta}{4   N_{\mathrm{w}}}}}.
\end{align}
\noindent Consequently, $N_{\mathrm{w}}=o\left({m}/{\log m}\right)$, $N_{\mathrm{w}}=\omega(1)$, and $m=\omega(1)$ yield  $\lim\limits_{n \to \infty}\mathbb{P}\left(\overline{\mathcal{B}}\right)  \leq \zeta/2$.
\paragraph{\textbf{Proof of~\eqref{eq:th328}}}\label{ap.7.5} Since $\mathcal{E} = \{d_{{\mathrm{w}_1},f_1}>\eta_1\} \cap \{\ell \beta' \leq d_{{\mathrm a},{\mathrm w}_1} <  \beta'\}$, 
\begin{align}
 \label{eq:102} \mathbb{P}(\overline{\mathcal{E}})    &\leq \mathbb{P}\left( d_{{\mathrm{w}_1},f_1}\leq \eta_1 \right) + \mathbb{P}\left( \beta' \leq  d_{{\mathrm a},{\mathrm w}_1} <  \ell \beta' \right).
\end{align}
\noindent Consider the first term on the RHS of~\eqref{eq:102}. Since $\eta_1 = \sqrt{\frac{\ln {\left(\frac{4}{4-\lambda}\right)}}{m \pi}}$, 
\begin{align}
\label{eq:112}\mathbb{P}\left( d_{{\mathrm{w}_1},f_1}\leq \eta_1 \right) \leq \left(1-\exp{(- m \pi\eta_1^2)}\right) \leq \frac{\lambda}{4}.
\end{align}
Consider the second term on the RHS of~\eqref{eq:102}. Since $\beta'=\Theta(1/\sqrt{N_{\mathrm{w}}})$, and $N_{\mathrm{w}}=\omega(1)$, for large enough $n$, $\beta'$ becomes small such that the semicircular region around Alice with radii $\beta'$ and $\ell\beta'$ are inside the unit square, and thus $\mathbb{P}(d_{{\mathrm a},{\mathrm w}_1} \geq \beta')=(1- {\pi\beta'^2}/{2})^{N_{\mathrm{w}}}$ and $\mathbb{P}(d_{{\mathrm a},{\mathrm w}_1} \geq \ell \beta')=(1- {\ell\pi \beta'^2}/{2})^{N_{\mathrm{w}}}$. Hence:
\begin{align}
\nonumber \mathbb{P}\left( \beta' \leq  d_{{\mathrm a},{\mathrm w}_1} <  \ell \beta' \right)\leq  \mathbb{P}(d_{{\mathrm a},{\mathrm w}_1} \geq  \beta')+1-\mathbb{P}(d_{{\mathrm a},{\mathrm w}_1} \geq  \ell  \beta')
=   \left(1- {\pi\beta'^2}/{2}\right)^{N_{\mathrm{w}}}+1-\left(1- {\pi\ell^2\beta'^2}/{2}\right)^{N_{\mathrm{w}}}.
\end{align}
\noindent Since $m=\omega(1)$, $N_{\mathrm{w}}=\omega(1)$, and $\beta'=\sqrt{\frac{2\ln{(8/\lambda)}}{\pi N_{\mathrm{w}}}}$, taking the limit of both sides 
yields
\begin{align}
\lim\limits_{n \to \infty} \mathbb{P}\left( \beta' \leq  d_{{\mathrm a},{\mathrm w}_1} <  \ell \beta' \right)   &\leq   e^{-\frac{\pi  \beta'^2 N_{\mathrm{w}}}{2}} + 1 -e^{-\frac{\pi \ell^2\beta'^2 N_{\mathrm{w}}}{2}}
\label{eq:th329}  =  \lambda/8 + 1 - (\lambda/8)^{\ell^2}=\lambda/4,
\end{align}
\noindent where the last step follows from~\eqref{eq:ell}. Combined with~\eqref{eq:112},~\eqref{eq:th333} is proved.
\paragraph{\textbf{Proof of~\eqref{eq:th333}}}\label{ap.8} 
Consider the RHS of~\eqref{eq:th3conv1}. Since $\mathcal{E}$ implies $ \ell \beta' \leq d_{{\mathrm a},{\mathrm w}_1} <  \beta'$, we replace $d_{{\mathrm a},{\mathrm w}_1}$ in the numerator with $\ell \beta'$ and in the denominator with $\beta'$ to achieve
\begin{align} 
\nonumber \mathbb{E}_{{\mathrm F},{\mathrm W}} \left[\left.\mathbb{P}_{\mathrm {MD}}(\boldsymbol{\sigma_{\mathrm{w}}^2},\boldsymbol{d_{\mathrm{a},\mathrm{w}}})\right|\mathcal{E}\right]
&\leq
\frac{4{ \frac{P_{\mathrm{a}}}{(\ell \beta')^\gamma} \mathbb{E}_{{\mathrm F},{\mathrm W}} [\sigma_{{\mathrm w}_1}^2|\mathcal{E}] }
}{n \left(\frac{P_{\mathrm{a}}}{\beta'^\gamma}-t \right)^2}  +  \frac{ 2\mathbb{E}_{{\mathrm F},{\mathrm W}} [\sigma_{{\mathrm w}_1}^4 |\mathcal{E}] }{n \left(\frac{P_{\mathrm{a}}}{\beta'^\gamma}-t \right)^2},\\
\nonumber &= \frac{4{ \frac{P_{\mathrm{a}}}{(\ell \beta')^\gamma} \mathbb{E}_{{\mathrm F},{\mathrm W}} [\sigma_{{\mathrm w}_1}^2|d_{{\mathrm{w}_1},f_1}>\eta_1] }
}{n \left(\frac{P_{\mathrm{a}}}{\beta'^\gamma}-t \right)^2}  +  \frac{ 2\mathbb{E}_{{\mathrm F},{\mathrm W}} [\sigma_{{\mathrm w}_1}^4 |d_{{\mathrm{w}_1},f_1}>\eta_1] }{n \left(\frac{P_{\mathrm{a}}}{\beta'^\gamma}-t \right)^2},
\end{align}
\noindent where the last step is true since Willie's noise is independent of his location. 
\paragraph{\textbf{Proof of~\eqref{eq:th3c3.6}}}~\label{ap.9} Define the event 
$$\mathcal{G} = \left\{d_{{\mathrm w}_2,{\mathrm f}_2}<\sqrt{\frac{2 \ln{(1/\tau)}}{N_{\mathrm{w}}\pi}}\right\} \cap \left\{d_{{\mathrm b},{\mathrm w}_2}<\sqrt{\frac{2 \ln{(1/\tau)}}{ N_{\mathrm{w}}\pi}}\right\}.$$ 
\noindent From the triangle inequality, when $\mathcal{G}$ occurs, $d_{{\mathrm b},{\mathrm f}_2}<d_{{\mathrm w}_2,{\mathrm f}_2}+d_{{\mathrm b},{\mathrm w}_2}<2\sqrt{\frac{2 \ln{(1/\tau)}}{\pi N_{\mathrm{w}}}}$. Hence, $\mathbb{P}\left(\mathcal{F} | \mathcal{G}\right)=1$. By the law of total probability:
\begin{align}
\label{eq:th337} \mathbb{P}\left(\mathcal{F}\right)
=  \mathbb{P}\left(\mathcal{F} | \mathcal{G}\right) \mathbb{P}(\mathcal{G}) +   \mathbb{P}\left(\mathcal{F} | \overline{\mathcal{G}}\right) \mathbb{P}(\overline{\mathcal{G}})\geq \mathbb{P}(\mathcal{G}).
\end{align}
\noindent Consider $\mathbb{P}(\mathcal{G})$. Since the locations of Willies are independent of the locations of friendly nodes,
\begin{align}
\label{th301}  \mathbb{P}(\mathcal{G}) =\mathbb{P}\left(d_{{\mathrm w}_2,{\mathrm f}_2}<\sqrt{{2 \ln{(1/\tau)}}/{ (N_{\mathrm{w}}\pi)}}\right) \mathbb{P} \left(d_{{\mathrm b},{\mathrm w}_2}<\sqrt{{2 \ln{(1/\tau)}}/{( N_{\mathrm{w}}\pi})}\right).
\end{align}
\noindent Consider the first term on the RHS of~\eqref{th301}. By~\eqref{eq:poisson}, $m=\omega(1)$, $N_{\mathrm{w}}=\omega(1)$, and $N_{\mathrm{w}}=o(m/\log{m})$, 
\begin{align}
\label{th301.1} \mathbb{P}\left(d_{{\mathrm w}_2,{\mathrm f}_2}<\sqrt{{2 \ln{(1/\tau)}}/{(\pi N_{\mathrm{w}}})}\right) = 1-e^{-\frac{2 m \ln{(1/\tau)}}{N_{\mathrm{w}}}} \to 1 \text{ as }n\to \infty.
\end{align}
\noindent Next, consider the second term on the RHS of~\eqref{th301}. Note that when $x<\frac{1}{2}$, $\mathbb{P} \left(d_{{\mathrm b},{\mathrm w}_2}<x\right)=1-\left(1-{\pi x^2}/{2 }\right)^{N_{\mathrm{w}}}$. Since $N_{\mathrm{w}}=\omega(1)$, for large enough $n$, $\sqrt{\frac{2 \ln{(1/\tau)}}{ N_{\mathrm{w}}\pi}}<1/2$ and thus
\begin{align}
\label{th301.2} \mathbb{P} \left(d_{{\mathrm b},{\mathrm w}_2}<\sqrt{{2 \ln{(1/\tau)}}/{(\pi  N_{\mathrm{w}}})}\right) =1-\left(1-{\ln{(1/\tau)}}/{ N_{\mathrm{w}}}\right)^{N_{\mathrm{w}}} \to 1-\tau \text{ as }n \to \infty.
\end{align}
\noindent By~\eqref{eq:th337}-\eqref{th301.2},~\eqref{eq:th3c3.6} is proved.

\paragraph{\textbf{Proof of high probability results}}~\label{ap.10} Assume the locations of Willie and the friendly nodes are fixed. Define the event 
$$\mathcal{K}=\{\frac{m^{\gamma/2}}{\sigma_{\mathrm{w}}^2}\leq c_0\}\cap \{d_{{\mathrm a},{\mathrm w}}>\psi\},$$
\noindent where $c_0=\frac{4 \alpha_0 \sqrt{2} \epsilon \psi^{\gamma}}{c}$, and $\alpha_0>1$ is arbitrary. By the law of total probability, the probability of covertness is
\begin{align}
\label{eq:100} \mathbb{P}\left(\mathbb{P}_{\mathrm e}^{(\mathrm w)}(\sigma_{\mathrm{w}}^2,d_{\mathrm{a},\mathrm{w}}) \geq  {1}/{2}-\epsilon\right) \geq \mathbb{P}(\mathcal{K}) \mathbb{P}\left(\mathbb{P}_{\mathrm e}^{(\mathrm w)}(\sigma_{\mathrm{w}}^2,d_{\mathrm{a},\mathrm{w}}) \geq {1}/{2}-\epsilon|\mathcal{K}\right). 
\end{align}
\noindent Consider the first term on the RHS of~\eqref{eq:100}. Note that $\sigma_{\mathrm{w}}^2$ is independent of $d_{{\mathrm a},{\mathrm w}}$. By~\eqref{eq:poisson},
$$\mathbb{P}\left(\frac{m^{\gamma/2}}{\sigma_{\mathrm{w}}^2}\leq c_0\right) = 1-e^{-\pi c_0^{2/\gamma} P_{\mathrm{f}}^{2/\gamma}}.$$ 
Recall that $\psi =\sqrt{\frac{\epsilon}{2\pi}}$, and thus $$\mathbb{P}(d_{{\mathrm a},{\mathrm w}}>\psi)=1-\pi \psi^2/2=1-\epsilon/4.$$ \noindent Consequently, 
\begin{align}
\label{eq:200} \mathbb{P}(\mathcal{K})=\mathbb{P}\left({m^{\gamma/2}}/{\sigma_{\mathrm{w}}^2}\leq c_0\right)\mathbb{P}(d_{{\mathrm a},{\mathrm w}}>\psi)=\left(1-\exp{(-\pi c_0^{2/\gamma} P_{\mathrm{f}}^{2/\gamma})}\right)\left(1-\epsilon/4\right).
\end{align}
\noindent Now, consider the second term on the RHS of~\eqref{eq:100}. Observe
\begin{align}
\nonumber \mathbb{P}\left(\mathbb{P}_{\mathrm e}^{(\mathrm w)}(\sigma_{\mathrm{w}}^2,d_{\mathrm{a},\mathrm{w}}) \geq \frac{1}{2}-\epsilon\Big|\mathcal{K}\right)
 \nonumber &=\mathbb{P}\left(\mathbb{P}_{\mathrm e}^{(\mathrm w)}(\sigma_{\mathrm{w}}^2,d_{\mathrm{a},\mathrm{w}})\geq \frac{1}{2}-\epsilon\frac{c_0 d_{{\mathrm a},{\mathrm w}}^{\gamma}}{c_0 d_{{\mathrm a},{\mathrm w}}^{\gamma}}\bigg|\mathcal{K}\right) \\ &\stackrel{(i)}{\geq}  \mathbb{P}\left(\mathbb{P}_{\mathrm e}^{(\mathrm w)}(\sigma_{\mathrm{w}}^2,d_{\mathrm{a},\mathrm{w}}) 
\geq \frac{1}{2}-\frac{\epsilon d_{{\mathrm a},{\mathrm w}}^{\gamma} m^{\gamma/2}}{c_0 \sigma_{\mathrm{w}}^2 d_{{\mathrm a},{\mathrm w}}^{\gamma}} \bigg|\mathcal{K}\right),\\
\nonumber &\stackrel{(j)}{\geq} \mathbb{P}\left(\mathbb{P}_{\mathrm e}^{(\mathrm w)}(\sigma_{\mathrm{w}}^2,d_{\mathrm{a},\mathrm{w}}) 
\geq \frac{1}{2}-\frac{\epsilon \psi^{\gamma} m^{\gamma/2}}{c_0 \sigma_{\mathrm{w}}^2 d_{{\mathrm a},{\mathrm w}}^{\gamma}} \bigg|\mathcal{K}\right),\\
\nonumber &\stackrel{(k)}{=} \mathbb{P}\left(\mathbb{P}_{\mathrm e}^{(\mathrm w)} (\sigma_{\mathrm{w}}^2,d_{\mathrm{a},\mathrm{w}})
\geq  \frac{1}{2}- \sqrt{\frac{1}{8}} \frac{ c m^{\gamma/2}}{2 \sigma_{\mathrm w}^2 d_{{\mathrm a},{\mathrm w}}^{\gamma}}\bigg| \mathcal{K}\right),\\
\label{eq:300} &\geq \mathbb{P}\left(\mathbb{P}_{\mathrm e}^{(\mathrm w)}(\sigma_{\mathrm{w}}^2,d_{\mathrm{a},\mathrm{w}}) 
\geq  \frac{1}{2}- \sqrt{\frac{1}{8}} \frac{ c m^{\gamma/2}}{2 \alpha_0 \sigma_{\mathrm w}^2 d_{{\mathrm a},{\mathrm w}}^{\gamma}}\bigg| \mathcal{K}\right),
\end{align}
where $(i)$ is true since when $\mathcal{K}$ occurs, $\frac{m^{\gamma/2}}{\sigma_{\mathrm{w}}^2}\leq c_0$, and $(j)$ is true since when $\mathcal{K}$ occurs, $d_{{\mathrm a},{\mathrm w}}>\psi$,  $(k)$ is true since $c_0=\frac{4 \sqrt{2} \epsilon \psi^{\gamma}}{c}$, and the last step is true since $\alpha_0>1$. Similar to the approach leading to~\eqref{eq:power} and~\eqref{eq:expcond13}, we can show that if Alice sets her average symbol power $P_{\mathrm{a}} \leq \frac{  c m^{\gamma/2}}{ \alpha_0\sqrt{n}}$, then $\mathbb{P}_{\mathrm e}^{(\mathrm w)}   \geq  \frac{1}{2}- \sqrt{\frac{1}{8}} \frac{  c m^{\gamma/2}}{2 \alpha_0 \sigma_{\mathrm w}^2 d_{{\mathrm a},{\mathrm w}}^{\gamma}}$. Consequently,~\eqref{eq:300} yields
 $$\mathbb{P}(\mathbb{P}_{\mathrm e}^{(\mathrm w)}(\sigma_{\mathrm{w}}^2,d_{\mathrm{a},\mathrm{w}}) \geq \frac{1}{2}-\epsilon|\mathcal{K})=1.$$
 \noindent Combined with By~\eqref{eq:100},~\eqref{eq:200}
\begin{align}
\label{eq:301} \mathbb{P}\left(\mathbb{P}_{\mathrm e}^{(\mathrm w)}(\sigma_{\mathrm{w}}^2,d_{\mathrm{a},\mathrm{w}}) \geq  \frac{1}{2}-\epsilon\right) \geq  \left(1-e^{-\pi c_0^{2/\gamma} P_{\mathrm{f}}^{2/\gamma}}\right)\left(1-\epsilon/4\right).
\end{align}
\noindent Consider $e^{-\pi c_0^{2/\gamma} P_{\mathrm{f}}^{2/\gamma}}$ in~\eqref{eq:301}. Since $\alpha_0$ is arbitrary, we choose $\alpha_0$ large enough such that $e^{-\pi c_0^{2/\gamma} P_{\mathrm{f}}^{2/\gamma}} \leq  \epsilon/2$. Therefore, 
$$\mathbb{P}\left(\mathbb{P}_{\mathrm e}^{(\mathrm w)}(\sigma_{\mathrm{w}}^2,d_{\mathrm{a},\mathrm{w}}) \geq  \frac{1}{2}-\epsilon\right) \geq  (1-\epsilon/2) (1-\epsilon/4) \geq 1-\epsilon.$$

\begin{figure}
	\begin{center}
		\includegraphics[ 
		scale=0.6]{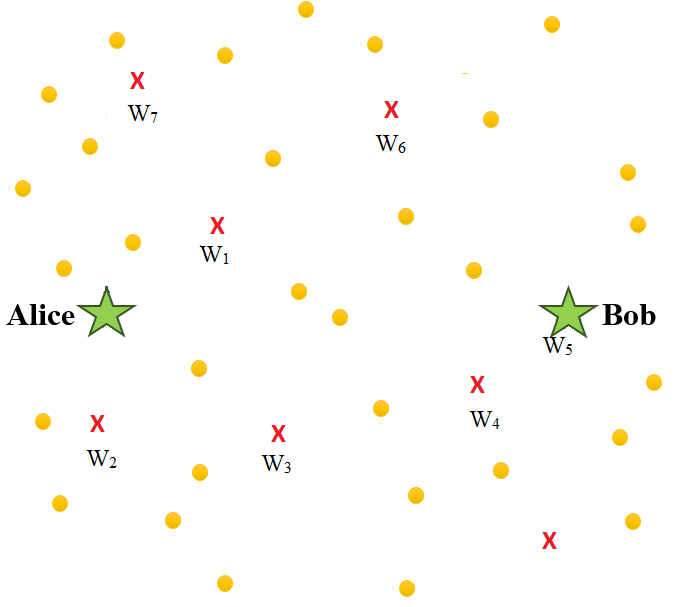}
	\end{center}
	\caption{System Configuration: Source node Alice wishes to communicate reliably and without detection to the intended receiver Bob at distance one (normalized) with the assistance of friendly nodes (represented by yellow nodes in the figure) and adversary nodes (represented by red nodes in the figure) distributed according to two-dimensional Poisson point processes with densities $m$ and $\lambda_N$, respectively.}
	\label{fig:SysMod3}
\end{figure}

 \paragraph{\textbf{Proof for the case where Willies are distributed according to a Poisson process}}  \label{ap.11}
 Instead of modeling Willies locations by a uniform distribution, here we model the locations of the Willies by a two-dimensional Poisson process (see Fig.~\ref{fig:SysMod3}), and consider the case of $\gamma>2$. Analogous to the strategy in Theorem~\ref{th:mwillie2p}, Alice and Bob's strategy is to turn on the closest friendly node to each Willie and keep all other friendly nodes off, whether Alice transmits or not.

\begin{mthm}{2.3} \label{th:mwillie2pp}
	When friendly nodes and collaborating Willies are independently distributed according to two-dimensional Poisson point processes with densities $m=\omega(1)$ and $\lambda_{N}={o}\left({m}/{\log{m}}\right))$, respectively, and Alice and Bob are a unit distance apart (see Fig.~\ref{fig:SysMod3}),	then Alice can reliably and covertly transmit $\mathcal{O}\left(\min\left\{n,\frac{m^{\gamma/2} \sqrt{n}}{\lambda_N^{\gamma}}\right\}\right)$ bits to Bob in $n$ channel uses. Conversely, if only the closest friendly node to each Willie is on and Alice attempts to transmit $\omega\left(\frac{\sqrt{n} m^{\gamma/2}}{ \lambda_N^\gamma}\right)$ bits to Bob in $n$ channel uses, there exists a detector that Willie can use to either detect her with arbitrarily low error probability $\mathbb{P}_{\mathrm e}^{(\mathrm w)}$ or Bob cannot decode the message with arbitrarily low error probability $\mathbb{P}_{\mathrm e}^{(\mathrm b)}$.
\end{mthm}  

We present the proof assuming $\lambda_N=\omega(1)$, as the proof for a finite $\lambda_N$ follows from it. In addition, according to the statement of Theorem~\ref{th:mwillie2pp}, if   $\lambda_N=\Omega\left(n^{\frac{1}{2\gamma}}\sqrt{m}\right)$, then Alice can reliably and covertly transmit $\mathcal{O}\left(1\right)$ bits to Bob in $n$ uses of channel, which is not of interest. Therefore, we present the proof assuming 
$\lambda_N={o}\left(\min\left\{\frac{m}{\log{m}},n^{\frac{1}{2\gamma}}\sqrt{m}\right\}\right)$.
\begin{proof}
	
	{\it (Achievability)} 
	
	\textbf{Construction:} The construction and Bob's decoding are the same as those of Theorem~\ref{th:mwillie2p}.
	
	\textbf{Analysis:} ({\em Covertness}) Consider a circle with radius $r$ around Alice. We first consider only the Willies in this circle and only the noise received from the closest nodes to each Willie in this region and we present a result which is valid for every $r>0$. Then, we let $r \to \infty$. 
	
	By~\eqref{eq:0}, when Willie applies the optimal hypothesis test to minimize his error probability, 
	\begin{align} 
	\label{eq:basic22} \mathbb{P}_{\mathrm e}^{(\mathrm w)}(\boldsymbol{\sigma_{\mathrm{w}}^2},\boldsymbol{d_{\mathrm{a},\mathrm{w}}},r) \geq \frac{1}{2}- \sqrt{\frac{1}{8} \mathcal{D}(\mathbb{P}_1(\boldsymbol{\sigma_{\mathrm{w}}^2},\boldsymbol{d_{\mathrm{a},\mathrm{w}}},r) || \mathbb{P}_0(\boldsymbol{\sigma_{\mathrm{w}}^2},r))}.
	\end{align}
	\noindent Here, $\mathbb{P}_{\mathrm e}^{(\mathrm w)}(\boldsymbol{\sigma_{\mathrm{w}}^2},\boldsymbol{d_{\mathrm{a},\mathrm{w}}},r)$ is Willies' probability of error when we only consider Willies in the circle of radius $r$ around Alice, $\boldsymbol{\sigma_{\mathrm{w}}^2}$ and $\boldsymbol{d_{\mathrm{a},\mathrm{w}}}$ are vectors containing  $\sigma_{\mathrm{w}_k}^2$ and $d_{\mathrm{a},\mathrm{w}_k}$, $\mathbb{P}_0(\boldsymbol{\sigma_{\mathrm{w}}^2},r)=\prod_{i=1}^{n}  \mathbb{P}_{0,i}(\boldsymbol{\sigma_{\mathrm{w}}^2},r)$ and  $\mathbb{P}_1(\boldsymbol{\sigma_{\mathrm{w}}^2},\boldsymbol{d_{\mathrm{a},\mathrm{w}}},r)=\prod_{i=1}^{n}  \mathbb{P}_{1,i}(\boldsymbol{\sigma_{\mathrm{w}}^2},\boldsymbol{d_{\mathrm{a},\mathrm{w}}},r)$ are the joint probability distributions of the Willies' channels observations when we only consider Willies within a circle of radius $r$ centered at Alice for the $H_0$ and $H_1$ hypotheses, respectively, where $\mathbb{P}_{0,i}({\boldsymbol{\sigma_{\mathrm{w}}^2}},r)=\prod_{k=1}^{N_{\mathrm{w}}}\mathbb{P}_{{\mathrm w}_k}^{(k)}({\sigma_{\mathrm{w}_k}^2})$ and $\mathbb{P}_{1,i}(\boldsymbol{\sigma_{\mathrm{w}}^2},\boldsymbol{d_{\mathrm{a},\mathrm{w}}},r)$ are the joint probability distribution of the $i^{\mathrm{th}}$ channel observation of the Willies when we only consider Willies within a circle of radius $r$ centered at Alice for $H_0$ and $H_1$ hypotheses, respectively. 

 Suppose Alice sets her average symbol power so that
	\begin{align}\label{eq:th3pf2}
	P_{\mathrm{a}} \leq \frac{c c_1  m^{\gamma/2}}{\sqrt{n} \lambda_N^{\gamma/2}},
	\end{align}
	\noindent where $c$ is given in~\eqref{eq:th3c} and 
	\begin{align}
	\label{eq:c1} c_1=\frac{\pi}{2} \left(\frac{4 \ln{\frac{2}{2-\epsilon}}}{\epsilon \pi}\right)^{\gamma/2-1}.
	\end{align}
	\noindent Similar to the approach leading to~\eqref{eq:dformulw}, we can show that
	\begin{align}
	\mathcal{D}(\mathbb{P}_1(\boldsymbol{\sigma_{\mathrm{w}}^2},\boldsymbol{d_{\mathrm{a},\mathrm{w}}},r) || \mathbb{P}_0(\boldsymbol{\sigma_{\mathrm{w}}^2},r)) 
	\label{eq:dformulw2}\leq  \frac{c^2 c_1^2  m^{\gamma}}{4 \lambda_N^{\gamma}} \left( \sum_{d_{{\mathrm a},{\mathrm w}_k}<r} \frac{1}{d_{{\mathrm a},{\mathrm w}_k}^{\gamma} \sigma_{{\mathrm w}_k}^2} \right)^2.
	\end{align}
	\noindent Define the event (see Fig.~\ref{fig:SysMod22})
	\begin{align}
	\nonumber \mathcal{A}' = \bigcap\limits_{k=1}^{\infty} \{d_{{\mathrm a},{\mathrm w}_k}>\kappa' \},
	\end{align}
	\noindent which occurs when all of the Willies are outside of the disk with radius 
	\begin{align}
	\label{eq:kappap}\kappa'=\sqrt{\frac{\ln{\frac{2}{2-\epsilon}}}{\pi \lambda_N}}
	\end{align} 
	\noindent centered at Alice. Then, we show in Appendix~\ref{ap.12} that when $n$ is enough large, for any $\epsilon>0$ Alice can achieve: 
	\begin{figure}
		\begin{center}
			\includegraphics[scale=0.7]{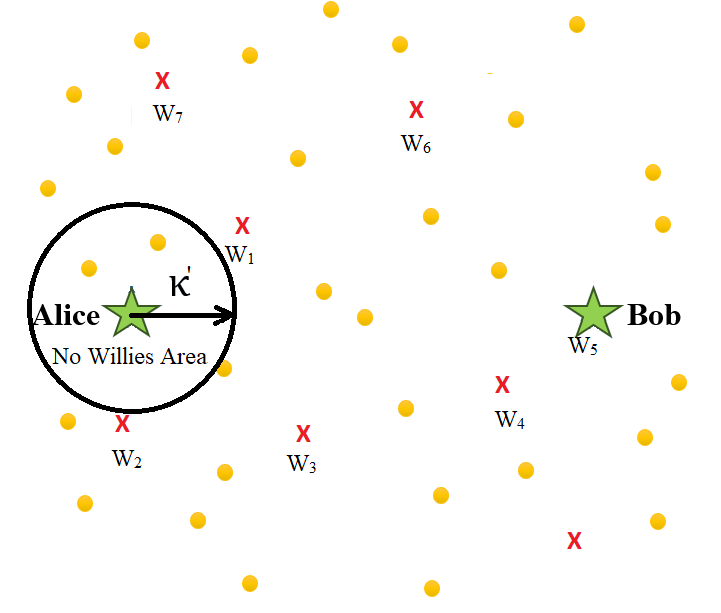}
		\end{center}
		\caption{Event $\mathcal{A}'$ is true when there is no Willie in the disk with radius $\kappa'$ centered at Alice, as shown above. Alice is only able to communicate covertly with intended receiver Bob if $\mathcal{A}'$ is true.}
		\label{fig:SysMod22}
	\end{figure}
	\noindent 
	\begin{align}\label{eq:basic32}
	\mathbb{E}_{{\mathrm F},{\mathrm W}}\left[\mathbb{P}_{\mathrm e}^{(\mathrm w)}(\boldsymbol{\sigma_{\mathrm{w}}^2},\boldsymbol{d_{\mathrm{a},\mathrm{w}}},r)\right|\mathcal{A}'] \geq \frac{1}{2}(1- \epsilon).
	\end{align}
	\noindent Since Willies are distributed according to a two-dimensional Poisson process with rate $\lambda_N$,
	\begin{align}
	\label {eq:th3a52} \mathbb{P}(\mathcal{A}') = e^{-\lambda_N \pi \kappa'^2}=e^{-\lambda_N \pi \frac{\ln{\frac{2}{2-\epsilon}}}{\pi \lambda_N}} = 1-\frac{\epsilon}{2},
	\end{align}
	\noindent By~\eqref{eq:basic32},~\eqref{eq:th3a52}, and the law of total expectation
	\begin{align} 
	\label{eq:120} \mathbb{E}_{{\mathrm F},{\mathrm W}} [ \mathbb{P}_{\mathrm e}^{(\mathrm w)}(\boldsymbol{\sigma_{\mathrm{w}}^2},\boldsymbol{d_{\mathrm{a},\mathrm{w}}},r)]   \geq  \mathbb{E}_{{\mathrm F},{\mathrm W}} \left[\left. \mathbb{P}_{\mathrm e}^{(\mathrm w)}(\boldsymbol{\sigma_{\mathrm{w}}^2},\boldsymbol{d_{\mathrm{a},\mathrm{w}}},r)\right|\mathcal{A}'\right]\; \mathbb{P}(\mathcal{A}') = \left(\frac{1}{2}- \frac{ \epsilon}{  2}\right)\left(1- \frac{ \epsilon}{  2}\right)\geq \frac{1}{2}-\epsilon,
	\end{align}
	\noindent Since $0\leq \mathbb{P}_{\mathrm e}^{(\mathrm w)}(\boldsymbol{\sigma_{\mathrm{w}}^2},\boldsymbol{d_{\mathrm{a},\mathrm{w}}},r)\leq 1$, by the dominated convergence theorem, $$\mathbb{E}_{{\mathrm F},{\mathrm W}} [ \lim\limits_{r \to \infty}  \mathbb{P}_{\mathrm e}^{(\mathrm w)}(\boldsymbol{\sigma_{\mathrm{w}}^2},\boldsymbol{d_{\mathrm{a},\mathrm{w}}},r)] = \lim\limits_{r \to \infty} 
	\mathbb{E}_{{\mathrm F},{\mathrm W}} [ \mathbb{P}_{\mathrm e}^{(\mathrm w)}(\boldsymbol{\sigma_{\mathrm{w}}^2},\boldsymbol{d_{\mathrm{a},\mathrm{w}}},r)].$$ 
	\noindent In addition, since the Willies use an optimal detector, their probability of error is a non-increasing function of $r$, i.e., considering more Willies for detection does not increase the probability of error. Therefore, we can use the monotone convergence theorem to show that 
	$$\mathbb{E}_{{\mathrm F},{\mathrm W}} [ \lim\limits_{r \to \infty}  \mathbb{P}_{\mathrm e}^{(\mathrm w)}(\boldsymbol{\sigma_{\mathrm{w}}^2},\boldsymbol{d_{\mathrm{a},\mathrm{w}}},r)] = 
	\mathbb{E}_{{\mathrm F},{\mathrm W}} [ \mathbb{P}_{\mathrm e}^{(\mathrm w)}(\boldsymbol{\sigma_{\mathrm{w}}^2},\boldsymbol{d_{\mathrm{a},\mathrm{w}}})].$$
	\noindent Consequently,~\eqref{eq:120} yields
	\begin{align}
	\label{eq:121} \lim\limits_{n \to \infty}\mathbb{P}_{\mathrm e}^{(\mathrm w)} =  \lim\limits_{r,n \to \infty}
	\mathbb{E}_{{\mathrm F},{\mathrm W}} [\mathbb{P}_{\mathrm e}^{(\mathrm w)}(\boldsymbol{\sigma_{\mathrm{w}}^2},\boldsymbol{d_{\mathrm{a},\mathrm{w}}},r)]\geq \frac{1}{2}-\epsilon,   
	\end{align}
	and thus, communication is covert as long as $P_{\mathrm{a}} = \mathcal{O}\left(\frac{m^{\gamma/2}}{\sqrt{n} \lambda_N^{\gamma/2}}\right)$.
	
	({\em Reliability}) Next, we calculate the number of bits that Alice can send to Bob covertly and reliably. Consider arbitrarily $\zeta > 0$.  We show that Bob can achieve $\mathbb{P}_{\mathrm e}^{(\mathrm b)} < \zeta$ as $n \to \infty$, where $\mathbb{P}_{\mathrm e}^{(\mathrm b)}$ is Bob's ML decoding error probability averaged over all possible codewords and the locations of friendly nodes and Willies. 
	
	Consider a circle with radius $r'=\lambda_N$ around Bob. Let $\sigma_{\mathrm b}^2(r')$ be Bob's noise power disregarding the jammers of Willies outside of this circle of radius $r'$ centered at Bob. Then: 
	\begin{align}
	\label{eq:130} \sigma_{\mathrm b}^2(\lambda_N) \leq \sigma_{{\mathrm b},0}^2 + \sum_{d_{{\mathrm b},{\mathrm w}_k}<\lambda_N} \frac{P_{\mathrm{f}}}{d_{{\mathrm b},{\mathrm f}_k}^\gamma}, 
	\end{align}
	where $d_{{\mathrm b},{\mathrm f}_k}$ is the distance between Bob and the closest friendly node to the $k^{\mathrm{th}}$ Willie ($W_k$), and the inequality becomes equality when each Willie has a distinct closest friendly node. By~\eqref{eq:Pebasic0} and~\eqref{eq:th3pf2}, Bob's probability of error disregarding the jammers of Willies outside of this circle of radius $r$ centered at Bob is 
	\begin{align}
	\label{eq:Pebasic12} \mathbb{P}_{\mathrm e}^{(\mathrm b)} \left(\sigma_{\mathrm b}^2(\lambda_N)\right)  & \leq
	2^{nR- \frac{n}{2} \log_2 \left(1+ \frac{ c  m^{\gamma/2} }{2 \sqrt{n} \sigma_{\mathrm b}^2(\lambda_N) \lambda_N^{\gamma/2}} \right)}.
	\end{align}
	\noindent Suppose Alice sets $R=\min{\{R_0,1\}}$, where
	\begin{align}
	\label{eq:Alicesrate3}R_0&=\frac{1}{4} \log_2\left(1+  \frac{c''  m^{\gamma/2} } {4 \lambda_N^{\gamma} \sqrt{n} } \right),\\
	\label{eq:cpp} c''&=c\frac {  (\ln{\frac{1}{1-\zeta/2}})^{\gamma/2-1} \left(\gamma-2\right)  }{2^{\gamma+5}P_{\mathrm{f}} \pi^{\gamma/2}},
	\end{align}
	\noindent and $c$ and $c_1$ are defined in~\eqref{eq:th3c} and~\eqref{eq:c1}, respectively. By the law of total expectation,
	\begin{align}\label{eq:EPEBobth412}
	\mathbb{E}_{{\mathrm F},{\mathrm W}}[\mathbb{P}_{\mathrm e}^{(\mathrm b)}\left(\sigma_{\mathrm b}^2(\lambda_N)\right)] 
	\leq \mathbb{E}_{{\mathrm F},{\mathrm W}}\left[\mathbb{P}_{\mathrm e}^{(\mathrm b)}\left( \sigma_{\mathrm b}^2(\lambda_N)\right)\Big|\frac{c''\sigma_{\mathrm b}^2(\lambda_N)}{ c \lambda_N^{\gamma/2}} \leq 1 \right] +   \mathbb{P}\left(\frac{c''\sigma_{\mathrm b}^2(\lambda_N)}{c \lambda_N^{\gamma/2}} > 1\right).
	\end{align}
	\noindent Consider the first term on the RHS of~\eqref{eq:EPEBobth412}. We show in Appendix~\ref{ap.13} that since $m=\omega(1)$, $\lambda_N=\omega(1)$, and $\lambda_N=o\left(n^{\frac{1}{2\gamma}}\sqrt{m}\right)$,
	\begin{align} \label{eq:th3h32}
	&\lim\limits_{n \to \infty} \mathbb{E}_{{\mathrm F},{\mathrm W}}\left[\mathbb{P}_{\mathrm e}^{(\mathrm b)}\left(\sigma_{\mathrm b}^2(\lambda_N)\right)\Big|\frac{c'' \sigma_{\mathrm b}^2(\lambda_N)}{c \lambda_N^{\gamma/2}} \leq 1 \right] = 0.
	\end{align}
	\noindent To upper bound the second term on the RHS of~\eqref{eq:EPEBobth41}, we define the event
	\begin{align}
	\label{eq:Bp} \mathcal{B}'= \bigcap_{k=1}^{\infty} \{ d_{{\mathrm b},{\mathrm w}_k}>2 \delta'\} \bigcap_{d_{{\mathrm b},{\mathrm w}_k}<\lambda_N} \{ d_{{\mathrm w}_k,{\mathrm f}_k}\leq \delta'\} \bigcap \{N_{\mathrm{w}}(\lambda_N)\leq 2 \pi \lambda_N^3\},
	\end{align}
	\noindent where 
	\begin{align}
	\label{eq:deltap}\delta'=\sqrt{\frac{\ln{\frac{1}{1-\zeta/2}}}{4 \pi  \lambda_N}},
	\end{align}
	 \noindent and $N_{\mathrm{w}}(\lambda_N)$ is the number of Willies in the circle of radius $r=\lambda_N$ centered at Bob.  
	\begin{figure}
		\begin{center}
			\includegraphics[ scale=0.6]{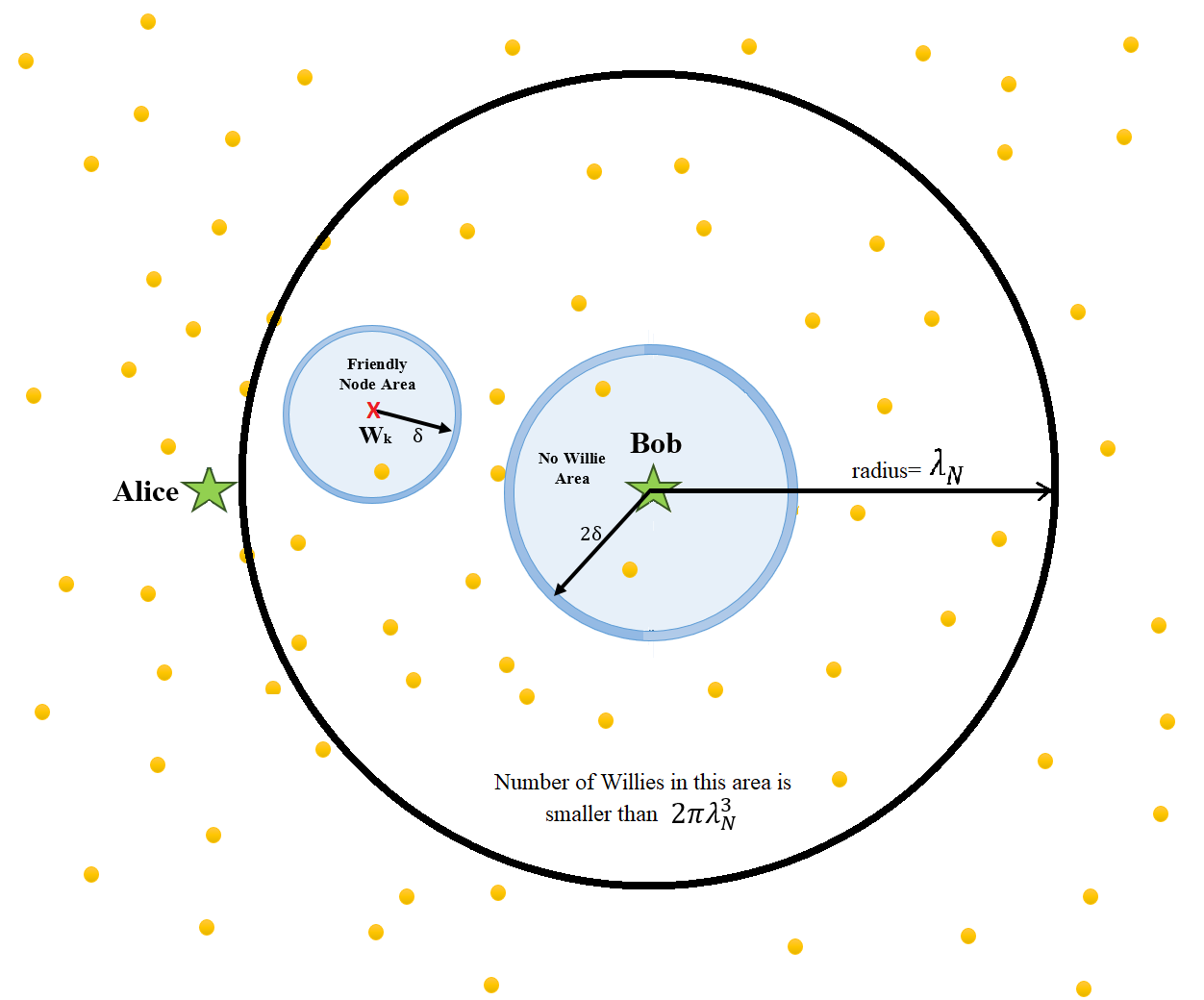}
		\end{center}
		\caption{Event $\mathcal{B}'$ occurs when there is no Willie in the disk with radius $2 \delta'$ centered Bob, the distance between each Willie $W_k$ and the closest friendly node to him is smaller than $\delta'$ if $W_k$ is within the circle of radius $r=\lambda_N$ centered at Bob, i.e., $\left\{ 2d_{{\mathrm w}_k,{\mathrm f}_k} \leq \delta' \}\cap \{d_{{\mathrm b},{\mathrm w}_k}>2 \delta'\right\}$ for $1\leq k \leq \lambda_N$, and and the number of Willies in the circle of radius $r'=\lambda_N$ centered at Bob is smaller than $\pi 2 \lambda_N^3$.}
		\label{fig:SysMod42}
	\end{figure}
	\noindent Event $\mathcal{B}'$ occurs when 
	\begin{enumerate}
		\item There is no Willie in the disk with radius $2 \delta'$ around Bob;
		\item For all Willies $W_k$ in circle of radius $r=\lambda_N$ around Bob, the distance between $W_k$ and the closest friendly node to $W_k$ is smaller than $\delta'$ (see Fig.~\ref{fig:SysMod42}); and,
		\item The number of Willies in the circle of radius $r=\lambda_N$ centered at Bob is larger than $\pi \lambda_N^2/2$.
	\end{enumerate}
	\noindent The law of total probability yields
	\begin{align} \nonumber
	\mathbb{P}\left(\frac{c'' \sigma_{\mathrm b}^2(r)}{c \lambda_N^{\gamma/2}} > 1 \right) &\leq \mathbb{P}\left(\frac{c'' \sigma_{\mathrm b}^2(r)}{c  \lambda_N^{\gamma/2}} > 1\bigg|\mathcal{B}'\right)
	+\mathbb{P}\left(\bar{\mathcal{B}'}\right).
	\end{align}
	\noindent We show in Appendix~\ref{ap.14} that since $\lambda_N=\omega(1)$,
	\begin{align}\label{eq:th3h42}
	&\lim\limits_{n\to \infty} \mathbb{P}\left( \frac{c''\sigma_{\mathrm b}^2(r)} {c \lambda_N^{\gamma/2}}  >  1\bigg|\mathcal{B}'\right)=0,
	\end{align}
	\noindent and in Appendix~\ref{ap.15} that since $\lambda_N=\omega(1)$ and $\lambda_N=o\left({m}/{\log m}\right)$,
	\begin{align}
	\label{eq:th3h52}
	&\lim\limits_{n\to \infty} \mathbb{P}\left(\bar{\mathcal{B}'}\right)  \leq \zeta/2.
	\end{align}
	\noindent Thus,~\eqref{eq:EPEBobth412}-\eqref{eq:th3h52} yield 
	\begin{align}
	\label{eq:123}\lim\limits_{n \to \infty} \mathbb{E}_{{\mathrm F},{\mathrm W}}[\mathbb{P}_{\mathrm e}^{(\mathrm b)}\left(\sigma_{\mathrm b}^2(\lambda_N)\right)] < \zeta,
	\end{align}
	for any $0<\zeta<1$. Since $0\leq \mathbb{P}_{\mathrm e}^{(\mathrm b)}({\sigma_{\mathrm{b}}^2}(\lambda_N))\leq 1$, and $\lambda_N=\omega(1)$ by the dominated convergence theorem, 
	$$\lim\limits_{n \to \infty} 
	\mathbb{E}_{{\mathrm F},{\mathrm W}} [   \mathbb{P}_{\mathrm e}^{(\mathrm b)}({\sigma_{\mathrm{b}}^2}(\lambda_N)]=\mathbb{E}_{{\mathrm F},{\mathrm W}} [ \lim\limits_{n \to \infty} \mathbb{P}_{\mathrm e}^{(\mathrm b)}({\sigma_{\mathrm{b}}^2}(\lambda_N)].$$ 
	\noindent Note that if Bob's noise increases, then his probability of error will increase. Therefore, by the monotone convergence theorem, 
	$$\mathbb{E}_{{\mathrm F},{\mathrm W}} [ \lim\limits_{n \to \infty} \mathbb{P}_{\mathrm e}^{(\mathrm b)}({\sigma_{\mathrm{b}}^2}(\lambda_N)]=\mathbb{E}_{{\mathrm F},{\mathrm W}} [  \mathbb{P}_{\mathrm e}^{(\mathrm b)}( {\sigma_{\mathrm{b}}^2})].$$ \noindent Hence,
\begin{align}
\label{eq:122} \lim\limits_{n \to \infty }\mathbb{P}_{\mathrm e}^{(\mathrm b)} = \lim\limits_{n \to \infty }
\mathbb{E}_{{\mathrm F},{\mathrm W}} [  \mathbb{P}_{\mathrm e}^{(\mathrm b)}({\sigma_{\mathrm{b}}^2}(\lambda_N)].   
\end{align}
\noindent By~\eqref{eq:123} and~\eqref{eq:122}, 
\begin{align}
\nonumber \mathbb{P}_{\mathrm e}^{(\mathrm b)} < \zeta,
\end{align}
\noindent for all $\zeta>0$, and thus the communication is reliable. 

	({\em Number of Covert Bits}) Similar to the analysis of Theorem~\ref{th:mwillie2p}, we can show that Bob receives $\mathcal{O}\left(\min\left\{n,\frac{m^{\gamma/2} \sqrt{n}}{\lambda_N^{\gamma}}\right\}\right)$ bits in $n$ channel uses.
	
	{\it (Converse)} The converse follows from that of Theorem~\ref{th:mwillie2p} assuming that the closest friendly node to each Willie is on and the Willies know this. Similarly, we can show that  the signal received by the closest Willie to Alice ($W_1$) is sufficient to detect Alice's communication. The converse of Theorem~\ref{th:mwillie2p} was based on upper-bounding $W_1$'s received noise power by that of the case where all friendly nodes are on. The same upper bound is applicable here as well. Furthermore, for the converse of Theorem~\ref{th:mwillie2p} we defined events $\mathcal{E},\mathcal{F}$, and $\mathcal{G}$ based on $N_{\mathrm{w}}$, the number of Willies in the unit box; however, here, the corresponding events are defined based on the density of Willies, $\lambda_N$.  
\end{proof}

\paragraph{\textbf{Proof of~\eqref{eq:basic32}}} \label{ap.12} By~\eqref{eq:dformulw2},
\begin{align} 
\mathbb{E}_{{\mathrm F},{\mathrm W}}\left[\left.\sqrt{\frac{1}{8} \mathcal{D}\left(\mathbb{P}_1(\boldsymbol{\sigma_{\mathrm{w}}^2},\boldsymbol{d_{\mathrm{a},\mathrm{w}}},r) || \mathbb{P}_0(\boldsymbol{\sigma_{\mathrm{w}}^2},r)\right)}\right|\mathcal{A}'\right] &\nonumber  \leq { c c_1 \over {4 \sqrt{2} \lambda_N^{\gamma/2} }}  \mathbb{E}_{{\mathrm F},{\mathrm W}}\left[\left.\sum_{d_{{\mathrm a},{\mathrm w}_k}<r} \frac{m^{\gamma/2}}{d_{{\mathrm a},{\mathrm w}_k}^{\gamma} \sigma_{{\mathrm w}_k}^2}\right|\mathcal{A}'\right],\\
&\nonumber \stackrel{(l)}{=} { c c_1  \pi r^2 \lambda_N\over {4 \sqrt{2}  \lambda_N^{\gamma/2} }} \mathbb{E}_{{\mathrm F},{\mathrm W}}\left[\left. \frac{m^{\gamma/2}}{d_{{\mathrm a},{\mathrm w}_k}^{\gamma} \sigma_{{\mathrm w}_k}^2}\right|\mathcal{A}'\right],\\
&\nonumber \stackrel{(m)}{=} { c c_1  \pi r^2 \over {4 \sqrt{2}  \lambda_N^{\gamma/2-1} }}   \mathbb{E}_{W}\left[\left. \frac{1}{d_{{\mathrm a},{\mathrm w}_k}^{\gamma}}\right|\mathcal{A}'\right] \mathbb{E}_{\mathrm F}\left[ \frac{m^{\gamma/2}}{ \sigma_{{\mathrm w}_k}^2}\right],\\
&\label{eq:156} = { c c_1  \pi r^2 \over {4 \sqrt{2}  \lambda_N^{\gamma/2-1} }}   \mathbb{E}_{W}\left[\left. \frac{1}{d_{{\mathrm a},{\mathrm w}_k}^{\gamma}}\right|d_{{\mathrm a},{\mathrm w}_k}>\kappa'\right] \mathbb{E}_{\mathrm F}\left[ \frac{m^{\gamma/2}}{ \sigma_{{\mathrm w}_k'}^2}\right],
\end{align}
\noindent where $(l)$ follows from Wald's identity, $(m)$  is true because the locations of friendly nodes are independent of the locations of Willies, and the last step is true since Willies are distributed independently. Recall that $\mathbb{E}_{W}[\cdot]$ denotes expectation with respect to the locations of the Willies. 

Consider $\mathbb{E}_{W}\left[\left. \frac{1}{d_{{\mathrm a},{\mathrm w}_k}^{\gamma}}\right|d_{{\mathrm a},{\mathrm w}_k}>\kappa'\right]$ in~\eqref{eq:156}. For $x\leq r$, the pdf of $d_{{\mathrm a},{\mathrm w}_k}$ given $d_{{\mathrm a},{\mathrm w}_k}>\kappa'$ is 
\begin{align}
\label{eq:1012} \frac{d}{dx}\mathbb{P}\left(\kappa' \leq d_{{\mathrm a},{\mathrm w}_k}\leq x\right)  = \frac{d}{dx} \frac{\pi x^2- \pi \kappa'^2}{\pi r^2 - \pi \kappa'^2}=\frac{2 x}{ r^2 - \kappa'^2}.
\end{align}
\noindent Hence,
\begin{align}
\label{eq:62} \mathbb{E}_{W}\left[\left.  \frac{1}{d_{{\mathrm a},{\mathrm w}_k}^\gamma}\right|d_{{\mathrm a},{\mathrm w}_k} >  \kappa'\right]   =  \int\limits_{x=\kappa'}^{r} \frac{ 2 x}{(r^2-\kappa'^2) x^{\gamma}} dx = \frac{2}{r^2-\kappa'^2}  \frac{\kappa'^{2-\gamma}-r^{2-\gamma}}{\gamma-2}\leq  \frac{2\kappa'^{2-\gamma}}{(\gamma-2)(r^2-\kappa'^2)}.
\end{align}
\noindent For large enough $n$, $r^2\geq 2 \kappa'^2$; therefore,~\eqref{eq:62} yields:
\begin{align}
\label{eq:63} \mathbb{E}_{W}\left[\left.  \frac{1}{d_{{\mathrm a},{\mathrm w}_k}^\gamma}\right|d_{{\mathrm a},{\mathrm w}_k} >  \kappa'\right]   \leq \frac{4\kappa'^{2-\gamma}}{(\gamma-2)r^2}
\end{align}
\noindent By,~\eqref{eq:1},~\eqref{eq:156}, and~\eqref{eq:63}, for large enough $n$,
\begin{align} 
\mathbb{E}_{{\mathrm F},{\mathrm W}}\left[\left.\sqrt{\frac{1}{8} \mathcal{D}\left(\mathbb{P}_1(\boldsymbol{\sigma_{\mathrm{w}}^2},\boldsymbol{d_{\mathrm{a},\mathrm{w}}},r) || \mathbb{P}_0(\boldsymbol{\sigma_{\mathrm{w}}^2},r)\right)}\right|\mathcal{A}'\right] 
\label{eq:88} &\leq { c c_1  \pi r^2\over {4 \sqrt{2}  \lambda_N^{\gamma/2-1} }}  \frac{\Gamma \left(\gamma/2+1\right)}{2 P_{\mathrm{f}} \pi^{\gamma/2+1}}  \frac{2 \kappa'^{2-\gamma}}{(\gamma-2)r^2}  =\frac{\epsilon}{2},
\end{align}
\noindent where the last step follows from substituting the values of $c$ (given in~\eqref{eq:th3c}), $c_1$ (given in~\eqref{eq:c1}), and $\kappa'$ (given in~\eqref{eq:kappap}). By~\eqref{eq:basic22} and~\eqref{eq:88},~\eqref{eq:basic32} is proved.
\paragraph{\textbf{Proof of~\eqref{eq:th3h32}}}\label{ap.13} The proof follows that of~\eqref{eq:th3h3}, replacing $N_{\mathrm{w}}$ with $\lambda_N$.
\paragraph{\textbf{Proof of~\eqref{eq:th3h42}}}\label{ap.14} When $\mathcal{B}'$ is true, for Willies $W_k$ that are within the circle of radius $\lambda_N$ centered at Bob, $d_{{\mathrm b},{\mathrm w}_k}>2 \delta'$ and $2 \delta' >2 d_{{\mathrm w}_k,{\mathrm f}_k}$. Thus, $ - d_{{\mathrm w}_k,{\mathrm f}_k} >  -\frac{d_{{\mathrm b},{\mathrm w}_k}}{2} $. On the other hand, the triangle inequality yields  $d_{{\mathrm b},{\mathrm f}_k} \geq d_{{\mathrm b},{\mathrm w}_k}-d_{{\mathrm w}_k,{\mathrm f}_k}$. Thus, 
\begin{align}
\label{ineq:032}  d_{{\mathrm b},{\mathrm f}_k} > \frac{ d_{{\mathrm b},{\mathrm w}_k}}{2}.
\end{align}
\noindent When $\mathcal{B}'$ is true, multiplying both sides of~\eqref{eq:130} by $\frac{c''} {c \lambda_N^{\gamma/2}}$,  applying~\eqref{ineq:032}, and substituting the values of $c''$ and $\delta'$ given in~\eqref{eq:cpp} and~\eqref{eq:deltap} yield
\begin{align}
\nonumber \frac{c''} {c \lambda_N^{\gamma/2}}  \sigma_{\mathrm b}^2(\lambda_N) &\leq  \frac{c'' \sigma_{{\mathrm b},0}^2} {c \lambda_N^{\gamma/2}} + \frac{c''} {c \lambda_N^{\gamma/2}} \sum_{d_{{\mathrm b},{\mathrm w}_k}<\lambda_N}\frac{P_{\mathrm{f}}}{d_{{\mathrm b},{\mathrm f}_k}^\gamma}\\
\nonumber &< \frac{c'' \sigma_{{\mathrm b},0}^2} {c \lambda_N^{\gamma/2}} + \frac{c''} {c \lambda_N^{\gamma/2}} \sum_{d_{{\mathrm b},{\mathrm w}_k}<\lambda_N}\frac{P_{\mathrm{f}} 2^{\gamma}}{d_{{\mathrm b},{\mathrm w}_k}^\gamma}\frac{\delta'^{\gamma-2}}{\delta'^{\gamma-2}},\\
\label{eq:1.52}&= \frac{c'' \sigma_{{\mathrm b},0}^2} {c \lambda_N^{\gamma/2}} + \frac{\gamma-2} {2^{7-\gamma} \pi }\frac{1}{\lambda_N} \sum_{d_{{\mathrm b},{\mathrm w}_k}<\lambda_N}\frac{\delta'^{\gamma-2} }{d_{{\mathrm b},{\mathrm w}_k}^\gamma},
\end{align}
\noindent By~\eqref{eq:1.52},
\begin{align}
\label{eq:131}\mathbb{P}\left( \frac{c''\sigma_{\mathrm b}^2} {c \lambda_N^{\gamma/2}}  >  1\bigg|\mathcal{B}'\right) &\leq \mathbb{P}\left( \frac{c'' \sigma_{{\mathrm b},0}^2} {c \lambda_N^{\gamma/2}} +  \frac{\gamma-2} {2^{7-\gamma} \pi }\frac{1}{\lambda_N} \sum_{d_{{\mathrm b},{\mathrm w}_k}<\lambda_N}\frac{\delta'^{\gamma-2} }{d_{{\mathrm b},{\mathrm w}_k}^\gamma} >  1\bigg|\mathcal{B}'\right).
\end{align}
\noindent Consider $\frac{c'' \sigma_{{\mathrm b},0}^2} {c \lambda_N^{\gamma/2}}$ in~\eqref{eq:131}. Since $\lambda_N=\omega(1)$, for large enough $n$, $\frac{c'' \sigma_{{\mathrm b},0}^2} {c \lambda_N^{\gamma/2}} \leq \frac{1}{2}$. Thus, 
\begin{align}
\nonumber \lim\limits_{n\to \infty}\mathbb{P}\left( \frac{c''\sigma_{\mathrm b}^2} {c \lambda_N^{\gamma/2}}  >  1\bigg|\mathcal{B}'\right) &\leq \lim\limits_{n\to \infty} \mathbb{P}\left( \frac{1}{2} + \frac{\gamma-2} {2^{7-\gamma} \pi }\frac{1}{\lambda_N} \sum_{d_{{\mathrm b},{\mathrm w}_k}<\lambda_N}\frac{\delta'^{\gamma-2} }{d_{{\mathrm b},{\mathrm w}_k}^\gamma} >  1\bigg|\mathcal{B}'\right),\\
 \nonumber &= \lim\limits_{n\to \infty} \mathbb{P}\left( \frac{\gamma-2} {2^{7-\gamma} \pi }\frac{1}{\lambda_N} \sum_{d_{{\mathrm b},{\mathrm w}_k}<\lambda_N}\frac{\delta'^{\gamma-2} }{d_{{\mathrm b},{\mathrm w}_k}^\gamma} > \frac{1}{2}\bigg|\mathcal{B}'\right),\\
 \nonumber &= \lim\limits_{n\to \infty} \mathbb{P}\left(\frac{1}{\lambda_N} \sum_{d_{{\mathrm b},{\mathrm w}_k}<\lambda_N}\frac{\delta'^{\gamma-2} }{d_{{\mathrm b},{\mathrm w}_k}^\gamma} >  \frac{\pi 2^{6-\gamma}  }{\gamma-2} \bigg|\mathcal{B}'\right),\\
 \label{eq:132} &\leq \lim\limits_{n\to \infty} \mathbb{P}\left(\frac{2 \pi \lambda_N^2}{N_{\mathrm{w}}(\lambda_N)} \sum_{d_{{\mathrm b},{\mathrm w}_k}<\lambda_N}\frac{\delta'^{\gamma-2} }{d_{{\mathrm b},{\mathrm w}_k}^\gamma} >  \frac{\pi 2^{6-\gamma}  }{\gamma-2} \bigg|\mathcal{B}'\right),\\
 \label{eq:1122} &= \lim\limits_{n\to \infty} \mathbb{P}\left(\frac{1}{{N_{\mathrm{w}}(\lambda_N)}} \sum_{d_{{\mathrm b},{\mathrm w}_k}<\lambda_N}\frac{\delta'^{\gamma-2} }{d_{{\mathrm b},{\mathrm w}_k}^\gamma} >  \frac{ 2^{5-\gamma}  }{(\gamma-2)\lambda_{N}^2} \bigg|\mathcal{B}'\right),
\end{align}
\noindent where~\eqref{eq:132} is true since when $\mathcal{B}'$ occurs, $N_{\mathrm{w}}(\lambda_N)<2 \pi \lambda_N^3$, and thus $1/\lambda_N < {2 \pi \lambda_N^2}/{N_{\mathrm{w}}}$. Next, we upper bound $\alpha'=\mathbb{E}_{{\mathrm F},{\mathrm W}}\left[ \frac{\delta'^{\gamma-2}}{d_{{\mathrm b},{\mathrm w}_k}^\gamma} \bigg| \mathcal{B}'\right]$ and then apply the WLLN to show that the RHS of~\eqref{eq:1122} tends to zero as $n \to \infty$. Since the locations of Willies are independent of the locations of friendly nodes and $\lambda_N=\omega(1)$, for large enough $n$,
\begin{align}
\label{eq:th3p20142} \alpha'= \mathbb{E}_{{\mathrm F},{\mathrm W}}\left[ \frac{\delta'^{\gamma-2}}{d_{{\mathrm b},{\mathrm w}_k}^\gamma} \bigg| d_{{\mathrm w}_k,{\mathrm f}_k}\leq \delta' \cap  d_{{\mathrm b},{\mathrm w}_k}>2 \delta'\right] &= \mathbb{E}_{{\mathrm F},{\mathrm W}}\left[ \frac{\delta'^{\gamma-2}}{d_{{\mathrm b},{\mathrm w}_k}^\gamma} \bigg|   d_{{\mathrm b},{\mathrm w}_k}>2 \delta'\right]\leq   \frac{ 2^{4-\gamma} }{(\gamma-2)\lambda_N^2} 
\end{align}
\noindent where the last step follows from the arguments leading to~\eqref{eq:62}. By the WLLN and $\lambda_N=\omega(1)$, for all $\epsilon'>0$, $\mathbb{P}\left(\frac{1}{N_{\mathrm{w}}(\lambda_N)} \sum\limits_{d_{{\mathrm b},{\mathrm w}_k}<\lambda_N}  \frac{\delta'^{\gamma-2} }{d_{{\mathrm b},{\mathrm w}_k}^\gamma}-\alpha'\geq \epsilon' \Bigg| \mathcal{B}'\right)=0$, as ${n \to \infty}$. Let $\epsilon'=\alpha'$,
\begin{align}
\label{eq:th4502} \lim\limits_{n \to \infty}\mathbb{P}\left(\frac{1}{N_{\mathrm{w}}(\lambda_N)} \sum_{d_{{\mathrm b},{\mathrm w}_k}<\lambda_N}  \frac{\delta'^{\gamma-2} }{d_{{\mathrm b},{\mathrm w}_k}^\gamma}\geq 2 \alpha' \Bigg| \mathcal{B}'\right)=0.
\end{align}
\noindent Applying the upper bound in~\eqref{eq:th3p20142} to~\eqref{eq:th4502} yields
\begin{align}
\label{eq:1022}\lim\limits_{n \to \infty}\mathbb{P}\left(\frac{1}{N_{\mathrm{w}}(\lambda_N)} \sum_{d_{{\mathrm b},{\mathrm w}_k}<\lambda_N}  \frac{\delta'^{\gamma-2} }{d_{{\mathrm b},{\mathrm w}_k}^\gamma}\geq  \frac{ 2^{5-\gamma}}{(\gamma-2)\lambda_{N}^2} \Bigg| \mathcal{B}'\right)=0.
\end{align}
\noindent By~\eqref{eq:1122} and~\eqref{eq:1022},~\eqref{eq:th3h42} is proved.

\paragraph{\textbf{Proof of~\eqref{eq:th3h52}}}\label{ap.15} Define the events
\begin{align}
{\mathcal{B}'}_1&=\bigcap\limits_{d_{{\mathrm b},{\mathrm w}_k}<r}  \{d_{{\mathrm w}_k,{\mathrm f}_k}\leq \delta'\},\\
{\mathcal{B}'}_2&={\bigcap\limits_{k=1}^{\infty}  \{d_{{\mathrm b},{\mathrm w}_k}> 2 \delta'\}}\\
{\mathcal{B}'}_3&=\{N_{\mathrm{w}}(\lambda_N) \leq 2 \pi \lambda_N^3\}.
\end{align}
\noindent By~\eqref{eq:Bp}, 
\begin{align}
\overline{\mathcal{B}'}=\overline{\mathcal{B}'}_1\cup\overline{\mathcal{B}'}_2\cup\overline{\mathcal{B}'}_3
\end{align}
\noindent Next, we upper bound the probability of the events $\overline{\mathcal{B}'}_1$, $\overline{\mathcal{B}'}_2$, and $\overline{\mathcal{B}'}_3$. Observe:
\begin{align}
\label{eq:133} \mathbb{P}\left(\overline{\mathcal{B}'_1}\right)=\mathbb{P}\left(\bigcup\limits_{d_{{\mathrm b},{\mathrm w}_k}<\lambda_N}  \{d_{{\mathrm w}_k,{\mathrm f}_k}> \delta'\} \right) \leq   \sum\limits_{d_{{\mathrm b},{\mathrm w}_k}<\lambda_N} \mathbb{P}\left(d_{{\mathrm w}_k,{\mathrm f}_k}> \delta\right)=\sum\limits_{k'=0}^{\infty}\mathbb{P}\left({N_{\mathrm{w}}(\lambda_N)}=k'\right)k' \mathbb{P}\left(d_{{\mathrm w}_k,{\mathrm f}_k}> \delta\right)
\end{align}
\noindent Note that $\mathbb{P}\left(d_{{\mathrm w}_k,{\mathrm f}_k}> \delta\right)$ is the same for all Willies, and that by~\eqref{eq:poisson}, $\mathbb{P}\left(d_{{\mathrm w}_k,{\mathrm f}_k}> \delta\right) = e^{-m \pi \delta'^2}$. In addition,  $\sum\limits_{k'=0}^{\infty}\mathbb{P}\left({N_{\mathrm{w}}(\lambda_N)}=k'\right) k' = \pi \lambda_N^3$ is the expected value of $N_{\mathrm{w}}(\lambda_N)$. Hence,~\eqref{eq:133} yields:
\begin{align}
\nonumber\mathbb{P}\left(\overline{\mathcal{B}'}_1\right) 
&\leq   \pi  \lambda_N^3  e^{-m \pi \delta'^2}= \pi    e^{3 \ln{(\lambda_N)}-m \pi \delta'^2}= \pi    e^{3 \ln{(\lambda_N)}-\frac{m}{4 \lambda_N}  {{\ln{\frac{1}{1-\zeta/2}}}}}.
\end{align}
\noindent where the last step is true since  $\delta'=\sqrt{\frac{\ln{\frac{1}{1-\zeta/2}}}{4 \pi  \lambda_N}}$. Because $\lambda_N=o\left({m}/{\log m}\right)$, $\lambda_N=\omega(1)$, and $m=\omega(1)$, 
\begin{align}
\label{eq:134}\lim\limits_{n \to \infty}\mathbb{P}\left(\overline{\mathcal{B}'}_1\right)=0
\end{align}
\noindent Now, consider $\mathbb{P}\left(\overline{\mathcal{B}'}_2\right)$. Since Willies are distributed according to a two-dimensional Poisson process and $\delta'=\sqrt{\frac{\ln{\frac{1}{1-\zeta/2}}}{4 \pi  \lambda_N}}$,
\begin{align}
\mathbb{P}\left(\overline{\mathcal{B}'}_2\right) 
 &\label{eq:135} =  1-\mathbb{P}\left(\bigcap\limits_{k=1}^{\infty} d_{{\mathrm b},{\mathrm w}_k}> 2 \delta\right)=  1-e^{-4 \pi \lambda_N \delta'^2}=1-e^{-4 \pi \lambda_N \frac{\ln{\frac{1}{1-\zeta/2}}}{4 \pi  \lambda_N}}=\zeta/2.
\end{align}
\noindent Consider $\mathbb{P}\left(\overline{\mathcal{B}'}_3\right)$. Since the average number of Willie in the circle of radius $\lambda_N$ around Bob is $\pi \lambda_N^2 \lambda_N = \pi \lambda_N^3$, the WLLN yields:
\begin{align}
\label{eq:136}\lim\limits_{n \to \infty}\mathbb{P}\left(\overline{\mathcal{B}'}_3\right) = 
\lim\limits_{n \to \infty}\mathbb{P}\left(N_{\mathrm{w}}(\lambda_N)> 2 \pi \lambda_N^3\right) = 0
\end{align}
\noindent Consequently, by~\eqref{eq:134}-\eqref{eq:136}, $\lim\limits_{n \to \infty}\mathbb{P}\left(\overline{\mathcal{B}'}\right)  \leq \zeta/2$.
\bibliographystyle{ieeetr}

\begin{thebibliography}{10}

\bibitem{soltani2014covert}
R.~Soltani, B.~Bash, D.~Goeckel, S.~Guha, and D.~Towsley, ``Covert single-hop
  communication in a wireless network with distributed artificial noise
  generation,'' in {\em Communication, Control, and Computing (Allerton), 2014
  52nd Annual Allerton Conference on}, pp.~1078--1085, IEEE, 2014.

\bibitem{snowden}
``Edward {Snowden}: Leaks that exposed {US} spy programme.''
  \url{http://www.bbc.com/news/world-us-canada-23123964}, Jan 2014.

\bibitem{nichols2001wireless}
R.~K. Nichols, P.~Lekkas, and P.~C. Lekkas, {\em Wireless security}.
\newblock McGraw-Hill Professional Publishing, 2001.

\bibitem{lopez2008wireless}
J.~L{\'o}pez and J.~Zhou, {\em Wireless sensor network security}, vol.~1.
\newblock Ios Press, 2008.

\bibitem{miller2001facing}
S.~K. Miller, ``Facing the challenge of wireless security,'' {\em Computer},
  vol.~34, no.~7, pp.~16--18, 2001.

\bibitem{arbaugh2003wireless}
W.~A. Arbaugh, ``Wireless security is different,'' {\em Computer}, vol.~36,
  no.~8, pp.~99--101, 2003.

\bibitem{hadian2016privacy}
M.~Hadian, X.~Liang, T.~Altuwaiyan, and M.~M. Mahmoud, ``Privacy-preserving
  mhealth data release with pattern consistency,'' in {\em Global
  Communications Conference (GLOBECOM), 2016 IEEE}, pp.~1--6, IEEE, 2016.

\bibitem{hadian2018privacy}
M.~Hadian, T.~Altuwaiyan, X.~Liang, and W.~Li, ``Privacy-preserving voice-based
  search over mhealth data,'' {\em Smart Health}, 2018.

\bibitem{takbiri2017limits}
N.~Takbiri, A.~Houmansadr, D.~L. Goeckel, and H.~Pishro-Nik, ``Limits of
  location privacy under anonymization and obfuscation,'' in {\em Information
  Theory (ISIT), 2017 IEEE International Symposium on}, pp.~764--768, IEEE,
  2017.

\bibitem{takbiri2017fundamental}
N.~Takbiri, A.~Houmansadr, D.~L. Goeckel, and H.~Pishro-Nik, ``Fundamental
  limits of location privacy using anonymization,'' in {\em Information
  Sciences and Systems (CISS), 2017 51st Annual Conference on}, pp.~1--6, IEEE,
  2017.

\bibitem{simon94ssh}
M.~K. Simon, J.~K. Omura, R.~A. Scholtz, and B.~K. Levitt, {\em Spread Spectrum
  Communications Handbook}.
\newblock McGraw-Hill, 1994.

\bibitem{bash_isit2012}
B.~Bash, D.~Goeckel, and D.~Towsley, ``Square root law for communication with
  low probability of detection on {AWGN} channels,'' in {\em Information Theory
  Proceedings (ISIT), 2012 IEEE International Symposium on}, pp.~448--452, July
  2012.

\bibitem{bash_jsac2013}
B.~Bash, D.~Goeckel, and D.~Towsley, ``Limits of reliable communication with
  low probability of detection on {AWGN} channels,'' {\em Selected Areas in
  Communications, IEEE Journal on}, vol.~31, pp.~1921--1930, September 2013.

\bibitem{jaggi_isit2013}
P.~H. Che, M.~Bakshi, and S.~Jaggi, ``Reliable deniable communication: Hiding
  messages in noise,'' in {\em Information Theory Proceedings (ISIT), 2013 IEEE
  International Symposium on}, pp.~2945--2949, July 2013.

\bibitem{kadhe2014reliable}
S.~Kadhe, S.~Jaggi, M.~Bakshi, and A.~Sprintson, ``Reliable, deniable, and
  hidable communication over multipath networks,'' in {\em Information Theory
  (ISIT), 2014 IEEE International Symposium on}, pp.~611--615, IEEE, 2014.

\bibitem{bash_isit2013}
B.~Bash, S.~Guha, D.~Goeckel, and D.~Towsley, ``Quantum noise limited optical
  communication with low probability of detection,'' in {\em Information Theory
  Proceedings (ISIT), 2013 IEEE International Symposium on}, pp.~1715--1719,
  July 2013.

\bibitem{hou2014effective}
J.~Hou and G.~Kramer, ``Effective secrecy: Reliability, confusion and
  stealth,'' in {\em Information Theory (ISIT), 2014 IEEE International
  Symposium on}, pp.~601--605, 2014.

\bibitem{bash_isit2014}
B.~A. {Bash}, D.~{Goeckel}, and D.~{Towsley}, ``{LPD Communication when the
  Warden Does Not Know When},'' in {\em Information Theory Proceedings (ISIT),
  2014 IEEE International Symposium on}.

\bibitem{sobers2017covert}
T.~V. Sobers, B.~A. Bash, S.~Guha, D.~Towsley, and D.~Goeckel, ``Covert
  communication in the presence of an uninformed jammer,'' {\em IEEE
  Transactions on Wireless Communications}, 2017.

\bibitem{bash2015hiding}
B.~A. Bash, D.~Goeckel, D.~Towsley, and S.~Guha, ``Hiding information in noise:
  Fundamental limits of covert wireless communication,'' {\em IEEE
  Communications Magazine}, vol.~53, no.~12, pp.~26--31, 2015.

\bibitem{bloch2016covert}
M.~R. Bloch, ``Covert communication over noisy channels: A resolvability
  perspective,'' {\em IEEE Transactions on Information Theory}, vol.~62, no.~5,
  pp.~2334--2354, 2016.

\bibitem{soltani2015covert}
R.~Soltani, D.~Goeckel, D.~Towsley, and A.~Houmansadr, ``Covert communications
  on poisson packet channels,'' in {\em 2015 53rd Annual Allerton Conference on
  Communication, Control, and Computing (Allerton)}, pp.~1046--1052, IEEE,
  2015.

\bibitem{soltani2016allerton}
R.~Soltani, D.~Goeckel, D.~Towsley, and A.~Houmansadr, ``Covert communications
  on renewal packet channels,'' in {\em 2016 54th Annual Allerton Conference on
  Communication, Control, and Computing (Allerton)}, IEEE, 2016.

\bibitem{soltani2017towards}
R.~Soltani, D.~Goeckel, D.~Towsley, and A.~Houmansadr, ``Towards provably
  invisible network flow fingerprints,'' in {\em 2017 51st Asilomar Conference
  on Signals, Systems, and Computers}, pp.~258--262, Oct 2017.

\bibitem{gupta_kumar}
P.~Gupta and P.~Kumar, ``The capacity of wireless networks,'' {\em Information
  Theory, IEEE Transactions on}, vol.~46, pp.~388--404, Mar 2000.

\bibitem{francheschetti}
M.~Franceschetti, O.~Dousse, D.~Tse, and P.~Thiran, ``Closing the gap in the
  capacity of wireless networks via percolation theory,'' {\em Information
  Theory, IEEE Transactions on}, vol.~53, pp.~1009--1018, March 2007.

\bibitem{jsac_goeckel}
D.~Goeckel, S.~Vasudevan, D.~Towsley, S.~Adams, Z.~Ding, and K.~Leung,
  ``Artificial noise generation from cooperative relays for everlasting secrecy
  in two-hop wireless networks,'' {\em Selected Areas in Communications:
  Special Issue on Advances in Military Communications and Networking, IEEE
  Journal on}, vol.~29, pp.~2067--2076, December 2011.

\bibitem{vasudevan2010security}
S.~Vasudevan, D.~Goeckel, and D.~F. Towsley, ``Security-capacity trade-off in
  large wireless networks using keyless secrecy,'' in {\em Proceedings of the
  eleventh ACM international symposium on Mobile ad hoc networking and
  computing}, pp.~21--30, ACM, 2010.

\bibitem{capar_infocom}
C.~Capar, D.~Goeckel, B.~Liu, and D.~Towsley, ``Secret communication in large
  wireless networks without eavesdropper location information,'' in {\em
  INFOCOM, 2012 Proceedings IEEE}, pp.~1152--1160, March 2012.

\bibitem{capar_ciss}
C.~Capar and D.~Goeckel, ``Network coding for facilitating secrecy in large
  wireless networks,'' in {\em Information Sciences and Systems (CISS), 2012
  46th Annual Conference on}, pp.~1--6, March 2012.

\bibitem{cormen2009introduction}
T.~H. Cormen, {\em Introduction to algorithms}.
\newblock MIT press, 2009.

\bibitem{lin1983error}
S.~Lin and D.~Costello, ``Error control coding: Fundamentals and
  applications,'' 1983.

\bibitem{klref}
F.~Nielsen and R.~Nock, ``Clustering multivariate normal distributions,'' in
  {\em Emerging Trends in Visual Computing} (F.~Nielsen, ed.), vol.~5416 of
  {\em Lecture Notes in Computer Science}, pp.~164--174, Springer Berlin
  Heidelberg, 2009.

\bibitem{abadir2005matrix}
K.~M. Abadir and J.~R. Magnus, {\em Matrix algebra}, vol.~1.
\newblock Cambridge University Press, 2005.

\bibitem{matrixDing}
J.~Ding and A.~Zhou, ``Eigenvalues of rank-one updated matrices with some
  applications,'' {\em Applied Mathematics Letters}, vol.~20, no.~12, pp.~1223
  -- 1226, 2007.

\bibitem{arkin2015optimal}
E.~Arkin, Y.~Cassuto, A.~Efrat, G.~Grebla, J.~S. Mitchell, S.~Sankararaman, and
  M.~Segal, ``Optimal placement of protective jammers for securing wireless
  transmissions in a geographic domain,'' in {\em Proceedings of the 14th
  International Conference on Information Processing in Sensor Networks},
  pp.~37--46, ACM, 2015.

\bibitem{wu2012optimum}
J.~Wu and N.~Sun, ``Optimum sensor density in distortion-tolerant wireless
  sensor networks,'' {\em IEEE transactions on wireless communications},
  vol.~11, no.~6, pp.~2056--2064, 2012.

\bibitem{mukherjee2014principles}
A.~Mukherjee, S.~A.~A. Fakoorian, J.~Huang, and A.~L. Swindlehurst,
  ``Principles of physical layer security in multiuser wireless networks: A
  survey,'' {\em IEEE Communications Surveys and Tutorials}, vol.~16, no.~3,
  pp.~1550--1573, 2014.

\bibitem{sankararaman2014optimization}
S.~Sankararaman, K.~Abu-Affash, A.~Efrat, S.~D. Eriksson-Bique, V.~Polishchuk,
  S.~Ramasubramanian, and M.~Segal, ``Optimization schemes for protective
  jamming,'' {\em Mobile Networks and Applications}, vol.~19, no.~1,
  pp.~45--60, 2014.

\bibitem{Sankararaman:2012:OSP:2248371.2248383}
S.~Sankararaman, K.~Abu-Affash, A.~Efrat, S.~D. Eriksson-Bique, V.~Polishchuk,
  S.~Ramasubramanian, and M.~Segal, ``Optimization schemes for protective
  jamming,'' in {\em Proceedings of the Thirteenth ACM International Symposium
  on Mobile Ad Hoc Networking and Computing}, MobiHoc '12, (New York, NY, USA),
  pp.~65--74, ACM, 2012.

\bibitem{elsawy2013stochastic}
H.~ElSawy, E.~Hossain, and M.~Haenggi, ``Stochastic geometry for modeling,
  analysis, and design of multi-tier and cognitive cellular wireless networks:
  A survey,'' {\em IEEE Communications Surveys \& Tutorials}, vol.~15, no.~3,
  pp.~996--1019, 2013.

\bibitem{andrews2010primer}
J.~G. Andrews, R.~K. Ganti, M.~Haenggi, N.~Jindal, and S.~Weber, ``A primer on
  spatial modeling and analysis in wireless networks,'' {\em IEEE
  Communications Magazine}, vol.~48, no.~11, 2010.

\bibitem{haenggi2012stochastic}
M.~Haenggi, {\em Stochastic geometry for wireless networks}.
\newblock Cambridge University Press, 2012.

\bibitem{moltchanov2012distance}
D.~Moltchanov, ``Distance distributions in random networks,'' {\em Ad Hoc
  Networks}, vol.~10, no.~6, pp.~1146--1166, 2012.

\end{thebibliography}

\end{document}